\documentclass[12pt]{book}
\usepackage{amssymb}
\usepackage{amsmath}
\usepackage{amsthm}
\usepackage{graphics}
\usepackage{epsfig,amssymb,latexsym,verbatim}
\usepackage{graphicx}
\usepackage{multirow}
\usepackage{multicol}
\usepackage{float}
\usepackage{booktabs}
\usepackage{fancyhdr}
\usepackage{enumerate,verbatim}
\usepackage[sort&compress]{natbib}
\usepackage{rotating}
\usepackage[lined,algonl,boxed]{algorithm2e}
\usepackage{graphicx}
\usepackage{hyperref}
\usepackage{subfigure}
\usepackage{graphicx}
\usepackage{ulem}
\usepackage{url}
\usepackage{threeparttable}
\usepackage{xr}
\usepackage{multirow}
\usepackage{multicol}
\usepackage{wrapfig}
\usepackage{lscape}
\usepackage{rotating}
\usepackage{epstopdf}
\usepackage{subfigure}

\floatstyle{ruled}
\newfloat{algorithm}{tbp}{loa}
\floatname{algorithm}{Algorithm}
\DeclareMathOperator*{\argmin}{arg\,min}

\newcommand{\thetheorem}{{\thesection. \arabic{theorem}}}
\newcommand{\thelemma}{{\thesection. \arabic{lemma}}}
\newcommand{\theproposition}{{\thesection. \arabic{proposition}}}
\newcommand{\thecorollary}{{\thesection. \arabic{corollary}}}
\newtheorem{theorem}{{\sc Theorem}}
\newtheorem{lemma}{{\sc Lemma}}
\newtheorem{corollary}{{\sc Corollary}}
\newtheorem{proposition}{{\sc Proposition}}
\newtheorem{remark}{{\sc Remark}}
\begin{document}

\renewcommand{\baselinestretch}{1.2}
\markboth{\hfill{\footnotesize\rm Jiangyan Wang, Guanqun Cao, Li Wang and Lijian Yang}\hfill}
{\hfill {\footnotesize\rm SCB for Stationary Covariance Function of Dense Functional Data} \hfill}
\renewcommand{\thefootnote}{} $\ $

\fontsize{10.95}{14pt plus.8pt minus .6pt}\selectfont
\vspace{0.8pc} \centerline{\Large\bf Simultaneous Confidence Band for  Stationary}
\centerline{\Large\bf  Covariance Function of Dense Functional
Data} \vspace{.4cm}
\centerline{Jiangyan Wang$^{a}$, Guanqun Cao$^{b}$, Li Wang$^{c}$ and Lijian Yang$^{d}$}
\vspace{.4cm}
\centerline{\it $^{a}$Nanjing Audit University, China, $^{b}$Auburn
University, USA,}
\centerline{\it $^{c}$Iowa State University, USA, and $^{d}$Tsinghua University, China} \vspace{%
.55cm} \fontsize{9}{11.5pt plus.8pt minus .6pt}\selectfont
\footnote{\emph{Address for correspondence}: Li Wang (lilywang@iastate.edu) and Lijian Yang (yanglijian@tsinghua.edu.cn)}

\begin{quotation}
\noindent \textit{Abstract:} The inference via simultaneous confidence band is studied for stationary covariance function of dense functional data. A two-stage estimation procedure is proposed based on spline approximation, the first stage involving estimation of all the individual trajectories and the second  stage involving  estimation of the covariance function through smoothing the empirical covariance function. The proposed covariance estimator is smooth and as efficient as the oracle estimator when all individual trajectories are known. An asymptotic simultaneous confidence band (SCB) is developed for the true covariance function, and the coverage probabilities are shown to be asymptotically correct. Intensive simulation experiments are conducted to demonstrate the performance of the proposed estimator and SCB. The proposed method is also illustrated with a real data example.

\vspace{9pt} \noindent \textit{Key words and phrases:} Confidence band,
Covariance function, Functional data, Stationary.
\end{quotation}

\fontsize{10.95}{14pt plus.8pt minus .6pt}\selectfont

\thispagestyle{empty}

\setcounter{chapter}{1} \setcounter{equation}{0} \renewcommand{%
\theequation}{\arabic{equation}} \renewcommand{\thesection}{\arabic{section}}%
\setcounter{section}{0} 

\section{Introduction}

\label{SEC:intro}

Since \cite{RD91} first coined the term ``functional data analysis'' (FDA), recent years have seen numerous publications emerging in the FDA theory, methods and applications, making it an important area in statistics research. Motivated by specific problems and complex data collected in modern experiments, such as \cite{RS05}, \cite{HE15}, considerable efforts have been made to analyze functional data. The estimation for population mean function and principal component in functional data has been extensively studied, for instance, \cite{C00}, \cite{CYT12}, \cite{FV06}, \cite{GGC13}, \cite{HMW06} and so on.

Related to the smoothness, the second-order structure of random functions can be depicted by the covariance, thus the covariance function is another indispensable ingredient in many areas, such as longitudinal analysis, spatial statistics, and Bayesian hierarchical modeling, see \cite{CLW13}, \cite{DV98}, \cite{HFH94}, \cite{YZCD16} and \cite{YGLW10}. In this sense, \cite{CWLY16} proposed a simultaneous confidence envelope of covariance function for functional data;  \cite{HKR13} proposed a consistent estimator for the long-run covariance operator of stationary time series; \cite{PSS10} considered the estimation of integrated covariance functions, which is required to construct asymptotic confidence intervals and significance tests for the mean vector in the context of stationary random fields. Since the covariance function measures stronger association among variables that are closer to each other, the employment of covariance function is considerably highlighted in spatial data analysis when the geometric structure of the surface is rough and self-similar. A common situation is that the observations are specified via a Gaussian process whose finite-dimensional joint distributions are determined by a valid covariance function; see, for instance, \cite{CLW13}.

Let $\left\{\eta (x),x\in \chi \right\} $ be a stochastic process defined on a compact interval $\chi $, with $\mathrm{E}\int_{\chi}\eta^{2}(x)dx<+\infty $. It is covariance stationary if $G(x,x^{\prime})=C(|x-x^{\prime}|)$, where
\begin{equation}
\label{EQ:Ch}
G(x,x^{\prime})=\mathrm{Cov}\left\{\eta (x),\eta (x^{\prime}) \right\},\quad x,x^{\prime}\in \chi.
\end{equation}
Consider a collection of $n$ trajectories $\left\{\eta_{i}(x)\right\}_{i=1}^{n} $, which are i.i.d realizations of $\eta(x) $, with mean and covariance functions, say $m(x)=\mathrm{E} \{\eta (x)\}$, $G(x,x^{\prime})=\mathrm{Cov}\left\{\eta (x),\eta \left(x^{\prime}\right) \right\} $, respectively. The trajectories $\left\{\eta _{i}(x)\right\}_{i=1}^{n}$ are decomposed as $\eta_{i}(x)=m(x)+Z_{i}(x)$, where $Z_{i}(x)$ can be viewed as a small-scale variation of $x$ on the $i$th trajectory, and is assumed to be a weakly stationary process with $\mathrm{E}Z_{i}(x)=0$ and covariance $G(x,x^{\prime})=\mathrm{Cov}\left\{Z_{i}(x),Z_{i}\left(x^{\prime}\right)\right\} $.

According to classical functional analysis FDA settings, for $G\left(\cdot,\cdot\right) $, there exist eigenvalues $\lambda _{1}\geq\lambda _{2}\geq \cdots \geq 0$ and corresponding eigenfunctions $\left\{\psi _{k}\right\}_{k=1}^{\infty }$, the latter being an orthonormal basis of $L^{2}\left(\chi\right) $, such that $\sum_{k=1}^{\infty}\lambda_{k}<\infty $, $G\left(x,x^{\prime}\right) =\sum_{k=1}^{\infty }\lambda_{k}\psi _{k}(x)\psi_{k}\left(x^{\prime}\right) $, and $\int G\left(x,x^{\prime}\right) \psi_{k}\left(x^{\prime}\right)dx^{\prime}=\lambda _{k}\psi _{k}(x)$. The standard process $\eta (x)$, $x\in \chi$, then allows the well-known Karhunen-Lo\`{e}ve $L^{2}$\ representation $\eta(x)=m(x)+\sum_{k=1}^{\infty}\xi _{k}\phi _{k}(x)$, in which the random coefficients $\xi _{k}$, called functional principal component (FPC) scores, are uncorrelated with each other of mean $0$\ and variance $1$. The rescaled eigenfunctions, $\phi _{k}$, called {FPC}, satisfy that $\phi _{k}=\sqrt{\lambda _{k}}\psi _{k}$ and $\int \left\{\eta (x) -m(x) \right\} \phi_{k}(x) dx=\lambda _{k}\xi _{k}$, for $k\geq 1$. The $i$th process $\eta _{i}(x)$, $x\in \chi$, is written as $\eta _{i}(x)=m(x)+\sum_{k=1}^{\infty }\xi _{ik}\phi _{k}(x) $, in which the
FPC scores $\left\{\xi _{ik}\right\} _{k=1}^{\infty }$, $i=1,\ldots,n$, are i.i.d copies of $\left\{\xi _{k}\right\} _{k=1}^{\infty }$. Although the sequences $\left\{\lambda _{k}\right\} _{k=1}^{\infty }$, $\left\{\phi_{k}\left(\cdot \right) \right\} _{k=1}^{\infty }$\ and $\left\{\xi_{ik}\right\} _{k=1}^{\infty }$ exist mathematically, they are either unknown or unobservable.

The actual observed functional data are noisy sampled points from trajectories $\left\{ \eta _{i}(x)\right\} _{i=1}^{n}$. Let $\{(Y_{ij},X_{ij}),~1\leq i\leq n,~1\leq j\leq N\}$ be repeated measurements on a random sample of $n$\ experimental units, where $Y_{ij}$ is the response observed on the $i$th unit at value $X_{ij}$ of the variable $x$. The observed data can be modeled as
\begin{equation*}
Y_{ij}=\eta _{i}\left( X_{ij}\right) +\sigma \left( X_{ij}\right)
\varepsilon _{ij}=m(X_{ij})+Z_{i}\left( X_{ij}\right) +\sigma \left(
X_{ij}\right) \varepsilon _{ij}, ~1\leq i\leq n, ~1\leq j\leq N,
\end{equation*}
where $\varepsilon _{ij}$, independent of $Z_{i}(\cdot )$'s, are {i.i.d} random errors with mean $0$ and variance $1$, and $\sigma ^{2}\left( \cdot \right) $ is the variance function of the measurement errors. For the data considered in this paper, without loss of generality, $\eta _{i}(\cdot )$ is assumed to be recorded on a regular grid in $\chi =[0,1]$, and $X_{ij}=x_{j}=$ $j/N$, $1\leq j\leq N$. This type of functional data was considered in \cite{LH10}, \cite{CSD09} and \cite{CWLY16}, among others. Consequently, our observed data can be written as
\begin{equation}
Y_{ij}= m(j/N)+Z_{i}\left( j/N\right) +\sigma \left( j/N\right) \varepsilon
_{ij},~1\leq i\leq n,~1\leq j\leq N.  \label{DEF:model j/N}
\end{equation}

It would not be a far stretch if the sample points for the $i$th subject $Y_{ij}$ admit the structure of a nonstationary or locally stationary time series, as in \cite{FO09} and \cite{SFJ10}. One may further ask if these random observations at regular grid points would even admit the structure of stationary time series. \cite{CWLY16}, for instance, concluded that the Tecator near-infrared spectra data is nonstationary based on the simultaneous confidence envelope for the covariance function. There are, however, interesting functional data for which the covariance function exhibits stationarity, because a closer relationship between the geometric structures and covariance function relies on the stationary assumption. In particular, the stationary random processes or fields are prominent in the analysis of 1D and 2D signals; see, for instance, the important spatial covariance model studied in Mat\'{e}rn random fields, stationary multivariate time series and the stationary spectral-space statistics studied in {physics} such as \cite{TG17}. As a fundamental issue, the study of covariance structure in stationary stochastic processes can be applied to a wide range of areas such as hydrosciences and geostatistics.

{Typically, it is difficult to interpret the covariance function in the case of FDA and longitudinal data analysis. The estimation strategies of the covariance function generally fall into two categories: direct smoothing and mixed-effects type of approaches, depending on whether there is a requirement of covariance reconstruction. In the literature, the mixed-effects type approaches have been considered intensively. For functional data, the functional principal component (FPC) analysis has become one of the first-line methods; see, for instance, the nonparametric estimation of covariance functions: \cite{HMW06}, \cite{LH10} and \cite{CWLY16}, among others. For longitudinal data, \cite{JHS00} considered reduced rank spline mixed-effects models to describe the modes of variation; \cite{YMW05b} estimated the covariance structure and the FPC score by a conditioning step;  \cite{PP09b} proposed a geometric approach within the framework of marginal maximum likelihood estimation by requiring the trajectories are i.i.d. Gaussian processes.}

However, to our best knowledge, the direct smoothing study is far from obvious in the FDA setting. This motivates us to develop an approach of direct smoothing for the covariance function. In this paper, we consider a nonparametric estimation of the covariance structure, which is useful either as a guide to the formulation of a parametric model or as the basis for formal inference without imposing parametric assumptions. Our estimation procedure is carried out by spline approximation, where the first step involves the estimation of the $i$th trajectory and the mean function, based on dense functional data,  which is a vital feature that we can borrow strength; the second step estimates the covariance function through smoothing using the residuals of the first step via direct smoothing. The proposed covariance estimator is smooth and as efficient as the oracle estimator when all trajectories $\eta_{i}(\cdot)$ and the mean $m(\cdot)$ are known.

After estimating the covariance function, our next concern is to provide an inferential tool to further examine the covariance structure. Although a straightforward way is to conduct a hypothesis test, it is not well developed as other FDA methods, due to the difficulty of the infinite-dimensionality of the functional space. {The existing methods mainly focus on testing of the mean functions for functional data, such as the pointwise t-test provided by \cite{RS05}. However, the hypothesis test for covariance receives relatively little attention even though a global conclusion is often more desirable in real data analysis.} In this line, \cite{GZZ18} proposed a supremum-norm based test for the equality of several covariance functions. However, it is a general-purpose smoother that is not designed specifically for covariance operators and it ignores that the smoothness of trajectories in FDA setting, hence the simple averaging of the observations is insufficient to meet the manifold needs in reality.

To surmount these challenges, we develop an asymptotic simultaneous confidence band (SCB), which can be used to test the adequacy and validity of certain covariance models. Specifically, the null hypothesis is $H_{0}: C(h)=C\left(h;\mathbf{\theta}\right)$ for some $\mathbf{\theta }\in \Theta$. An SCB is an intuitive and theoretically reliable tool for global inference of functions. {For example,  in the FDA framework, \cite{CWLY16} proposed SCBs for the covariance functions,  and \cite{GGC13} derived a corrected SCB using principal component. 
The contribution of this paper is twofold. First, it provides the methodology and asymptotic theory for the estimation of the covariance $C(\cdot)$ in the framework of stationary dense functional data under mild assumptions; second, the estimator of $C(\cdot)$ is accompanied by a procedure for constructing asymptotic SCB.}

The rest of the paper is organized as follows. In Section \ref{Sec:cov-est}, we introduce the two-stage B-spline estimation procedure for the covariance function. Section \ref{Sec:asymptotics} shows that the proposed estimator is as efficient as if all the $n$ trajectories $\eta_{i}(\cdot)$ and the mean function $m(\cdot)$ are known over the entire data range. Section \ref{Sec:bands} presents the asymptotic SCB for the covariance function, and describes the implementation of the SCB. Section \ref{sec:simulation} carries out intensive simulation studies to evaluate the finite sample performance of the proposed SCB. The methodology is verified by a real data example in Section \ref{Sec:real data}.  Technical lemmas
and proofs are deferred to Appendices A and B. More simulation studies are
carried out in Appendix C. Additional analysis of real data is given in Appendix D.

%
\setcounter{chapter}{2} \renewcommand{\theproposition}{{2.%
\arabic{proposition}}} \renewcommand{\thesection}{\arabic{section}} %
\renewcommand{\thesubsection}{2.\arabic{subsection}} \setcounter{lemma}{0} %
\setcounter{section}{1} \setcounter{theorem}{0} \setcounter{proposition}{0} %
\setcounter{corollary}{0}

\section{B-spline covariance function estimation}

\label{Sec:cov-est}

In this section, we describe the estimation procedure for the covariance
function $C(\cdot)$. If the small-scale variation of $x$, $Z_{i}(x)=\eta
_{i}(x) -m(x) $, $1\leq i\leq n$, $x\in ${$[0,1]$}, on the $i$th trajectory
could be observed, one would estimate the covariance as
\begin{equation}
\widetilde{C}(h) =\frac{1}{1-h}\int_{0}^{1-h}\frac{1}{n}%
\sum_{i=1}^{n}Z_{i}(x)Z_{i}(x+h)dx,\quad h\in \left[ 0,h_{0}\right] ,
\label{DEF:C-tilde}
\end{equation}%
where $h_{0}\in \left(0,1\right) $ is a pre-specified upper limit.

Since {$\{Z_{i}(x)\}_{i=1}^{n}$, $x\in ${$[0,1]$}, are} unobserved, the
above estimator $\widetilde{C}(h) $ is ``infeasible" in practice. In this
paper, we propose to estimate the covariance function based on the following
residuals
\begin{equation}
\widehat{Z}_{i}(x)=\widehat{\eta}_{i}(x) -\widehat{m}(x),\quad 1\leq i\leq
n,\quad x\in {[0,1],}  \label{DEF:Z-residuals}
\end{equation}
where $\widehat{\eta}_{i}(x)$ and $\widehat{m}(x)$ are the estimators of $%
\eta _{i}(x)$ and $m(x)$.

In such case, a sample-based consistent estimator can be employed, such as
the spline smoother proposed in \cite{CYT12}. Denote {by} $\left\{
t_{\ell}\right\} _{\ell=1}^{J_{s}}$ a sequence of equally-spaced points, $%
t_{\ell}=\ell/\left( J_{s}+1\right) $, $1\leq \ell\leq J_{s}$, $%
0<t_{1}<\cdots <t_{J_{s}}<1$, called interior knots{,} which divide the
interval $[0,1]$\ into $\left( J_{s}+1\right) $ equal subintervals $I_{0}=%
\left[ 0,t_{1}\right) $, $I_{\ell}=\left[ t_{\ell},t_{\ell+1}\right) $, $%
\ell=1,\ldots ,J_{s}-1$, $I_{J_{s}}=\left[ t_{J_{s}},1\right] $. For any
positive integer $p$, let $t_{1-p}=\cdots =t_{0}=0$ and $1=t_{J_{s}+1}=%
\cdots =t_{J_{s}+p}$ be auxiliary knots. Let $\mathcal{S}^{\left( p-2\right)
}=\mathcal{S}^{\left( p-2\right) }\left[ 0,1\right] $ be the polynomial
spline space of order $p$ on $I_{\ell}$, $\ell=0,\ldots ,J_{s}$, which
consists {of} all $\left( p-2\right) $ times continuously differentiable
functions {on $[0,1]$} that are polynomials of degree $\left( p-1\right) $ {%
on subintervals $I_\ell$, $\ell=0,\ldots,J_s$}. Following the notation in
\cite{deB01}, we denote by $\{B_{\ell,p}(x),1\leq \ell\leq J_{s}+p\}$ the $p$%
th order B-spline basis functions of $\mathcal{S}^{\left( p-2\right) }$,
hence $\mathcal{S}^{\left( p-2\right) }=\left\{ \left.
\sum_{\ell=1}^{J_{s}+p}\lambda _{\ell,p}B_{\ell,p}(x)\right\vert \lambda
_{\ell,p}\in \mathbb{R},x\in \lbrack 0,1]\right\} $.

The $i$th unknown trajectory $\eta _{i}(x) $ is estimated by using the
following formula
\begin{equation}
\widehat{\eta}_{i}\left(\cdot \right) =\argmin_{g\left(\cdot \right) \in
\mathcal{S}^{\left(p-2\right)
}}\sum_{j=1}^{N}\left\{Y_{ij}-g\left(x_{j}\right) \right\} ^{2}.
\label{EQ:eta-i-hat}
\end{equation}%
One can then estimate the unknown mean function $m\left(\cdot\right) $ {as}
\begin{equation}
\widehat{m}(x) =n^{-1}\sum_{i=1}^{n}\widehat{\eta}_{i}(x),  \label{EQ:mhat}
\end{equation}
{and obtain} the covariance estimator
\begin{equation}
\widehat{C}(h) =\frac{1}{1-h}\int_{0}^{1-h}\frac{1}{n}\sum_{i=1}^{n}
\widehat{Z}_{i}(x)\widehat{Z}_{i}(x+h)dx,\quad h\in \left[ 0,h_{0}\right] .
\label{EQ:C-hat}
\end{equation}

\setcounter{chapter}{3} \renewcommand{\thetheorem}{{\arabic{theorem}}} %
\renewcommand{\thelemma}{{\arabic{lemma}}} \renewcommand{%
\theproposition}{{\arabic{proposition}}} \renewcommand{\thesection}{%
\arabic{section}} \renewcommand{\thesubsection}{3.\arabic{subsection}} %
\setcounter{section}{2}

\section{Asymptotic Properties}
\label{Sec:asymptotics}

This section studies the asymptotic properties for the proposed estimators.


\subsection{Assumptions}
\label{subsec:assumptions}

To study the asymptotic properties of the two-step spline estimator $%
\widehat{C}(\cdot )$, one needs some assumptions. Throughout the paper, 
for sequences $a_{n}$ and $b_{n}$, denote $a_{n}\asymp b_{n}$ 
if $a_{n}$ and $b_{n}$ are asymptotically equivalent. For
any function $\varphi (x)$ defined on a domain $\mathcal{\chi } $, denote $%
\left\Vert \varphi \right\Vert_{\infty}=\sup_{x\in \mathcal{\chi }%
}\left\vert \varphi (x) \right\vert $, and $\varphi ^{(q)}(x)$ its $q$th
order derivative with respect to $x$. For any $L^{2}$ integrable functions $%
\phi (x) $ and $\varphi(x)$, $x\in \mathcal{\chi }$, define their
theoretical inner product as $\left\langle \phi ,\varphi \right\rangle
=\int_{\mathcal{\chi }}\phi (x) \varphi (x) dx$, and the empirical inner
product as $\left\langle \phi ,\varphi \right\rangle
_{N}=N^{-1}\sum_{j=1}^{N}\phi \left(j/N\right) \varphi \left(j/N\right)$.
The related theoretical and empirical norms are $\left\Vert \phi \right\Vert
_{2}^{2}=\left\langle \phi ,\phi \right\rangle $, $\left\Vert \phi
\right\Vert _{2,N}^{2}=\left\langle \phi ,\phi \right\rangle _{N}$.

For a non-negative integer $q$ and a real number $\mu \in \left(0,1\right] $%
, write $\mathcal{H}^{\left(q,\mu \right) }[0,1]$ as the space of $\mu $-H%
\"{o}lder continuous functions, i.e.,
\begin{equation*}
\mathcal{H}^{\left(q,\mu \right) } [0,1] =\left\{\varphi :[0,1]\rightarrow%
\mathbb{\ R}\left\vert \left\Vert \varphi \right\Vert _{q,\mu }=\sup_{x,y\in
[0,1],x\neq y}\left\vert \frac{\varphi ^{\left(q\right) }(x) -\varphi
^{\left(q\right) }\left(y\right) }{\left\vert x-y\right\vert ^{\mu }}%
\right\vert <+\infty \right. \right\} .
\end{equation*}

We next introduce some technical assumptions.

\begin{enumerate}
\item[(A1)] There exist an integer $q>0$ and a constant $\mu \in \left(0,1%
\right] $, such that the regression function $m\left(\cdot \right) \in
\mathcal{H}^{\left(q,\mu \right) }\left[ 0,1\right]$. {In the following, one
denotes $p^*=q+\mu$.}

\item[(A2)] The standard deviation function $\sigma (\cdot )\in \mathcal{H}%
^{\left( 0,\nu \right) }[0,1]$ for positive index $\nu \in (0,1]$ and for
some constants $M_{\sigma },$ $M_{0}>0$, $\sup_{x\in \lbrack 0,1]}\sigma
(x)\leq M_{\sigma }$, $\sup_{h\in \left[ 0,h_{0}\right] }\left\vert
C(h)\right\vert \leq M_{0} $.

\item[(A3)] There exists a constant $\theta >0$, such that as $N\rightarrow
\infty $, $n=n\left( N\right) \rightarrow \infty $, $n=\mathcal{O}\left(
N^{\theta }\right) $.

\item[(A4)] The rescaled FPCs $\phi _{k}\left( \cdot \right) \in \mathcal{H}%
^{\left( q,\mu \right) }\left[ 0,1\right] $ with $\sum_{k=1}^{\infty
}\left\Vert \phi _{k}\right\Vert _{q,\mu }<+\infty $, $\sum_{k=1}^{\infty
}\left\Vert \phi _{k}\right\Vert _{\infty }<+\infty $; for increasing
positive integers $\left\{ k_{n}\right\} _{n=1}^{\infty }$, as $n\rightarrow
\infty $, $\sum_{k_{n}+1}^{\infty }\left\Vert \phi _{k}\right\Vert _{\infty
}={\ \scriptstyle{\mathcal{O}}}(n^{-1/2})$ and $k_{n}=\mathcal{O}\left(
n^{\omega }\right) $ for some $\omega >0$.

\item[(A5)] There are positive constants $c_{1}, c_{2} \in \left( 0,+\infty \right)$, $\gamma _{1},~\gamma
_{2}\in \left( 1,+\infty \right) ,~\beta _{1},~\beta _{2}\in \left(
0,1/2\right) $, and iid $N\left( 0,1\right) $ variables $\left\{
U_{ij,\varepsilon }\right\} _{i=1,j=1}^{n,N}$, $\left\{ U_{ik,\xi }\right\}
_{i=1,k=1}^{n,k_{n}}$ such that $\min \left\{ {\frac{2\left( 1-\beta
_{2}\right) p^{\ast }}{3+\left( 1+\beta _{1}\right) p^{\ast }},}\frac{%
2\left( \nu -\beta _{2}\right) }{\left( 1+\beta _{1}\right) }\right\} ${$%
>\theta $ for the index }$\nu ${\ in Assumption (A2), }$p^{\ast }$ in
Assumption (A1){, and}
\begin{align*}
\Pr \left\{ \max_{1\leq k\leq k_{n}}\max_{1\leq t\leq n}\left\vert
\sum_{i=1}^{t}\xi _{ik}-\sum_{i=1}^{t}U_{ik,\xi }\right\vert >n^{\beta
_{1}}\right\} & <c_{1}n^{-\gamma _{1}}, \\
\Pr \left\{ \max_{1\leq i\leq n}\max_{1\leq t\leq N}\left\vert
\sum_{j=1}^{t}\varepsilon _{ij}-\sum_{j=1}^{t}U_{ij,\varepsilon }\right\vert
>N^{\beta _{2}}\right\} & <c_{2}N^{-\gamma _{2}}.
\end{align*}

\item[(A5')] The {iid} variables $\left\{ \varepsilon _{ij}\right\} _{i\geq
1,j\geq 1}$ are independent of $\left\{ \xi _{ik}\right\} _{i\geq 1,k\geq 1}$%
. The number of distinct distributions for all FPC scores $\left\{ \xi
_{ik}\right\} _{i\geq 1,k\geq 1}$\ is finite. There exist constants $%
r_{1}>4+2\omega $, $r_{2}>4+2\theta $, for $\omega $ in Assumption (A4) and $%
\theta $ in Assumption (A3), such that $\mathrm{E}|\varepsilon
_{11}|^{r_{2}} $ and $\mathrm{E}|\xi _{1k}|^{r_{1}},k=1,2,\ldots $ are
finite.

\item[(A6)] The spline order $p\geq p^{\ast }$, the number of interior knots
$J_{s}\asymp N^{\gamma }d_{N}$ for some $\tau >0$ with $d_{N}+d_{N}^{-1}=\mathcal{%
O}\left( \log ^{\tau }N\right) $ as $N\rightarrow \infty $, and for $p^{\ast
}$ in Assumption (A1), $\nu $ in Assumption (A2), $\theta $ in Assumption
(A3), $\beta _{1},\beta _{2}$ and $\gamma _{1}$ in Assumption (A5)
\begin{equation*}
\max \left\{ \frac{5\theta }{4p^{\ast }},\frac{\theta +\left( \gamma
_{1}+1+\omega \right) ^{-1}8\theta \beta _{1}}{2p^{\ast }}, 1-\nu \right\}
<\gamma <1-\frac{\theta }{2}-\beta _{2}-\frac{\theta }{2}\beta _{1}\text{.}
\end{equation*}
\end{enumerate}

Assumptions (A1)--(A2) are standard in the literature, see \cite{CYT12} and
\cite{SY09} for instance. In particular, (A1) and (A4) control the size of
the bias of the spline smoother for $m(\cdot)$ and $\phi_k(\cdot)$.
Assumption (A2) ensures the variance function is a uniformly bounded
function. Assumption (A3) regulates that {sample size} $n$ grows as a
fractional power $\theta$ of $N$, the number of observations per subject.
The bounded smoothness of the principal components is guaranteed in
Assumption (A4). Assumption (A5) provides a strong approximation of
estimation errors and FPC scores. Assumption (A5') is an elementary
assumption to guarantee the high level Assumption (A5). {It is noteworthy
that} the smoothness of our estimator is controlled by the knots of the
splines. Assumption (A6) specifies the requirement that the number of knots
has to meet {for the B-spline smoothing}.

\begin{remark}
\label{REM:parameters}
These assumptions are mild conditions that can be satisfied in many practical situations. One simple and reasonable setup for the above parameters $q$, $\mu $, $\theta $, $p$, $\gamma $ can be as follows: $q+\mu =p^{\ast }=4$, $\nu =1$, $\theta =1$, $p=4$ (cubic spline), $\gamma =3/8$, $d_{N}\asymp \log \log N$. These constants are used as defaults in implementing the method; see Section \ref{Sec:bands}.
\end{remark}


\subsection{Oracle efficiency}

We now show that the proposed two-step estimator $\widehat{C}(\cdot)$
defined in (\ref{EQ:C-hat}) is oracle-efficient, i.e., it is as efficient as
if all trajectories $\eta_i(\cdot)$ are known over the entire data range. To
begin with, we first investigate the asymptotic property of the infeasible
covariance estimator $\widetilde{C}(h)$. Denote by $\Delta (h)=\widetilde{C}%
(h) -C(h)$, $h \in [0,h_0]$.

According to the definition of $C(h) $ and $\widetilde{C}(h)$ in (\ref{EQ:Ch}%
) and (\ref{DEF:C-tilde}), one has
\begin{align*}
C(h) &=\sum_{k=1}^{\infty }\sum_{k^{\prime}=1}^{\infty }\mathrm{E}\left(\xi
_{ik}\xi _{ik^{\prime}}\right) \frac{1}{1-h}\int_{0}^{1-h}\phi _{k}(x) \phi
_{k^{\prime}}\left(x+h\right) dx \\
&=\frac{1}{1-h}\int_{0}^{1-h}\sum_{k=1}^{\infty }\phi _{k}(x) \phi _{k}(x+h)
dx, \\
\widetilde{C}(h) &=\frac{1}{n(1-h)}\sum_{i=1}^{n}\sum_{k=1}^{\infty
}\sum_{k^{\prime}=1}^{\infty }\xi _{ik}\xi _{ik^{\prime}}\int_{0}^{1-h}\phi
_{k}(x) \phi _{k^{\prime}}(x+h) dx.
\end{align*}
Thus,
\begin{equation*}
\Delta (h) =\frac{1}{1-h}\sum_{k,k^{\prime}=1}^{\infty }\left(\bar{\xi}%
_{\cdot kk^{\prime}}-\delta _{kk^{\prime}}\right) \int_{0}^{1-h}\phi _{k}(x)
\phi _{k^{\prime}}\left(x+h\right) dx,
\end{equation*}%
where $\bar{\xi}_{\cdot kk^{\prime}}=n^{-1}\sum_{i=1}^{n}\xi _{ik}\xi
_{ik^{\prime}}$, and $\delta _{kk^{\prime}}=1$ for $k=k^{\prime}$ and $0$
otherwise.

Then the asymptotic mean squared error of the infeasible covariance
estimator $\widetilde{C}(\cdot)$ is provided in Theorem \ref{THM:MSE} below.

\begin{theorem}
\label{THM:MSE} Under Assumptions (A1)--(A6), $\sup_{h \in \left[0,h_{0}\right]}\vert n\mathrm{E}\left\{\Delta (h)\right\}
^{2}-\Xi (h)\vert=o(1)$, in which
\begin{align}
\Xi (h) =&\sum_{k,k^{\prime}=1}^{\infty }\left\{\frac{1}{1-h}%
\int_{0}^{1-h}\phi _{k}(x) \phi _{k^{\prime}}(x+h) dx\right\} ^{2}  \notag \\
&+\sum_{k,k^{\prime}=1}^{\infty }\left\{\frac{1}{1-h}\int_{0}^{1-h}\phi
_{k}(x) \phi _{k^{\prime}}(x+h) dx\right\} \left\{\frac{1}{1-h}%
\int_{0}^{1-h}\phi _{k^{\prime}}(x) \phi _{k}(x+h) dx\right\}  \notag \\
&+\sum_{k=1}^{\infty }\left(\mathrm{E}\xi _{1k}^{4}-3\right) \left\{\frac{1}{%
1-h}\int_{0}^{1-h}\phi _{k}(x) \phi _{k}\left(x+h\right) dx\right\} ^{2}.
\label{EQ:Variance MSE}
\end{align}
\end{theorem}

\begin{remark}
By rewriting $\Xi (h)$, one has
\begin{align*}
\Xi (h)=& \sum_{k=1}^{\infty }\left( \mathrm{E\xi _{1k}^{4}-1}\right)
\left\{ \frac{1}{1-h}\int_{0}^{1-h}\phi _{k}(x)\phi _{k}(x+h)dx\right\} ^{2}
\\
& +\sum_{k<k^{\prime }}^{\infty }\left[ \frac{1}{1-h}\left\{
\int_{0}^{1-h}\phi _{k}(x)\phi _{k^{\prime }}(x+h)dx+\int_{0}^{1-h}\phi
_{k^{\prime }}(x)\phi _{k}\left( x+h\right) dx\right\} \right] ^{2}.
\end{align*}%
Following from (3.2) in \cite{CWLY16},
\begin{equation*}
V\left( x,x+h\right) =\sum_{k<k^{\prime }}^{\infty }\left\{ \phi _{k}(x)\phi
_{k^{\prime }}(x+h)+\phi _{k^{\prime }}(x)\phi _{k}(x+h)\right\}
^{2}+\sum_{k=1}^{\infty }\phi _{k}^{2}(x)\phi _{k}^{2}(x+h)\left( \mathrm{E}%
\xi _{1k}^{4}-1\right) ,
\end{equation*}%
thus, $\left( 1-h\right) ^{-1}\int_{0}^{1-h}V\left( x,x+h\right) dx\geq \Xi
(h)$, $h\in \left[ 0,h_{0}\right] $. {Therefore, if the covariance function
is stationary, the infeasible estimator $\widetilde{C}(\cdot )$ is more
efficient than the covariance estimator given in \cite{CWLY16}.}
\end{remark}

\begin{proposition}
\label{Pro: Uniform Ctilde-C} Under Assumptions (A1)--(A6), as $N\rightarrow
\infty $, $\sqrt{n}\Delta (\cdot )\rightarrow _{D}\zeta (\cdot )$, where $%
\zeta (\cdot )$ is a Gaussian process defined on $\left[ 0,h_{0}\right] $
such that {$\mathrm{E}\zeta (h)=0$, $\mathrm{E}\zeta ^{2}(h)=\Xi (h)$}, with
covariance function
\begin{align*}
\Omega \left( h,h^{\prime }\right) =& \mathrm{Cov}\left( \zeta (h),\zeta
\left( h^{\prime }\right) \right) =(1-h)^{-1}(1-h^{\prime })^{-1} \\
& \times \Bigg\{\int_{0}^{1-h}\int_{0}^{1-h^{\prime }}\sum_{k,k^{\prime
}=1}^{\infty }\phi _{k}(x)\phi _{k}\left( x^{\prime }\right) \phi
_{k^{\prime }}(x+h)\phi _{k^{\prime }}\left( x^{\prime }+h^{\prime }\right)
dxdx^{\prime } \\
& +\int_{0}^{1-h}\int_{0}^{1-h^{\prime }}\sum_{k,k^{\prime }=1}^{\infty
}\phi _{k}(x)\phi _{k}\left( x^{\prime }+h^{\prime }\right) \phi _{k^{\prime
}}\left( x+h\right) \phi _{k^{\prime }}\left( x^{\prime }\right)
dxdx^{\prime } \\
& +\int_{0}^{1-h}\int_{0}^{1-h^{\prime }}\sum_{k=1}^{\infty }\left( \mathrm{E%
}\xi _{1k}^{4}-3\right) \phi _{k}(x)\phi _{k}(x+h)\phi _{k}\left( x^{\prime
}\right) \phi _{k}\left( x^{\prime }+h^{\prime }\right) dxdx^{\prime }\Bigg\}%
,
\end{align*}%
for any $h,h^{\prime }\in \left[ 0,h_{0}\right] $.
\end{proposition}

The proof is deferred to the Appendix. Although the oracle smoother $%
\widetilde{C}(\cdot)$ enjoys the desirable theoretical property, it is not a
statistic since $Z_{i}(x)=\eta _{i}(x) -m(x) $ is unknown. According to
Proposition \ref{PROP:Uniform Chat-Ctilde} below, the price for using $%
\widehat{Z}_{i}(x)=\widehat{\eta}_{i}(x) -\widehat{m}(x)$ in place of $%
Z_{i}(x)$ in the covariance estimator is asymptotically negligible, that is,
two-step estimator $\widehat{C}(\cdot )$ is as efficient as the infeasible
estimator $\widetilde{C}(\cdot )$.

\begin{proposition}
\label{PROP:Uniform Chat-Ctilde} Under Assumptions (A1)--(A6), $
\sup_{h\in \left[ 0,h_{0}\right] }\vert \widehat{C}(h)- \widetilde{C}%
(h)\vert ={\scriptstyle{\mathcal{O}}}_{p}(n^{-1/2})$.
\end{proposition}

Combining the above two propositions, {we obtain the following result}.

\begin{theorem}
\label{THM:Chat} Under Assumptions (A1)--(A6),
$\sup_{h\in \left[ 0,h_{0}\right] }\vert \widehat{C}(h)-C(h)-\Delta
(h)\vert ={\scriptstyle{\mathcal{O}}}_{p}(n^{-1/2})$.

\end{theorem}

Theorem \ref{THM:Chat} indicates that $\Delta (h)$ is the leading term of $%
\widehat{C}(h)-C(h)$.

\setcounter{chapter}{4} \renewcommand{\thesection}{\arabic{section}} %
\renewcommand{\thetable}{{\arabic{table}}} \setcounter{table}{0} %
\renewcommand{\thefigure}{\arabic{figure}} \setcounter{figure}{0} %
\renewcommand{\thesubsection}{4.\arabic{subsection}} \setcounter{section}{3}

\section{Simultaneous confidence band}

\label{Sec:bands}

In this section, we construct the SCB for the covariance function $C(\cdot)$.


\subsection{Asymptotic SCB}
\label{subsec:asymCB}

Next theorem presents the asymptotic behavior of the maximum of the
normalized deviation of the covariance estimator $\widehat{C}(\cdot)$, which
sheds the lights on how to construct the asymptotic SCB for $C(\cdot)$.

\begin{theorem}
\label{THM:band} Under Assumptions (A1)--(A6), for any $\alpha \in \left(
0,1\right) $,
\begin{equation*}
\lim_{N\rightarrow \infty }\Pr \left\{ \sup_{h\in \left[ 0,h_{0}\right]
}n^{1/2}\left\vert \widehat{C}(h)-C(h)\right\vert \Xi (h)^{-1/2}\leq
Q_{1-\alpha }\right\} =1-\alpha ,
\end{equation*}%
\begin{equation*}
\lim_{N\rightarrow \infty }\Pr \left\{ n^{1/2}\left\vert \widehat{C}%
(h)-C(h)\right\vert \Xi (h)^{-1/2}\leq z_{1-\alpha /2}\right\} =1-\alpha
,\quad \forall h\in \left[ 0,h_{0}\right] ,
\end{equation*}%
where $Q_{1-\alpha }$ is the $100\left( 1-\alpha \right) ^{th}$ percentile
of the absolute maxima distribution of $\zeta (h)\Xi ^{-1/2}(h)$, while $%
z_{1-\alpha /2}$ is denoted as the $100\left( 1-\alpha /2\right) ^{th}$
percentile of the standard normal distribution, and $\zeta (h)$ is the mean
zero Gaussian process defined in Proposition \ref{Pro: Uniform Ctilde-C}.
\end{theorem}

Theorem \ref{THM:band} is a direct result of Propositions \ref{Pro: Uniform
Ctilde-C}, \ref{PROP:Uniform Chat-Ctilde} and Theorem \ref{THM:Chat}, thus
the proof is omitted.

\begin{corollary}
\label{COR:Asymp mhatx} Under Assumptions (A1)--(A6), an asymptotic $%
100\left(1-\alpha \right) \%$ exact {SCB} for $C(\cdot)$ is $\widehat{C}%
(h)\pm n^{-1/2}Q_{1-\alpha }\Xi ^{1/2}(h)$, $h\in \left[0,h_{0}\right]$.
While an asymptotic pointwise confidence band for $C(\cdot)$ is given by $%
\widehat{C}(h)\pm n^{-1/2}z_{1-\alpha /2}\Xi ^{1/2}(h)$, $h\in \left[0,h_{0}%
\right]$.
\end{corollary}

{Note that} the percentile $Q_{1-\alpha }$ and the variance function $\Omega
\left(h,h^{\prime}\right) $ have to be estimated from the data. These issues
are addressed in Section \ref{subsec:variance_percentile}.


\subsection{Knots selection}
\label{subsec:knots select}

In spline smoothing, the number of knots is often treated as an unknown tuning parameters, and the fitting results can be sensitive to it. Though in the literature there is no optimal method to choose $J_{s}$, we recommend the following two ways: (a) criterion-based selection strategies such as Generalized Cross Validation (GCV) and Bayesian Information Criterion (BIC), {and the candidate pool for $J_{s}$ is all the integers between $1$ and $J_{s^{\ast}}$, where $J_{s}^{\ast }=\min \left\{10,\lfloor n/4\rfloor \right\} $; (b) formula based selection strategies stated in Remark \ref{REM:parameters}, specifically, we seek $J_{s}$ that satisfies Assumption (C6) such that $J_{s}\asymp N^{\gamma}d_{N}$. In practice, the smoothness order $\left( q,\mu \right) $ of $m\left( \cdot \right) $ and $\phi_{k}\left( \cdot \right) $ are taken as default $(3,1)$ or $(4,0)$ with a matching spline order $p=4$ (cubic spline). Therefore, we suggest $J_{s}=\lfloor cN^{\gamma}\left\{ \log \log \left( N\right) \right\} ^{\gamma}\rfloor$ for some positive constant $c$. Note that the default of parameter $\gamma =3/8$ satisfies the condition given in Assumption (C6). In our extensive simulation studies, we find that $c=0.8$ is a good choice for the tuning parameter.  Both methods give very similar estimators and SCBs in our numerical studies.}


\subsection{FPC analysis}
\label{subsec:fpca}

We now describe how to obtain the covariance function $\widehat{G}%
\left(\cdot, \cdot\right) $, and its eigenfunctions $\widehat{\phi}%
_{k}(\cdot)$ and eigenvalues $\widehat{\lambda}_{k}$ in the FPC analysis.

We estimate $G(\cdot,\cdot)$ by
\begin{equation}  \label{EQ:Ghat}
\widehat{G}\left( x,x^{\prime }\right)= n^{-1}\sum_{i=1}^n\widehat{Z}_i(x)%
\widehat{Z}_i(x^{\prime })=\sum_{s=1}^{J_{s}+p}\sum_{s^{\prime }=1}^{J_{s}+p}%
\widehat{\beta }_{ss^{\prime }}B_{s,p}(x)B_{s^{\prime },p}\left( x^{\prime
}\right),
\end{equation}
where $\widehat{Z}_i$ is defined in (\ref{DEF:Z-residuals}) and $\widehat{%
\beta }_{ss^{\prime }}$'s are the coefficients.

{In FPC applications, it is typical to truncate the spectral decomposition
at an integer $\kappa $ to account for the some predetermined proportion of
the variance. For example, in our numerical studies below, $\kappa $ is
selected as the number of eigenvalues that can explain 95\% of the variation
in the data. Next, let $\mathbf{B}(x)=\left\{ B_{1,p}(x),\ldots
,B_{J_{s}+p,p}(x)\right\} ^{\top }$, and the $N\times \left( J_{s}+p\right) $
design matrix $\mathbf{B}$ for spline regression is
\begin{equation}
\mathbf{B}=\left\{ \mathbf{\mathbf{B}}\left( 1/N\right) \mathbf{,\ldots ,%
\mathbf{B}}\left( N/N\right) \right\} ^{\top }=\left(
\begin{array}{ccc}
B_{1,p}\left( 1/N\right) & \cdots & B_{J_{s}+p,p}\left( 1/N\right) \\
\vdots & \cdots & \vdots \\
B_{1,p}\left( N/N\right) & \cdots & B_{J_{s}+p,p}\left( N/N\right)%
\end{array}%
\right) .  \label{DEF:B}
\end{equation}%
Then for any $k=1,\ldots ,\kappa $, we consider the following spline
approximation for $\psi _{k}(\cdot )$: $\widehat{\psi }_{k}\left( x^{\prime
}\right) =\sum_{\ell =1}^{J_{s}+p}\widehat{\gamma }_{\ell k}B_{\ell
,p}\left( x^{\prime }\right) $, where $\widehat{\gamma }_{\ell k}$'s are
coefficients of B-spline estimator subject to} $\widehat{\mathbf{\gamma }}%
_{k}^{\top }\mathbf{B}^{\top }\mathbf{B\widehat{\gamma }}_{k}=1$ with $%
\widehat{\mathbf{\gamma }}_{k}=\left( \widehat{\gamma }_{1,k},\ldots ,%
\widehat{\gamma }_{J_{s}+p,k}\right) ^{\top }$. The estimates of
eigenfunctions and eigenvalues correspond ${\psi }_{k}$ and $\lambda _{k}$
can be obtained by solving the eigenequations,
\begin{equation}
\int \widehat{G}\left( x,x^{\prime }\right) \widehat{\psi }_{k}\left(
x^{\prime }\right) dx^{\prime }=\widehat{\lambda }_{k}\widehat{\psi }%
_{k}\left( x\right) ,\quad k=1,\ldots ,\kappa .  \label{EQ:Ghatx}
\end{equation}
According to (\ref{EQ:Ghat}), solving (\ref{EQ:Ghatx}) is equivalent to
solve the following:
${\mathbf{B}^{\top }(x)}\widehat{\mathbf{\beta }}\mathbf{B}^{\top }\mathbf{B}%
\widehat{\mathbf{\gamma }}_{k}=\widehat{\lambda }_{k}{\mathbf{B}^{\top }(x)}%
\widehat{\mathbf{\gamma }}_{k}$, $k=1,\ldots ,\kappa $,
where $\widehat{\mathbf{\beta }}^{\top}=(\widehat{\beta}_{s,s'})_{s,s'=1}^{J_{s}+p}$.

By simple algebra and Lemma 3.1 in \cite{WW09}, one needs to solve $\widehat{%
\mathbf{\beta }}\mathbf{B}^{\top }\mathbf{B}\widehat{\mathbf{\gamma }}_{k}=%
\widehat{\lambda }_{k}\widehat{\mathbf{\gamma }}_{k}$, for any $k=1,\ldots
,\kappa $. Consider the following Cholesky decomposition: $\mathbf{B}^{\top }%
\mathbf{B}=\mathbf{L}_{B}\mathbf{L}_{B}^{\top }$. Therefore, solving (\ref%
{EQ:Ghatx}) is equivalent to solve
$\widehat{\lambda }_{k}\mathbf{L}_{B}^{\top }\widehat{\mathbf{\gamma }}_{k}=%
\mathbf{L}_{B}^{\top }\widehat{\mathbf{\beta }}\mathbf{L}_{B}\mathbf{L}%
_{B}^{\top }\widehat{\mathbf{\gamma }}_{k}$, that is, $\widehat{\lambda }_{k}$ and $\mathbf{L}_{B}^{\top }\widehat{%
\mathbf{\gamma }}_{k}$, $k=1,\ldots ,\kappa $, are the eigenvalues and unit
eigenvectors of $\mathbf{L}_{B}^{\top }\widehat{\mathbf{\beta }}\mathbf{L}%
_{B}$. In other words, $\widehat{\mathbf{\gamma }}_{k}$ is obtained by
multiplying $\left( \mathbf{L}_{B}^{\top }\right) ^{-1}$ immediately after
the unit eigenvectors of $\mathbf{L}_{B}^{\top }\widehat{\mathbf{\beta }}%
\mathbf{L}_{B}$, hence $\widehat{\psi }_{k}\left( \cdot \right) $ is
obtained. Consequently, $\widehat{\phi }_{k}\left( x^{\prime }\right) =%
\widehat{\lambda }_{k}^{1/2}\widehat{\psi }_{k}\left( x^{\prime }\right) $.
Then, the $k$th {FPC} score of the $i$th curve can be estimated by a
numerical integration%
\begin{equation*}
\widehat{\xi }_{ik}=\frac{1}{N}\sum\limits_{j=1}^{N}\widehat{\lambda }%
_{k}^{-1}\left\{ Y_{ij}-\widehat{m}\left( \frac{j}{N}\right) \right\}
\widehat{\phi }_{k}\left( \frac{j}{N}\right) .
\end{equation*}


\subsection{Estimating the variance function $\Xi$ and the percentile $Q_{1-\protect\alpha}$}
\label{subsec:variance_percentile}

Notice the fact that (\ref{EQ:Variance MSE}) entails us to estimate the
variance function $\Xi (\cdot)$ by merely computing $\widehat{\xi}_{1k}^{4}$%
, $\widehat{C}(\cdot) $ and $\widehat{\phi}_{k}$. In practice, the following
estimator is employed
\begin{align*}
\widehat{\Xi}(h) =&\sum_{k,k^{\prime}=1}^{\kappa}\left(\frac{1}{1-h}%
\int_{0}^{1-h}\widehat{\phi}_{k}(x) \widehat{\phi}_{k^{\prime}}\left(x+h%
\right) dx\right)^{2} \\
&+\widehat{C}^{2}(h) +\sum_{k=1}^{\kappa}\left(\mathrm{E}\widehat{\xi}%
_{1k}^{4}-3\right) \left\{\frac{1}{1-h}\int_{0}^{1-h}\widehat{\phi}_{k}(x)
\widehat{\phi}_{k}(x+h) dx\right\} ^{2}.
\end{align*}

Next, to derive the percentile $Q_{1-\alpha }$, the Gaussian process is simulated
as follows
\begin{align*}
\widehat{\zeta }(h)=& \sum_{k\neq k^{\prime }}^{\kappa }\frac{1}{1-h}%
\int_{0}^{1-h}\epsilon _{kk^{\prime }}\widehat{\phi }_{k}(x)\widehat{\phi }%
_{k^{\prime }}(x+h)dx \\
& +\sum_{k=1}^{\kappa }\frac{1}{1-h}\int_{0}^{1-h}\epsilon _{k}\widehat{\phi
}_{k}(x)\widehat{\phi }_{k}(x+h)\left( \mathrm{E}\widehat{\xi }%
_{1k}^{4}-1\right) ^{1/2}dx,
\end{align*}%
where $\epsilon _{kk^{\prime }}$ and $\epsilon _{k}$ are independent
standard Gaussian random variables. Hence, $\widehat{\zeta }(h)$ is a zero
mean Gaussian process with variance function $\widehat{\Xi }(h)$ and
covariance function
\begin{align*}
\widehat{\Omega }& \left( h,h^{\prime }\right) =\mathrm{Cov}\left\{ \widehat{%
\zeta }(h),\widehat{\zeta }\left( h^{\prime }\right) \right\} \\
& =\frac{1}{1-h}\frac{1}{1-h^{\prime }}\int_{0}^{1-h}\int_{0}^{1-h^{\prime
}}\left\{ \sum_{k,k^{\prime }=1}^{\kappa }\widehat{\phi }_{k}(x)\widehat{%
\phi }_{k}\left( x^{\prime }\right) \widehat{\phi }_{k^{\prime }}(x+h)%
\widehat{\phi }_{k^{\prime }}\left( x^{\prime }+h^{\prime }\right) \right. \\
& \quad \left. +\sum_{k=1}^{\kappa }\left( \mathrm{E}\widehat{\xi }%
_{1k}^{4}-3\right) \widehat{\phi }_{k}(x)\widehat{\phi }_{k}\left(
x+h\right) \widehat{\phi }_{k}\left( x^{\prime }\right) \widehat{\phi }%
_{k}\left( x^{\prime }+h^{\prime }\right) \right\} dxdx^{\prime }+\widehat{C}%
(h)\widehat{C}\left( h^{\prime }\right) ,
\end{align*}%
for any $h,h^{\prime }\in \left[ 0,h_{0}\right] $. A large number of
independent realizations of $\widehat{\zeta }(h)$ are simulated, then the
maximal absolute deviation for each copy of $\widehat{\zeta }(h)\widehat{\Xi
}^{-1/2}(h)$ is taken. Eventually, $Q_{1-\alpha }$ is estimated by the
empirical percentiles of these maximum values.

\setcounter{chapter}{5} \renewcommand{\thesection}{\arabic{section}} %
\renewcommand{\thetable}{{\arabic{table}}} \setcounter{table}{0} %
\renewcommand{\thefigure}{\arabic{figure}} \setcounter{figure}{0} %
\renewcommand{\thesubsection}{5.\arabic{subsection}}

\section{Simulation Studies}
\label{sec:simulation}

To illustrate the finite-sample behavior of our confidence bands, we conduct simulation studies to illustrate the finite-sample performance of the proposed method.

\subsection{General study}
\label{subsec:general-eg}

The data are generated from the following model: $Y_{ij}=m\left( j/N\right) +\sum_{k=1}^{\infty }\xi _{ik}\phi _{k}\left(
j/N\right) +\sigma (j/N)\varepsilon _{ij}$, $1\leq j\leq N$, $1\leq i\leq n$, where $m(x)=\sin \{2\pi \left( x-1/2\right) \}$, $\varepsilon _{ij}$ are i.i.d standard normal variables, $\phi _{k}(x)=\sqrt{\lambda _{k}}\psi _{k}(x)$  with $\lambda _{k}=\left( 1/4\right) ^{\left[ k/2\right] }$, $\psi_{2k-1}(x)=\sqrt{2}\cos \left( 2k\pi x\right) $, $\psi _{2k}(x)=\sqrt{2}\sin \left( 2k\pi x\right) $, $k\geq 1$. {We consider both homogenous errors with $\sigma (x)=\sigma _{\epsilon}$ and strongly heteroscedastic errors with $\sigma (x)=\sigma_{\epsilon }\left\{ 5+\exp \left( x\right) \right\} ^{-1}\left\{ 5-\exp\left( x\right) \right\} $, where the noise level $\sigma _{\epsilon}=0.1,~0.5$. Since Assumption (C5) is satisfied, following \cite{CWLY16}, we truncate $\sum_{k=1}^{\infty }\xi _{ik}\phi _{k}\left( j/N\right) $ at $1000$. The number of curves $n=\lfloor cN^{\theta }\rfloor$ with $c=0.8$ and $\theta =1$, and the number of observations per curve $N$ is taken to be $50$, $100$ and $200$, respectively. Each simulation is repeated $500$ times. Throughout this section, the mean function is estimated by cubic splines, i.e., $p=4$, with the number of knots selected using the formula and GCV given in Section \ref{subsec:knots select}. }

First, we examine the accuracy of the proposed two-stage estimation procedure. The average mean squared error (AMSE) is computed to assess the performance of the covariance estimators $\widehat{C}(\cdot )$ and $\widetilde{C}(\cdot )$ defined in (\ref{EQ:C-hat}) and (\ref{DEF:C-tilde}), respectively. The AMSE of  $\widehat{G}(\cdot ,\cdot )$, the eigenvalue $\widehat{\lambda}_k$'s and the eigenfunction $\widehat{\phi }_k$'s are defined as
\begin{align*}
\text{AMSE}(\widehat{G})&=\frac{1}{500N^2}\sum_{s=1}^{500}\sum_{j,j^{\prime }=1}^{N}\left\{ \widehat{G}_{s}\left(\frac{j}{N},\frac{j^{\prime}}{N}\right)-G\left(\frac{j}{N},\frac{j^{\prime}}{N}\right)\right\} ^{2}, \\
\text{AMSE}(\widehat{\lambda})&= \frac{1}{500\kappa}\sum_{s=1}^{500} \sum_{k=1}^{\kappa
}(\widehat{\lambda }_{ks}-\lambda _{k}) ^{2},
\text{AMSE}(\widehat{\phi })=\frac{1}{500N\kappa}\sum_{s=1}^{500}\sum_{j=1}^{N}\sum_{k=1}^{\kappa }\left\{ (\widehat{\phi }_{ks}-\phi _{k})\left(\frac{j}{N}\right)\right\} ^{2},
\end{align*}
where $\widehat{G}_{s}$, $\widehat{\lambda }_{ks}$, $\widehat{\phi }_{ks}$ represent the values of the $s$-th replication of $\widehat{G}(\cdot ,\cdot)$, $\widehat{\lambda }_{k}$, $\widehat{\phi }_{k}$ in (\ref{EQ:Ghatx}), respectively. {Our simulation results based on homogeneous and heteroscedastic variance functions are listed in Tables \ref{TAB:AMSE-SCB-general} and \ref{TAB:hetero-AMSE-SCB-genaral}, respectively. One concludes that a lager noise level leads to a higher AMSE over all, and the AMSEs of the two estimators $\widehat{C}(\cdot )$ and $\widetilde{C}(\cdot )$ are very similar in each scenario. Moreover, the AMSE$(\widehat{\lambda})$ is getting smaller when $N$ is increasing in each scenario. The GCV method has smaller AMSE$(\widehat{\lambda})$  and AMSE$(\widehat{G})$ than the formula method does. The value of the AMSE for $\widehat{\phi}$ varies in each of the scenarios. When $N=50$, the AMSE$(\widehat{\phi})$ based on the formula method is smaller than that based on the GCV method, while a converse phenomenon is observed when $N=100, 200$.}

Tables \ref{TAB:AMSE-SCB-general} and \ref{TAB:hetero-AMSE-SCB-genaral} also present the empirical coverage rate (CR), i.e., the percentage of the event that the true curve $C(\cdot )$ is entirely covered by the SCB among all $500$ replications, respectively. As the sample size increases, the CR of the SCB becomes closer to the nominal confidence level, which shows a positive confirmation of Theorem \ref{THM:band}. In addition, the average widths (WD) of the bands are calculated and presented in columns $9$ and $11$ of Tables \ref{TAB:AMSE-SCB-general}--\ref{TAB:hetero-AMSE-SCB-genaral}. It is obvious that the width tends to be narrower when the sample size becomes larger and noise level $\sigma _{\epsilon}$ smaller.


\begin{table}[htbp]
\caption{Simulation results based on homogeneous errors with $\protect\sigma (x)=\protect\sigma _{\protect\epsilon }$: AMSE of estimators $\protect\widehat{C}$, $\protect\widetilde{C}$, $\protect\widehat{\protect\lambda}$, $\protect\widehat{G}$, $\protect\widehat{\protect\phi}$; CR (outside/inside of the parentheses is based on $\protect\widehat{C}$, $\protect\widetilde{C}$) and WD of SCBs based on $\protect\widehat{C}$. }
\label{TAB:AMSE-SCB-general}
\renewcommand*{\arraystretch}{0.5} \centering%
\vspace{0.2cm}
\resizebox{150mm}{30mm}{
\begin{tabular}{ccc ccc cc c ccc cc}
\toprule \multirow{3}{*}{$\sigma _{\epsilon }$} & \multirow{3}{*}{$N$} &
\multicolumn{5}{c}{AMSE} &  & \multicolumn{5}{c}{SCB} &     \\ \cline{3-7}\cline{9-13}
&  & \multirow{2}{*}{$\widehat{C}$} & \multirow{2}{*}{$\widetilde{C}$} & \multirow{2}{*}{$\widehat{\lambda}$} &\multirow{2}{*}{$\widehat{G}$} &  \multirow{2}{*}{$\widehat{\phi}$}&  &\multicolumn{2}{c}{$95\%$} &  &
\multicolumn{2}{c}{$99\%$} &    \\ \cline{9-10} \cline{12-13}
&  &  &  &  &  &   &  & CR & WD &  & CR & WD &    \\
\midrule
\multirow{1}{*}{$0.1$} & $50$ & $0.068$ & $0.065$ & $0.014$ & $0.130$ & $0.661$ &  & $0.866(0.892)$ & $1.25$ &  & $0.926(0.942)$ & $1.51$ &  \\
\multirow{1}{*}{Formula} & $100$ & $0.034$ & $0.035$ & $0.006$ & $0.051$ & $0.871$ & & $0.910(0.922)$ & $0.92$ &  & $0.970(0.970) $ & $1.11$ &    \\
& $200$ & $0.016$ & $0.016$ & $0.003$ & $0.025$ & $0.845$ &  & $0.958(0.962)$ & $0.67$ &  & $0.992(0.992) $ & $0.80$ &   \\
\midrule
\multirow{1}{*}{$0.1$} & $50$ & $0.065$ & $0.065$ & $0.009$ & $0.095$ & $0.806$ &  & $0.868(0.894)$ & $1.25$ &  & $0.940(0.948)$ & $1.51$ &  \\
\multirow{1}{*}{GCV} & $100$ & $0.035$ & $0.034$ & $0.005$ & $0.048$ & $0.790$ &  & $0.914(0.920)$ & $0.92$ &  & $0.968(0.970) $ & $1.11$ &    \\
& $200$ & $0.016$ & $0.016$ & $0.002$ & $0.025$ & $0.769$ &  & $0.960(0.960)$ & $0.67$ &  & $0.992(0.994) $ & $0.80$ &    \\
\midrule
\multirow{1}{*}{$0.5$} & $50$ & $0.070$ & $0.065$ & $0.014$ & $0.133$ & $0.651$ &  & $0.864(0.892)$ & $1.26$ &  & $0.920(0.938)$ & $1.53$ &   \\
\multirow{1}{*}{Formula} & $100$ & $0.035$ & $0.035$ & $0.006$ & $0.052$  & $0.876$ &  & $0.910(0.922)$ & $0.93$ &  & $0.966(0.970) $ & $1.11$ &    \\
& $200$ & $0.016$ & $0.016$ & $0.003$ & $0.025$ & $0.845$ &  & $0.956(0.960)$ & $0.67$ &  & $0.992(0.992) $ & $0.80$ &   \\
\midrule
\multirow{1}{*}{$0.5$} & $50$ & $0.070$ & $0.065$ & $0.012$ & $0.107$ & $0.871$ &  & $0.858(0.896)$ & $1.27$ &  & $0.918(0.940)$ & $1.53$ &   \\
\multirow{1}{*}{GCV} & $100$ & $0.036$ & $0.035$ & $0.006$ & $0.052$ & $0.826$ &  & $0.898(0.920)$ & $0.92$ &  & $0.958(0.964) $ & $1.11$ &   \\
& $200$ & $0.017$ & $0.016$ & $0.003$ & $0.026$ & $0.798$ &  & $0.946(0.960)$ & $0.67$ &  & $0.988(0.992) $ & $0.80$ &   \\
\bottomrule
\end{tabular}}
\end{table}

\begin{figure}[tbp]
\centering \vspace*{-0.4in}
\subfigure[$N=50$]{\label{fig:SCB-a}
\vspace*{-0.4in}
\includegraphics[trim={1cm 1.5cm 0cm 1cm}, height=2.5in,width=2.75in]{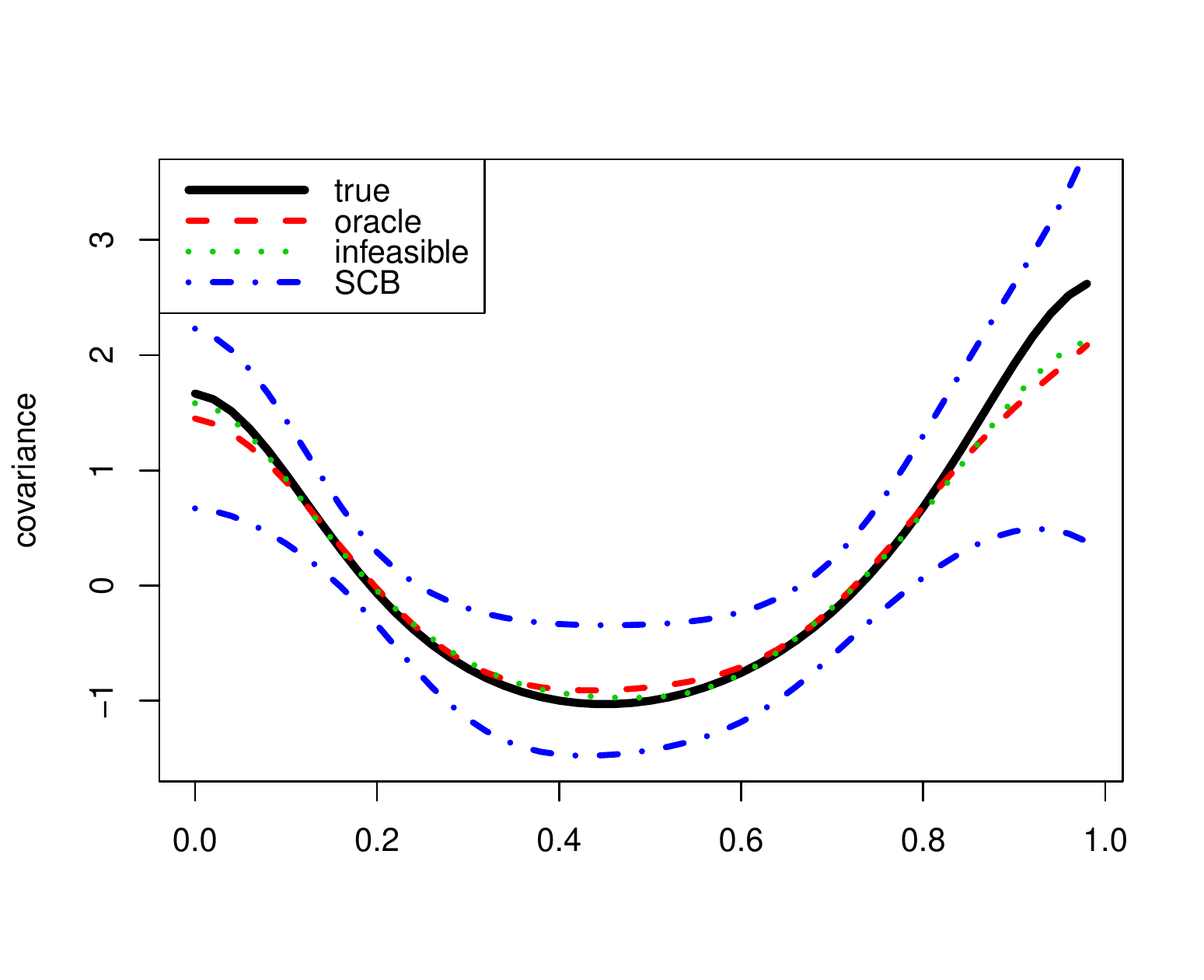}}
\subfigure[$N=200$]{\label{fig:SCB-b}
\vspace*{-0.4in}
\includegraphics[trim={1cm 1.5cm 0cm 1cm}, height=2.5in,width=2.75in]{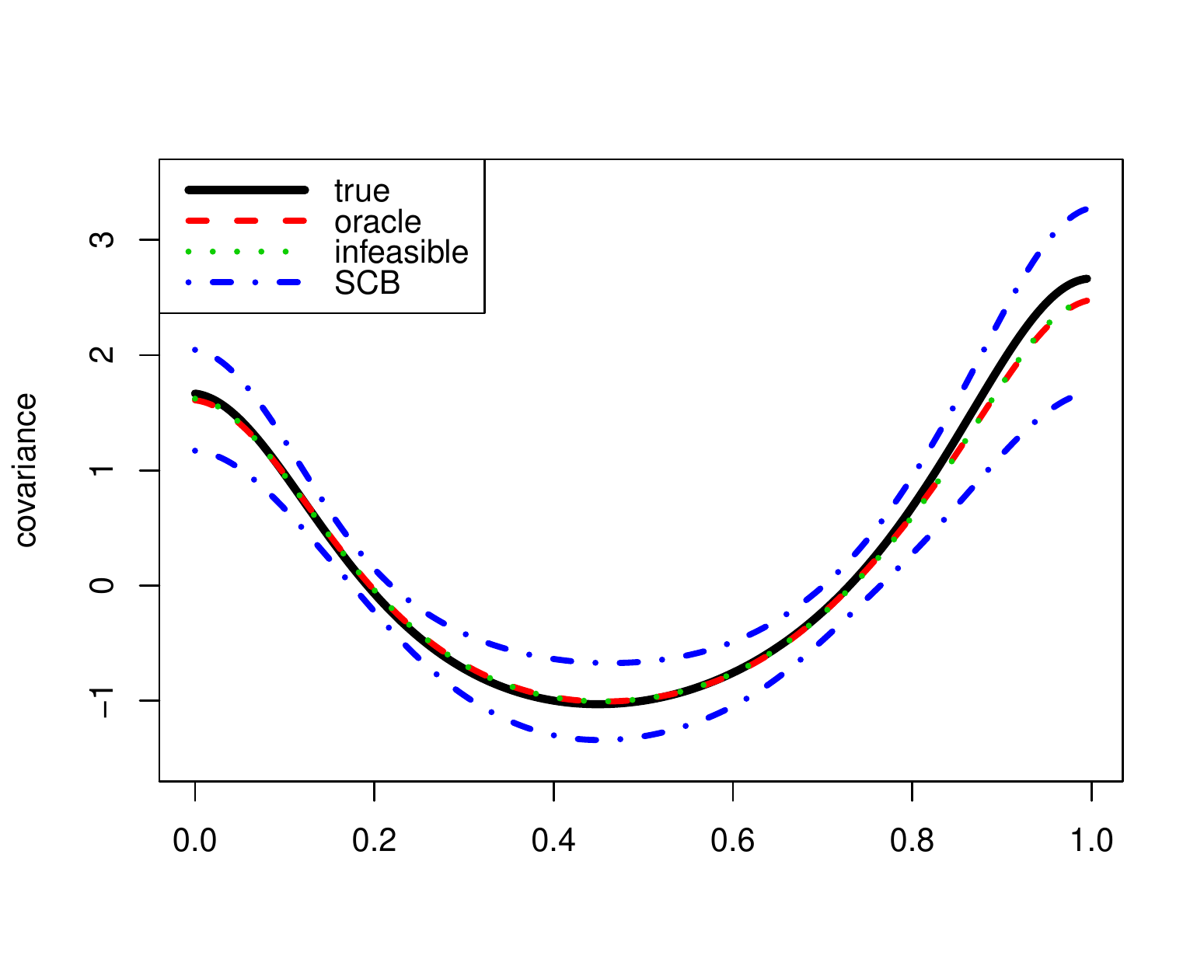}}
\vspace{-0.1in}
\caption{Plot of true covariance function (thick solid line), oracle
estimator $\protect\widehat{C}$ (dashed line) and the $99\%$ SCB
(dotted-dashed line), infeasible estimator $\protect\widetilde{C}$ (dotted
line) for the covariance function with $\protect\sigma_{\protect\epsilon%
}=0.1 $.}
\label{Fig:SCB sigma05N50}
\end{figure}

Overall, the performance of the SCB based on estimator $\widehat{C}$ is indistinguishable from the infeasible SCB based on estimator $\widetilde{C}$; and they approximate the nominal level as $N$ increases. The knots number selected by the GCV yield similar results as those of the formula. For visualization of actual estimation, Figure \ref{Fig:SCB sigma05N50} depicts the true covariance $C(\cdot)$, the spline covariance estimators $\widehat{C}(\cdot)$, as well as the $99\%$ SCB for $C(\cdot)$. They are all based on a typical run under the setting $N=50$, $N=200$ and $\sigma_{\epsilon}=0.1$. It is clear from Figure \ref{Fig:SCB sigma05N50} that the estimator $\widehat{C}(\cdot)$ is very close to the true covariance function $C(\cdot)$ and the true covariance function is entirely covered by the SCB.

\begin{table}[htbp]
\caption{Simulation results based on heteroscedastic errors with $\protect\sigma (x)=\protect\sigma _{\protect\epsilon }\frac{ 5-\exp \left( x\right)}{ 5+\exp\left( x\right)}$: AMSE of estimators $\protect\widehat{C}$, $\protect\widetilde{C}$, $\protect\widehat{\protect\lambda}$, $\protect\widehat{G}$, $\protect\widehat{\protect\phi}$; CR (outside/inside of the parentheses is based on $\protect\widehat{C}$, $\protect\widetilde{C}$), and WD of SCBs based on $\protect\widehat{C}$.}
\label{TAB:hetero-AMSE-SCB-genaral}
\renewcommand*{\arraystretch}{0.5} %
\centering\vspace{0.15cm}
\resizebox{150mm}{32mm}{
\begin{tabular}{ccc ccc ccc ccc cc}
\toprule \multirow{3}{*}{$\sigma _{\epsilon }$} & \multirow{3}{*}{$N$} &
\multicolumn{5}{c}{AMSE} &  & \multicolumn{5}{c}{SCB}  &  \\ \cline{3-7}\cline{9-13}
&  & \multirow{2}{*}{$\widehat{C}$} & \multirow{2}{*}{$\widetilde{C}$} & \multirow{2}{*}{$\widehat{\lambda}$} &\multirow{2}{*}{$\widehat{G}$} &  \multirow{2}{*}{$\widehat{\phi}$}&  &\multicolumn{2}{c}{$95\%$} &  &
\multicolumn{2}{c}{$99\%$} &    \\ \cline{9-10} \cline{12-13}
&  &  &  &  &  &   &  & CR & WD &  & CR & WD &   \\
\midrule
\multirow{1}{*}{$0.1$} & $50$ & $0.068$ & $0.065$ & $0.014$ & $0.130$ & $0.650$ &  &$0.862(0.890)$ & $1.25$ &  & $0.930(0.942)$ & $1.51$ &  \\
\multirow{1}{*}{Formula} & $100$ & $0.034$ & $0.035$ & $0.006$ & $0.051$  & $0.870$ &  &$0.912(0.922)$ & $0.92$ &  & $0.966(0.970) $ & $1.11$ &    \\
& $200$ & $0.016$ & $0.016$ & $0.003$ & $0.025$ & $0.859$ &  &$0.954(0.960)$ & $0.67$ &  & $0.988(0.990) $ & $0.80$ &   \\
\midrule
\multirow{1}{*}{$0.1$} & $50$ & $0.065$ & $0.065$ & $0.009$ & $0.095$ & $0.808$ &  &$0.868(0.896)$ & $1.25$ &  & $0.938(0.948)$ & $1.51$ &  \\
\multirow{1}{*}{GCV} & $100$ & $0.034$ & $0.035$ & $0.005$ & $0.048$ & $0.772$ &  & $0.918(0.920)$ & $0.92$ &  & $0.970(0.970) $ & $1.11$ &    \\
& $200$ & $0.016$ & $0.016$ & $0.002$ & $0.025$ & $0.772$ &  &$0.960(0.960)$ & $0.67$ &  & $0.992(0.994) $ & $0.80$ &   \\
\midrule
\multirow{1}{*}{$0.5$} & $50$ & $0.070$ & $0.065$ & $0.014$ & $0.133$ & $0.651$ &  & $0.864(0.890)$ & $1.26$ &  & $0.926(0.940)$ & $1.52$ &   \\
\multirow{1}{*}{Formula} & $100$ & $0.035$ & $0.035$ & $0.006$ & $0.051$ & $0.879$  &  &$0.916(0.920)$ & $0.92$ &  & $0.966(0.970) $ & $1.11$ &   \\
& $200$ & $0.016$ & $0.016$ & $0.003$ & $0.025$ & $0.847$ &  &$0.960(0.964)$ & $0.67$ &  & $0.988(0.990) $ & $0.80$ &   \\
\midrule
\multirow{1}{*}{$0.5$} & $50$ & $0.067$ & $0.065$ & $0.010$ & $0.098$ & $0.857$ &  &$0.868(0.898)$ & $1.26$ &  & $0.930(0.942)$ & $1.51$ &   \\
\multirow{1}{*}{GCV} & $100$ & $0.035$ & $0.034$ & $0.005$ & $0.049$ & $0.817$ &  & $0.912(0.922)$ & $0.92$ &  & $0.962(0.968) $ & $1.11$ &    \\
& $200$ & $0.016$ & $0.016$ & $0.003$ & $0.025$ & $0.783$ &  &$0.954(0.962)$ & $0.67$ &  & $0.992(0.994) $ & $0.80$ &    \\
\bottomrule
\end{tabular}}
\end{table}

\subsection{Spatial covariance models}
\label{subsec:spatial}

In order to compare the finite-sample performance of the proposed estimator to that of \cite{CWLY16}, we consider the following spatial covariance models:

\begin{itemize}
\item Spherical model (M1): $C(h;\sigma _{s}^{2},\theta _{s})=\sigma
_{s}^{2}\{1-\frac{3}{2}\frac{h}{\theta _{s}}+\frac{1}{2}(\frac{h}{\theta _{s}%
})^{3}\}I\left\{ h\leq \theta _{s}\right\}$;

\item Mat\'{e}rn model (M2): $C(h;\sigma _{s}^{2},\theta _{s},v)=\sigma
_{s}^{2}\left\{ \Gamma (v)\right\} ^{-1}2^{1-v}\left( 2\sqrt{v}h/\theta
_{s}\right) ^{v}\digamma _{v}\left( 2\sqrt{v}h/\theta _{s}\right) ,$ where $%
\Gamma $ is the gamma function, $\digamma _{v}$ is the modified Neumann
function;

\item Gaussian Model (M3): $C(h;\sigma _{s}^{2},\theta _{s})=\sigma
_{s}^{2}\exp (-h^{2}/\theta _{s}^{2})$.
\end{itemize}

In the parameterization (following \cite{BCG04} page 29) of the covariance structure, $\sigma _{s}^{2}$ is the sill and $\theta _{s}$ is the range parameter. In the simulation, we set $\sigma _{s}^{2}$ $=2$ for M1, M2 and M3, and choose $\theta _{s}=1$ for M1 and M2, $\theta _{s}=3$ for M3, while for M2, $v=3$. Since $C\left( h\right) \rightarrow 0$ as $h\rightarrow \infty $, in practice, we only numerically evaluate the covariance $C\left(h\right) $ over the \textquotedblleft effective range" defined as the distance beyond which the correlation between observations, $\rho(h)=C\left( h\right) /C\left( 0\right) $, is less than or equal to $0.05$. In such sense, we choose the compact interval $[0,s]$ to represent the ``effective range", where $s$ is the largest $h$ satisfying $ \rho (h)\leq 0.05$. An exception of this phenomenon is the spherical model that has an exact range $\left[ 0,\theta _{s}\right] $, i.e., $C\left(h\right) =0$ when $h=\theta _{s}$. To be consistent in our evaluation of the methods, we apply the \textquotedblleft effective range" to the spherical model as well.

Our data are generated from $Y_{ij}= m(x_j)+Z_{i}\left(x_j\right) +\sigma \left(x_j\right) \varepsilon
_{ij}$, where $m(x)=\sin \{2\pi \left( x-1/2\right) \}$, $\{x_j\}_{j=1}^{N}$ are equally spaced grid points over ``effective range'' $[0,s]$, $\varepsilon _{ij}$ $\sim N(0,1)$ are i.i.d variables, and the process $Z_{i}(\cdot )$ is generated from a zero mean Gaussian process. {We examine the performance of models containing homogeneous errors with $\sigma (x)=\sigma _{\epsilon }$ and heteroscedastic errors with $\sigma (x)=\sigma _{\epsilon}\frac{5-\exp \left( x/2\right) } {5+\exp \left(x/2\right)}$ for M1, and $\sigma (x)=\sigma _{\epsilon}\frac{30-\exp \left( x/2\right) } {30+\exp \left(x/2\right)} $ for M2 and M3. The results are similar to each other, so we only present the results with homogeneous errors. The number of curves $n=\lfloor 0.8N\rfloor$ with $N=50$, $100$ and $200$, and the noise levels are $\sigma _{\epsilon }=0.1,~0.5$. The mean function is estimated by cubic splines, i.e., $p=4$, with the number of knots selected using the formula given in Section \ref{subsec:knots select}. The GCV selected knots yield similar results but it is more time consuming, hence they are not summarized here.}

The AMSE of the covariance estimators $\widehat{C}$ and $\widetilde{C}$ are reported in columns 4--5 of Table \ref{TAB:AMSE-SCB-spatial}. The performance of the two estimators is very similar. Columns 6 and 8 present the empirical coverage rate {CR}, i.e., the percentage of the true curve $C(\cdot)$ entirely covered by the SCB, based on $95\%$ and $99\%$ confidence levels, respectively. As the sample size increases, the coverage probability of the SCB becomes closer to the nominal confidence level. In addition, the WDs of the bands are calculated and presented in columns 7 and 9 in Table \ref{TAB:AMSE-SCB-spatial}. It is obvious that the width tends to be narrower when the sample size becomes larger or $\sigma_{\epsilon}$ is smaller.

\begin{table}[htbp]
\caption{Simulation results based on homogeneous errors with $\protect\sigma (x)=\protect\sigma _{\protect\epsilon }$: AMSE of estimators $\protect\widehat{C}$, $\protect\widetilde{C}$; CR (outside/inside of the parentheses is based on $\protect\widehat{C}$ and $\protect\widetilde{C}$), and WD of SCBs based on $\protect\widehat{C}$.}
\label{TAB:AMSE-SCB-spatial}
\renewcommand*{\arraystretch}{0.55} %
\centering\vspace{0.15cm}
\resizebox{120mm}{40mm}{
\begin{tabular}{ccccccccccc}
\toprule
\multirow{3}{*}{$\sigma _{\epsilon}$} & \multirow{3}{*}{Model} & \multirow{3}{*}{$N$} & \multicolumn{2}{c}{AMSE} &  & \multicolumn{5}{c}{SCB} \\ \cline{4-5}\cline{7-11}
&  &  & \multirow{2}{*}{$\widehat{C}$} & \multirow{2}{*}{$\widetilde{C}$} &  & \multicolumn{2}{c}{$95\%$} &  & \multicolumn{2}{c}{$99\%$} \\  \cline{7-8} \cline{10-11}
&  &  &  &  &  & CR & WD &  & CR & WD \\
\midrule

\multirow{9}{*}{$0.1$} & \multirow{3}{*}{M1} & $50 $ & $0.082$ & $0.081$ &  & $0.910(0.918)$ & $1.37$ &  & $0.960(0.966)$ & $1.68$ \\
&  & $100$ & $0.040$ & $0.040$ &  & $0.920(0.926)$ & $0.99$ &  & $0.974(0.978) $ & $1.21$ \\
&  & $200$ & $0.019$ & $0.018$ &  & $0.946(0.952)$ & $0.72$ &  & $0.980(0.986) $ & $0.87$ \\ \cline{2-11}
&  \multirow{3}{*}{M2} & $50$ & $0.096$ & $0.095$ &  & $0.904(0.908)$ & $1.44$ &  & $0.950(0.954)$ & $1.78$ \\
&  & $100 $ & $0.048$ & $0.049$ &  & $0.926(0.924)$ & $1.05$ &  & $0.978(0.978)$ & $1.30$ \\
&  & $200$ & $0.022$ & $0.022$ &  & $0.958(0.958)$ & $0.76$ &  & $0.992(0.994)$ & $0.94$ \\  \cline{2-11}
&  \multirow{3}{*}{M3} & $50$ & $0.109$ & $0.109$ &  & $0.906(0.910)$ & $1.50$ &  & $0.954(0.958)$ & $1.86$ \\
&  & $100 $ & $0.055$ & $0.055$ &  & $0.922(0.928)$ & $1.09$ &  & $0.976(0.978)$ & $1.35$ \\
&  & $200$ & $0.025$ & $0.025$ &  & $0.960(0.958)$ & $0.79$ &  & $0.988(0.990)$ & $0.98$ \\
\midrule

\multirow{9}{*}{$0.5$} & \multirow{3}{*}{M1} & $50 $ & $0.082$ & $0.080$ &  & $0.896(0.912)$ & $1.38$ &  & $0.952(0.964)$ & $1.70$ \\
&  & $100$ & $0.040$ & $0.040$ &  & $0.920(0.928)$ & $0.99$ &  & $0.980(0.980)$ & $1.21$ \\
&  & $200$ & $0.019$ & $0.018$ &  & $0.938(0.946)$ & $0.72$ &  & $0.988(0.988)$ & $0.88$ \\ \cline{2-11}
&  \multirow{3}{*}{M2} & $50$ & $0.097$ & $0.096$ &  & $0.896(0.908)$ & $1.46$ &  & $0.946(0.958)$ & $1.80$ \\
&  & $100 $ & $0.048$ & $0.049$ &  & $0.914(0.930)$ & $1.06$ &  & $0.978(0.980)$ & $1.30$ \\
&  & $200$ & $0.022$ & $0.022$ &  & $0.954(0.958)$ & $0.77$ &  & $0.990(0.994)$ & $0.94$ \\  \cline{2-11}
&  \multirow{3}{*}{M3} & $50$ & $0.111$ & $0.111$ &  & $0.908(0.916)$ & $1.51$ &  & $0.952(0.962)$ & $1.88$ \\
&  & $100 $ & $0.055$ & $0.055$ &  & $0.912(0.924)$ & $1.10$ &  & $0.974(0.978)$ & $1.36$ \\
&  & $200$ & $0.025$ & $0.025$ &  & $0.958(0.956)$ & $0.79$ &  & $0.988(0.990)$ & $0.98$ \\
\bottomrule
\end{tabular}%
}
\end{table}

When the covariance structure is not necessarily stationary, \cite{CWLY16} proposed a tensor-product bivariate B-spline estimator $\widehat{G}^{\mathrm{TPS}}(x,x^{\prime})$ and a SCB for the covariance function $G(x,x^{\prime})=\mathrm{Cov}\left\{Z_{1}(x),Z_{1}\left(x^{\prime}\right) \right\} $.  Following the suggestion of one referee, to assess the accuracy of recovering $G(\cdot,\cdot)$, the covariance function estimators $\widehat{C}$ is also presented in 2D to make a comparison, say, $\widehat{G}^\mathrm{PROP}(x,x^{\prime})=\widehat{C}(\left\vert x-x^{\prime}\right\vert )$. In addition, the simultaneous confidence envelops (SCE) is constructed by using $\widehat{G}^\mathrm{PROP}(x,x^{\prime})$ and $\widehat{G}^{\mathrm{TPS}}(x,x^{\prime})$ are compared, named SCE-I and SCE-II, respectively.

Columns 4--5 of Table \ref{TAB:AMSE-SCB-formula} present the AMSEs of $\widehat{G}^{\mathrm{PROP}}(x,x^{\prime })$ and $\widehat{G}^{\mathrm{TPS}}(x,x^{\prime })$. The results of AMSEs indicate that $\widehat{G}^\mathrm{PROP}$ is more accurate than $\widehat{G}^{\mathrm{TPS}}$, while $\widehat{G}^{\mathrm{TPS}}$ usually gives larger AMSE. Columns 6--13 of Table \ref{TAB:AMSE-SCB-formula} report the CR and WD of SCE-I and SCE-II. One sees that the CRs of SCE-I are much closer to the nominal levels than those of SCE-II, and increasing the sample size helps to improve the {CR of the SCEs}
to their nominal levels. One also observes the widths of the SCE-I are much narrower than those of the SCE-II. These findings indicate our proposed SCE-I is more efficient than SCE-II when the true covariance function is stationary.

\begin{table}[htp]
\renewcommand*{\arraystretch}{0.5}
\caption{Simulation results based on homogeneous errors with $\protect\sigma (x)=\protect\sigma _{\protect\epsilon }$: AMSE of estimators $\widehat{G}^{\mathrm{PROP}}(\cdot,\cdot)$, $\widehat{G}^{\mathrm{TPS}}(\cdot,\cdot)$; CR and WD of SCE-I and SCE-II.}
\label{TAB:AMSE-SCB-formula}\centering%
\vspace{0.1cm}
\resizebox{140mm}{36mm}{\begin{tabular}{ccc ccc ccc ccc ccc}
\toprule
\multirow{3}{*}{$\sigma _{\epsilon}$} & \multirow{3}{*}{Model} & \multirow{3}{*}{$N$} & \multicolumn{2}{c}{AMSE} & & \multicolumn{4}{c}{SCE-I} & & \multicolumn{4}{c}{SCE-II} \\ \cline{4-5}\cline{7-10} \cline{12-15}
&  &  &  \multirow{2}{*}{$\widehat{G}^{\mathrm{PROP}}$} & \multirow{2}{*}{$\widehat{G}^{\mathrm{TPS}}$} & & \multicolumn{2}{c}{$95\%$} & \multicolumn{2}{c}{$99\%$} & & \multicolumn{2}{c}{$95\%$} & \multicolumn{2}{c}{$99\%$} \\ \cline{7-10} \cline{12-15}
&  &  &  &  &  & CR & WD & CR & WD & & CR & WD & CR & WD\\
\midrule

\multirow{9}{*}{$0.1$} & \multirow{3}{*}{M1} & $50$ & $0.079$ & $0.123$ & & $0.910$ & $1.40$ & $0.960$ & $1.71$ & & $0.744$ & $2.08$ & $0.840$ & $2.55$\\
&  & $100$ & $0.039$ & $0.063$ & & $0.920$ & $1.01$ & $0.974$ & $1.24$ & & $0.852$ & $1.64$ & $0.944$ & $2.02$\\
&  & $200$ & $0.018$ & $0.031$ & & $0.946$ & $0.73$ & $0.980$ & $0.90$ & & $0.904$ & $1.19$ & $0.964$ & $1.47$\\
\cline{2-15}

&  \multirow{3}{*}{M2} & $50$ & $0.096$ & $0.148$ & & $0.904$ & $1.50$ & $0.950$ & $1.86$ & & $0.682$ & $2.09$ & $0.816$ & $2.56$\\
&  & $100$ & $0.048$ & $0.072$ & & $0.926$ & $1.10$ & $0.978$ & $1.35$ & & $0.780$ & $1.62$ & $0.898$ & $2.00$\\
&  & $200$ & $0.022$ & $0.036$ & & $0.958$ & $0.79$ & $0.992$ & $0.98$ & & $0.926$ & $1.16$ & $0.976$ & $1.43$\\
\cline{2-15}

&  \multirow{3}{*}{M3} & $50$ & $0.114$ & $0.153$ & & $0.906$ & $1.57$ & $0.954$ & $1.95$ & & $0.724$ & $2.15$ & $0.824$ & $2.65$\\
&  & $100$ & $0.057$ & $0.074$ & & $0.922$ & $1.15$ & $0.976$ & $1.42$ & & $0.852$ & $1.52$ & $0.942$ & $1.86$\\
&  & $200$ & $0.026$ & $0.039$ & & $0.960$ & $0.83$ & $0.988$ & $1.21$ & & $0.882$ & $1.09$ & $0.956$ & $1.33$\\
\midrule

\multirow{9}{*}{$0.5$} & \multirow{3}{*}{M1} & $50$ & $0.079$ & $0.133$ & & $0.896$ & $1.41$ & $0.952$ & $1.73$ & & $0.740$ & $2.10$ & $0.858$ & $2.58$\\
&  & $100$ & $0.039$ & $0.064$ & & $0.920$ & $1.02$ & $0.980$ & $1.25$ & & $0.814$ & $1.64$ & $0.924$ & $2.03$\\
&  & $200$ & $0.018$ & $0.032$ & & $0.938$ & $0.74$ & $0.988$ & $0.90$ & & $0.896$ & $1.19$ & $0.968$ & $1.46$\\
 \cline{2-15}

&  \multirow{3}{*}{M2} & $50$ & $0.097$ & $0.146$ & & $0.898$ & $1.51$ & $0.946$ & $1.87$ & & $0.666$ & $2.13$ & $0.780$ & $2.60$\\
&  & $100$ & $0.048$ & $0.071$ & & $0.914$ & $1.10$ & $0.978$ & $1.35$ & & $0.768$ & $1.62$ & $0.906$ & $2.00$\\
&  & $200$ & $0.022$ & $0.036$ & & $0.954$ & $0.79$ & $0.990$ & $0.98$ & & $0.928$ & $1.16$ & $0.970$ & $1.43$\\
\cline{2-15}

&  \multirow{3}{*}{M3} & $50$ & $0.115$ & $0.161$ & & $0.908$ & $1.58$ & $0.952$ & $1.97$ & & $0.694$ & $2.16$ & $0.792$ & $2.65$\\
&  & $100$ & $0.057$ & $0.075$ & & $0.912$ & $1.15$ & $0.974$ & $1.42$ & & $0.814$ & $1.53$ & $0.928$ & $1.87$\\
&  & $200$ & $0.026$ & $0.037$ & & $0.958$ & $0.83$ & $0.988$ & $1.03$ & & $0.864$ & $1.09$ & $0.956$ & $1.33$\\
\bottomrule
\end{tabular}}%
\end{table}

\section{Real data analysis}
\label{Sec:real data}

\begin{figure}[tbp]
\centering
\subfigure[]{\label{fig:Hip-a}
\hspace*{0in} \includegraphics[trim={1cm 1cm 1cm 1cm}, width=2.35in, height=2.3in]{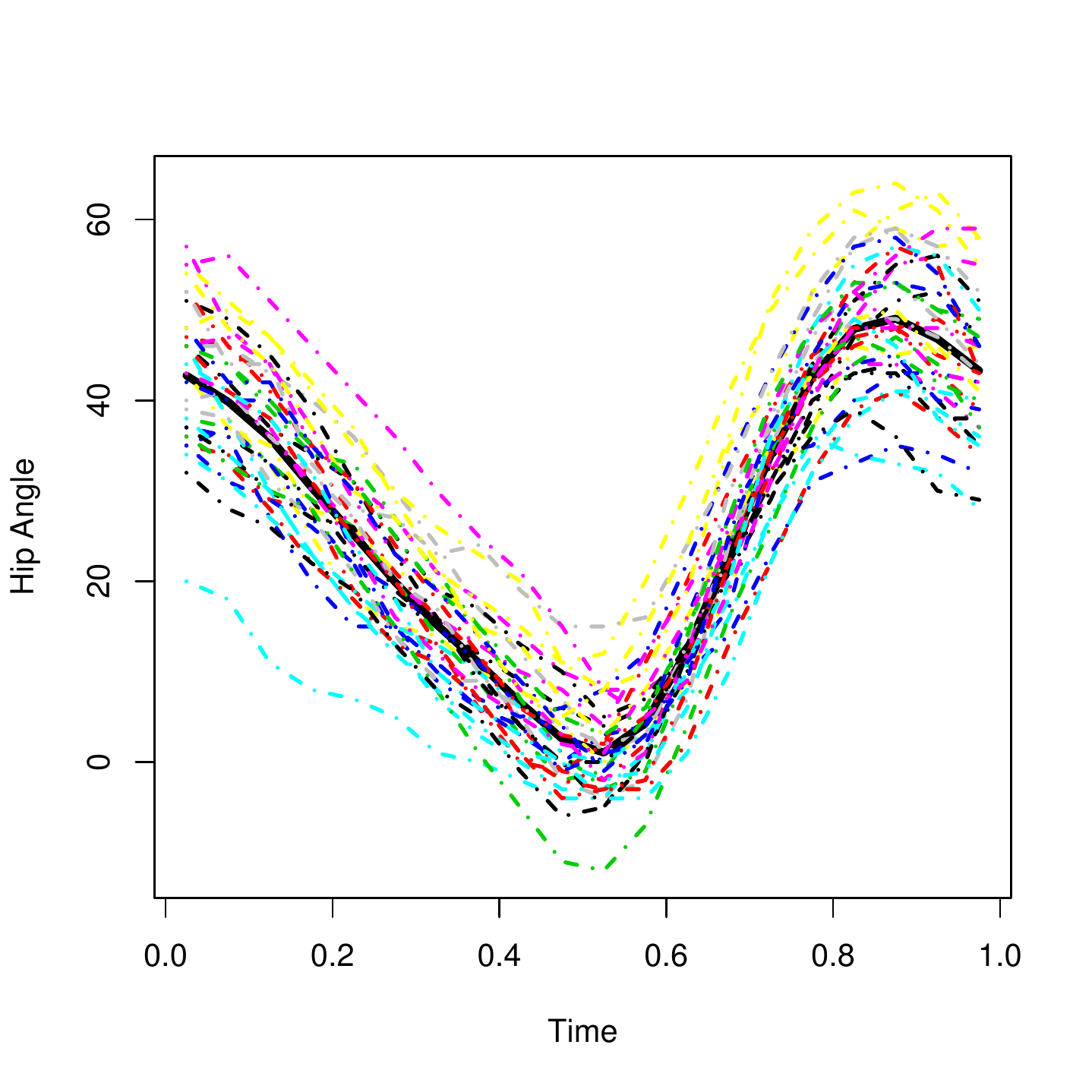}}
\hspace{0.1in}
\subfigure[]{\label{fig:Hip-b}
\hspace*{0in} \includegraphics[trim={1cm 1cm 1.5cm 1.5cm}, width=2.35in, height=2.3in]{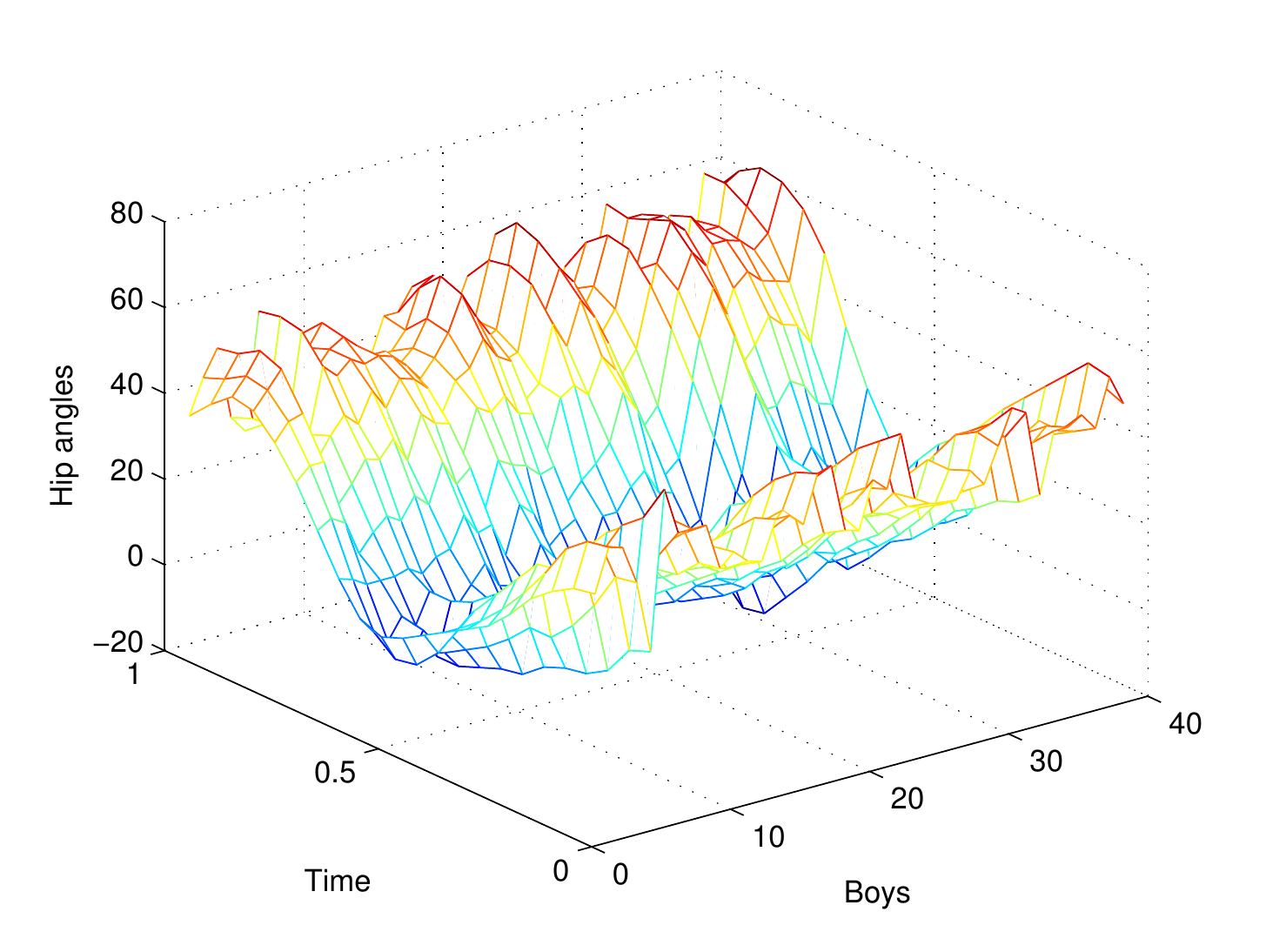}}
\hspace{0.1in} \vskip -.1in
\caption{(a) Hip angle data (dotted lines) with its mean function estimator
(solid line); (b) 3D plot.}
\label{Fig:Hip plot}
\end{figure}

To further illustrate our methodology, we first consider the modeling of the Gait Data collected by the Motion Analysis Laboratory at the Children's Hospital in San Diego, CA. We focus on the ``Hip Angle" functional dataset, which consists of the angles formed by the hip of each boy over his gait cycle. See \cite{OBWS89} for the details. In the study, the cycle begins and ends at the point where the heel of the limb under observation strikes the ground, which has been translated into values over $[0,1]$. There are measurements on $n=39$ samples (boys), where for each sample $N=20$ hip angles were recorded every $0.05$ second with time being measured on $[0,1]$. Denote by $Y_{ij}$ the hip angle of the $i$th boy at the time $x_{j}$, $j=1,\ldots ,N$ and $i=1,\ldots ,n$. Figure \ref{Fig:Hip plot} (a) shows hip curves together with their estimated mean curve, and Figure \ref{Fig:Hip plot} (b) describes the 3D shape of all curves, where ``time" is plotted on one axis and sample index on the other.

\begin{figure}[htbp]
\begin{center}
\subfigure[]{\label{fig:subfig:a}
\hspace*{0in} \includegraphics[trim={1cm 1cm 1.2cm 1cm}, width=2.3in,height=2.2in]{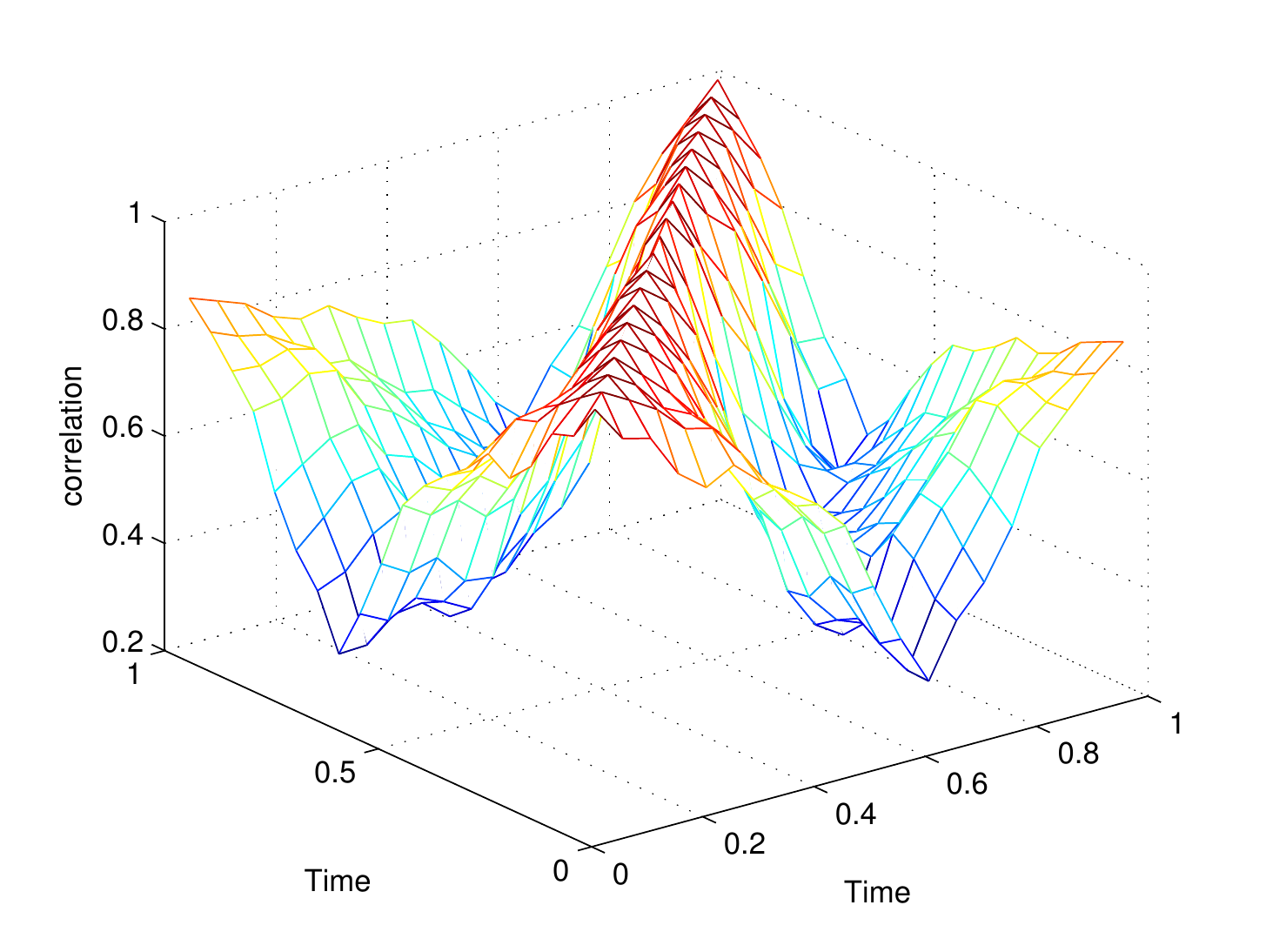}}
\hspace{0.05in}
\subfigure[]{\label{fig:subfig:b}
\hspace*{0in} \includegraphics[trim={1cm 1cm 1.2cm 1cm}, width=2.3in,height=2.2in]{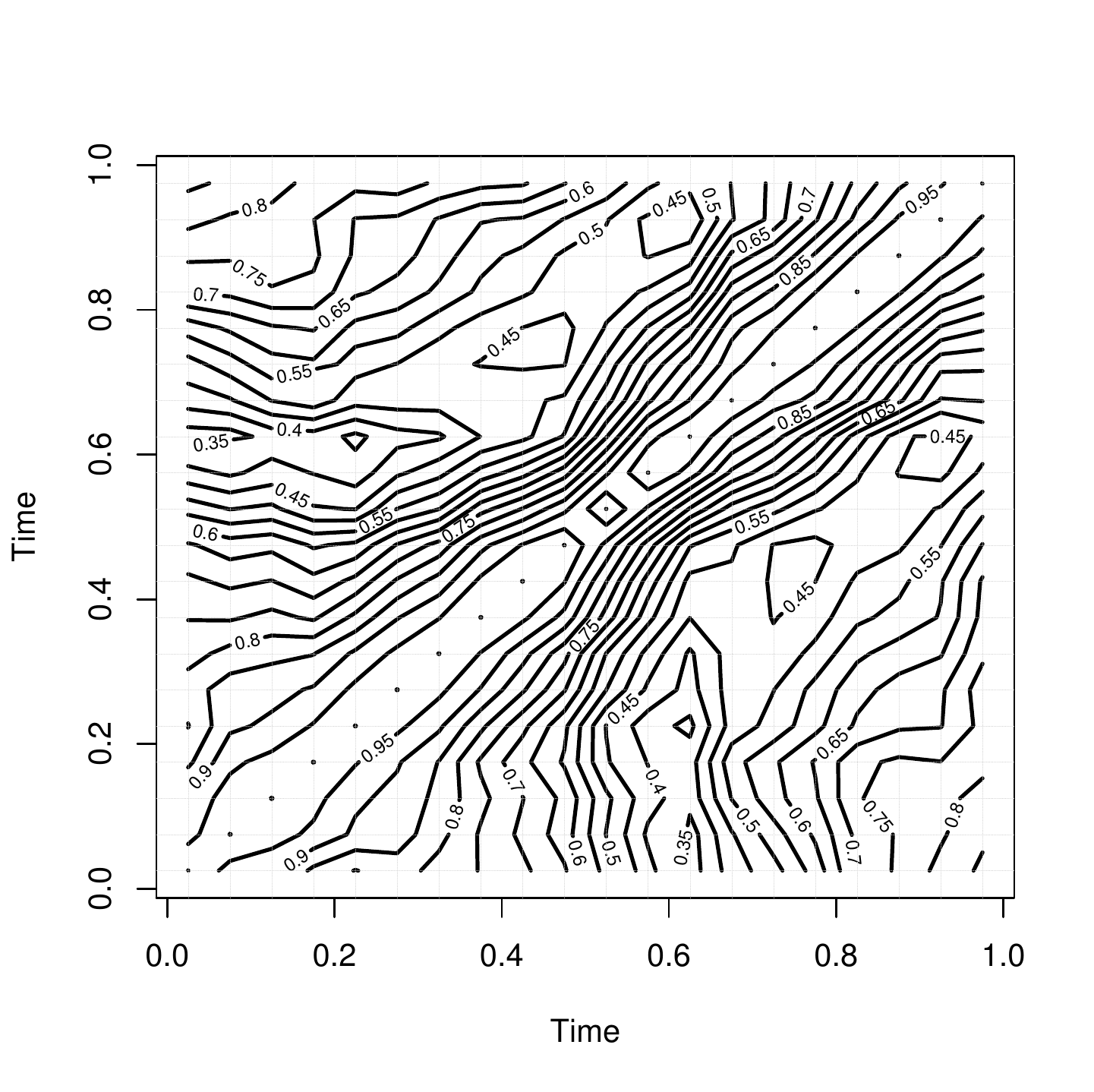}}
\hspace{0.05in}
\end{center} \vskip -.2in
\caption{(a): Unsmoothed sample correlation 3D plot for the hip angle data;
(b): Unsmoothed sample correlation contour plot for the hip angle data.}
\label{Fig:Hip-cor level 3D and contour plot}
\end{figure}

\begin{figure}[htbp]
\begin{center}
\subfigure[]{\label{fig:surface-a}
\hspace*{0in} \includegraphics[trim={1cm 0.5cm 1.2cm 1.2cm}, width=2.5in]{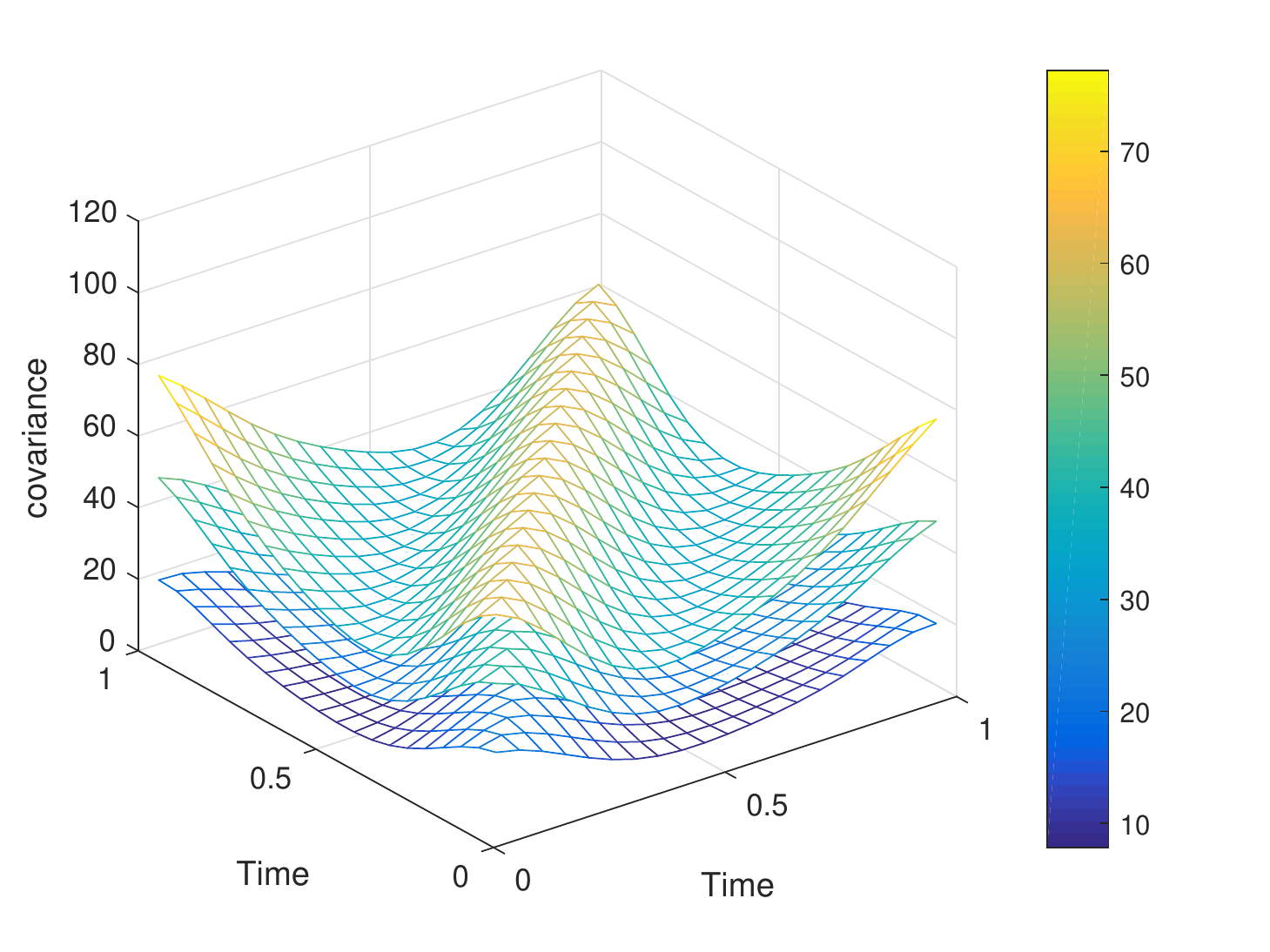}}
\hspace{0.1in}
\subfigure[]{\label{fig:surface-b}
\hspace*{0in} \includegraphics[trim={1cm 0.5cm 1.2cm 1.2cm}, width=2.5in]{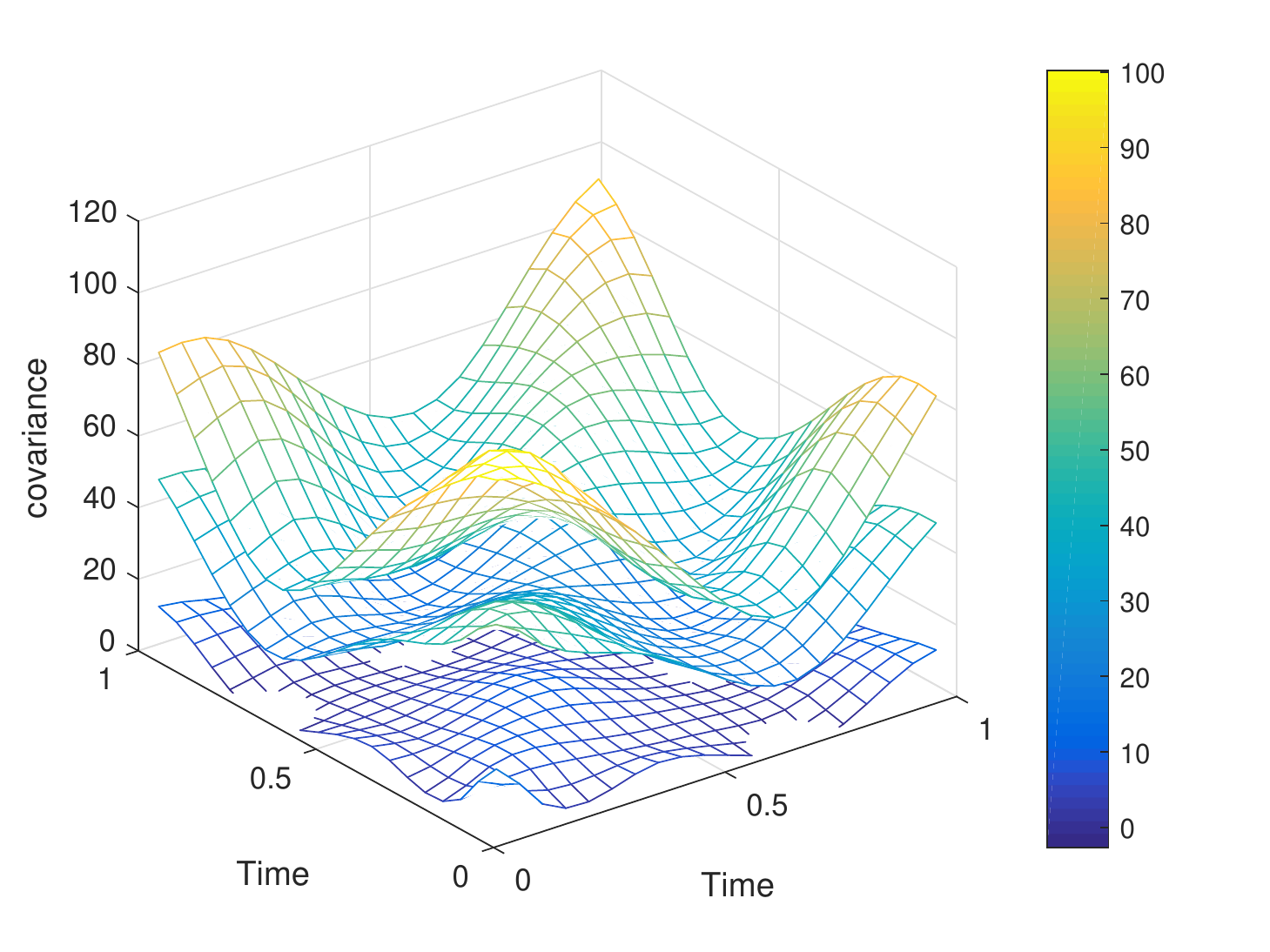}}
\hspace{0.1in}
\end{center} \vskip -.2in
\caption{(a): Covariance matrix estimator based on $\protect\widehat{G}^{\mathrm{PROP}} (x,x^{\prime})$ (middle) with $95\%$ SCE (up and below); (b): covariance matrix estimator $\protect\widehat{G}^{\mathrm{TPS}}(x,x^{\prime})$ (middle) of \protect\cite{CWLY16} with $95\%$ SCE (up and below).}
\label{Fig:surface}
\end{figure}

Figure \ref{Fig:Hip-cor level 3D and contour plot} (a) and (b) display the 3D and contour plots of the sample correlation of the hip data. From the plot, the contours are almost parallel to the main diagonal, indicating that the variation of the hip angles can be considered as an approximately stationary process. Figure \ref{Fig:surface} (a) shows a 3D plot of the proposed covariance matrix estimator $\widehat{G}^{\mathrm{PROP}}(x,x^{\prime })=\widehat{C}(|x-x^{\prime }|)$ with its asymptotic SCE. For comparison, the nonstationary covariance function estimator $\widehat{G}^{\mathrm{TPS}}$ and its SCE are also presented; see Figure \ref{Fig:surface} (b).

\begin{figure}[tbp]
\begin{center}
\subfigure[99\% SCB for M1]{\label{test-hip-case1-9999}
\hspace*{0in} \includegraphics[trim={1cm 0.8cm 1cm 2cm}, height=2.1in, width=1.9in]{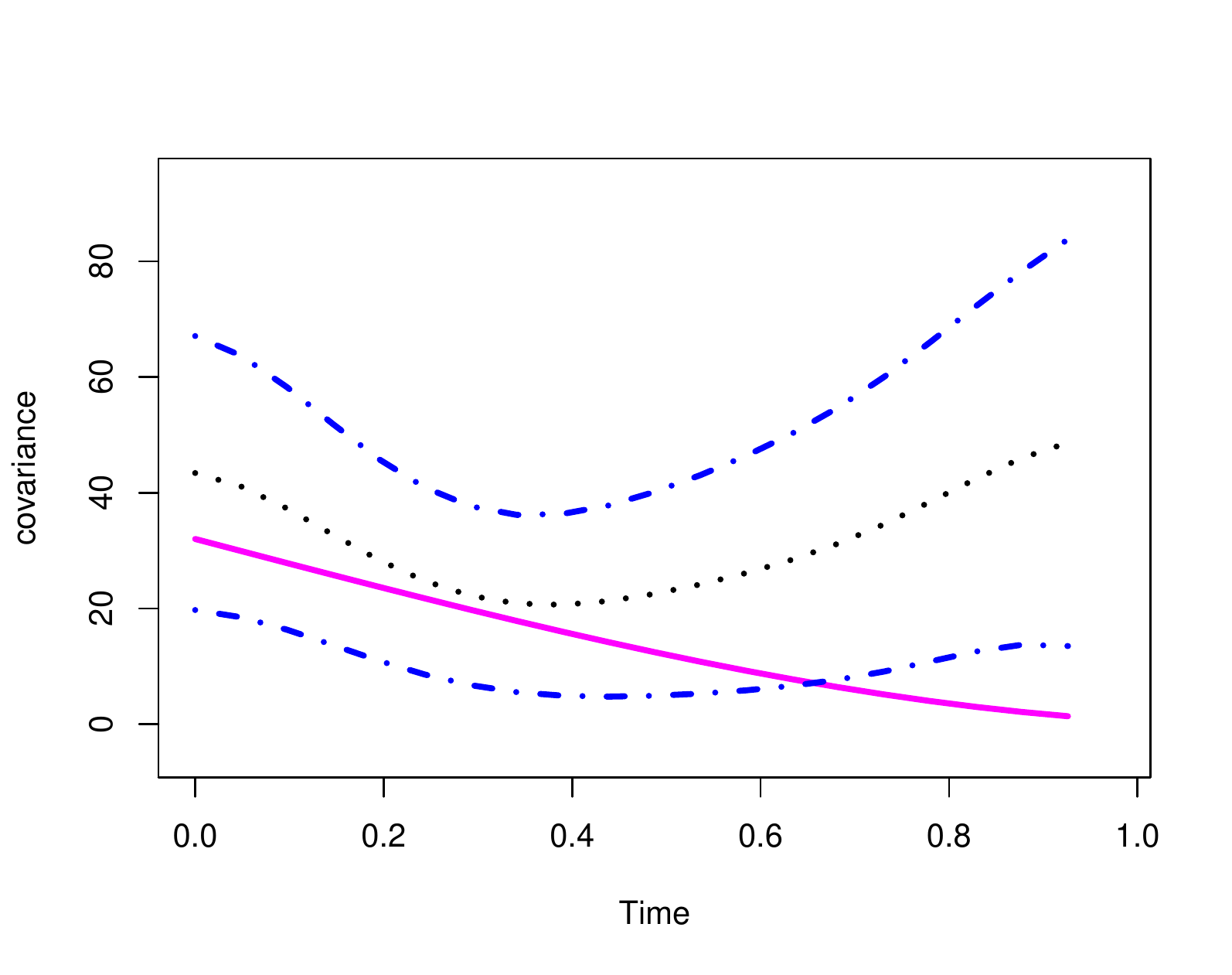}}
\subfigure[80\% SCB for M2]{\label{test-hip-case2-80}
\hspace*{0in} \includegraphics[trim={1cm 0.8cm 1cm 2cm}, height=2.1in, width=1.9in]{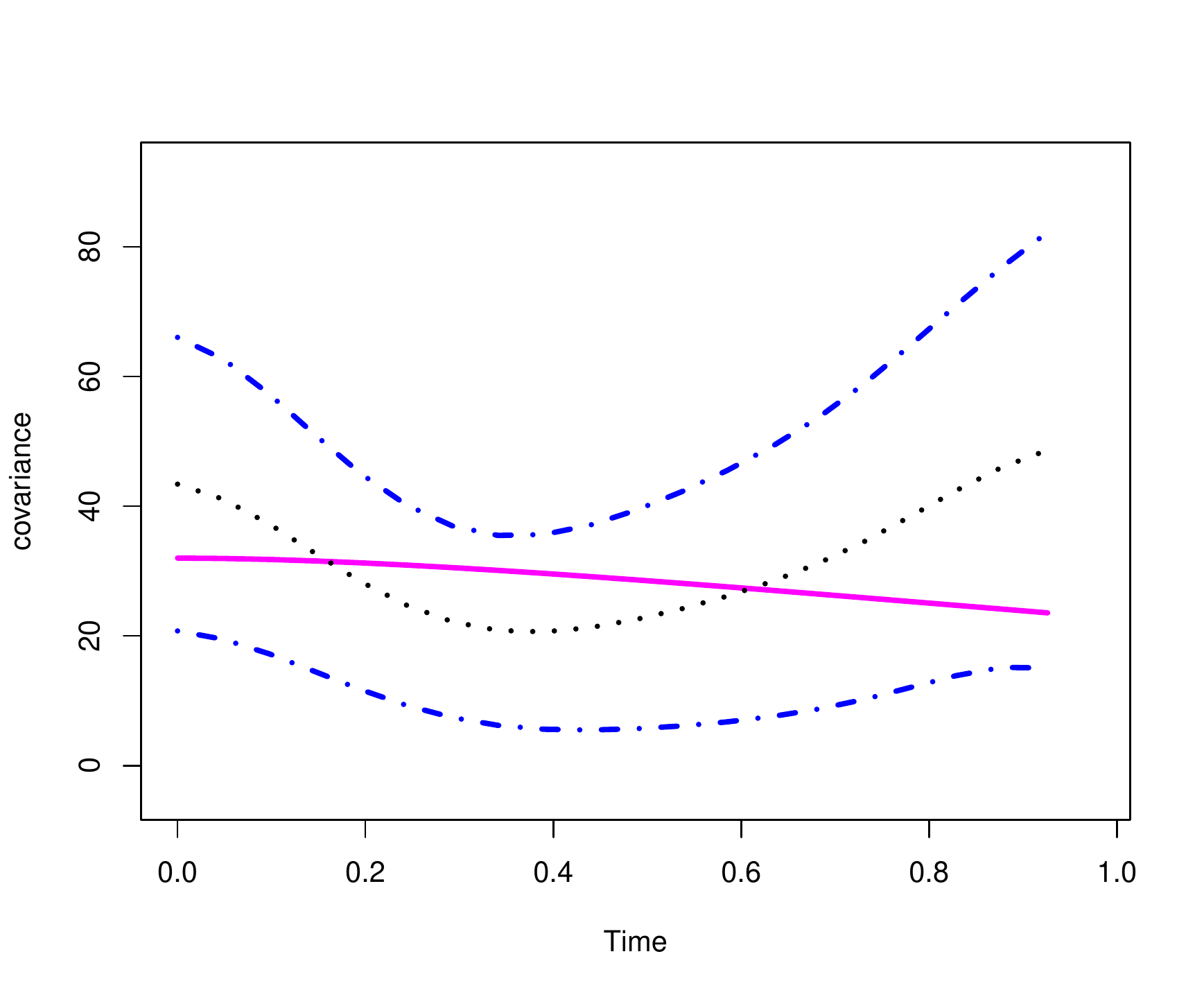}}
\subfigure[80\% SCB for M3]{\label{test-hip-case3-80}
\hspace*{0in} \includegraphics[trim={1cm 0.8cm 1cm 2cm}, height=2.1in, width=1.9in]{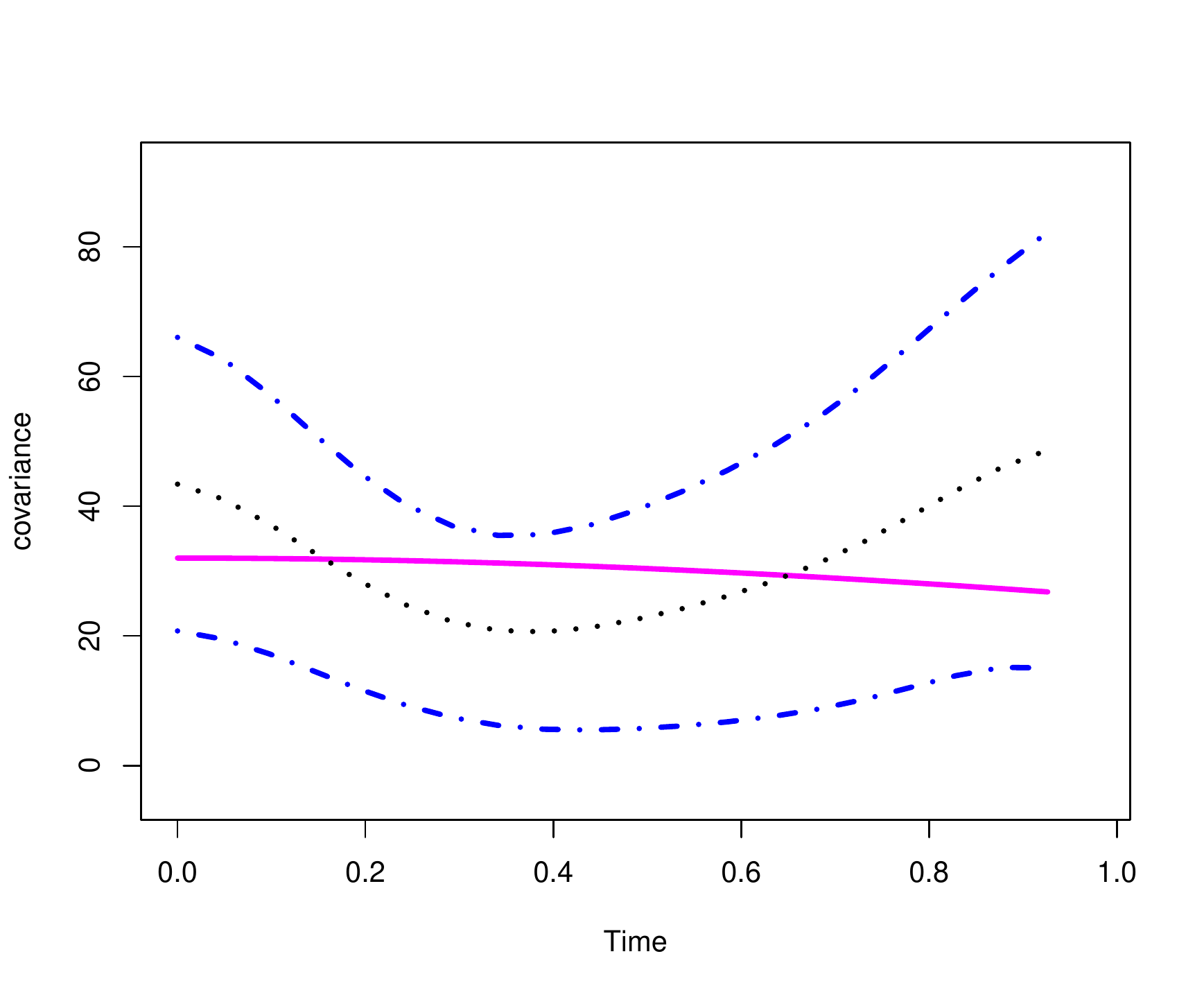}}
\end{center} \vskip -.2in
\caption{{\protect\small Covariance function $C\left( h\right)$ under $H_0$ {(thick solid line)}, B-spline covariance estimator $\protect\widehat{C}$ (dotted line), and the SCB based on $\protect\widehat{C}$ {(dotted-dashed line)} for the hip angle data. }}
\label{Fig:Hip-SCBs-test-S32}
\end{figure}

As mentioned in Section \ref{SEC:intro}, SCB is a very insightful and useful tool to examine the adequacy of certain parametric specifications of a covariance function. {Now we make use of the proposed SCB to test if this hip data has a parametric covariance form like M1, M2 or M3.} We set the null hypothesis $H_{0}$ for M1, M2 and M3 in the following:
\begin{align}
\text{M1 }H_{0} &\text{:}\text{ }C\left( h\right) =C\left( h;32,\theta
_{s}\right) =32\{1-1.5\left( h/\theta _{s}\right) +0.5(h/\theta
_{s})^{3}\}I\left\{ h\leq \theta _{s}\right\} \text{,}  \label{EQ:testM1} \\
\text{M2 }H_{0} &\text{:}\text{ }C\left( h\right) =C(h;32,\theta _{s},\nu
)=32\left\{ \Gamma \left( v\right) \right\} ^{-1}2^{1-v}\left\{ 2\sqrt{v}%
h/\theta _{s}\right\} ^{v}\digamma _{v}\left( 2\sqrt{v}h/\theta _{s}\right)
\text{,}  \label{EQ:TestM2} \\
\text{M3 }H_{0} &\text{:}\text{ }C\left( h\right) =C(h;32,\theta
_{s})=32\exp (-h^{2}/\theta _{s}^{2}),  \label{EQ:TestM3}
\end{align}%
where $\theta _{s}=1.12$ for M1 and M2, $v=1.2$ for M2 and $\theta _{s}=2.19$ for M3. In Figure \ref{Fig:Hip-SCBs-test-S32}, {the thick solid line is the covariance function $C\left( h\right)$ under $H_0$, the center dashed line is the B-spline estimator , and the dotted-dashed lines are the SCBs.} From Figure \ref{Fig:Hip-SCBs-test-S32} (a), one observes that even the $99\%$ SCB cannot contain $C\left( h;32,1.12\right) $, hence the null hypothesis in (\ref{EQ:testM1}) is rejected with $p$-value $< 0.01$. Figure \ref{Fig:Hip-SCBs-test-S32} (b) and (c) indicate that the $80\%$ SCB contains $C\left( h;32,1.2,1.12\right) $ and $C\left( h;32,2.19\right) $, the null hypothesis in (\ref{EQ:TestM2}) and (\ref{EQ:TestM3}) is not rejected with $p$-value $>0.2$.

\section*{Acknowledgment}
\label{Sec:acknowledgment}

This work is supported in part by National Natural Science Foundation of China awards NSFC 11771240, 11801272, Natural Science Foundation of Jiangsu BK20180820; Natural Science Foundation of the Higher Education Institutions of Jiangsu Province 17KJB110005, 19KJA180002, China Scholarship Council, the National Science Foundation grants DMS 1542332, DMS 1736470 and DMS 1916204. The authors are truly grateful to the editor, the associate editor, two reviewers, and Mr. Jie Li from Tsinghua University Center for Statistical Science for their constructive comments and suggestions that led to significant improvement of the paper.

\setcounter{chapter}{6} 
\renewcommand{\theequation}{A.\arabic{equation}} \renewcommand{%
\thesection}{A.\arabic{section}} \renewcommand{\thesubsection}{A.%
\arabic{subsection}} \renewcommand{\thetheorem}{A.\arabic{theorem}} %
\renewcommand{\thelemma}{A.\arabic{lemma}} \renewcommand{\theproposition}{A.%
\arabic{proposition}} \renewcommand{\thecorollary}{A.\arabic{corollary}} %
\renewcommand{\thefigure}{A.\arabic{figure}} \renewcommand{\thetable}{A.%
\arabic{table}} \setcounter{equation}{0} \setcounter{theorem}{0} %
\setcounter{lemma}{0} \setcounter{figure}{0} \setcounter{table}{0} %
\setcounter{proposition}{0} \setcounter{section}{0} %
\setcounter{subsection}{0} \markboth{}{}

\vskip .4in \noindent \textbf{\Large Appendices}

\section*{A. Technical Lemmas and Proofs of Propositions
\protect\ref{Pro: Uniform Ctilde-C} and \protect\ref{PROP:Uniform Chat-Ctilde}}

\label{APP:Lemmas} Throughout this section, $\mathcal{O}_{p}$ (or ${%
\scriptstyle{\mathcal{O}}}_{p}$) denotes a sequence of random variables of
certain order in probability. For instance, ${\scriptstyle{\mathcal{O}}}%
_{p}(n^{-1/2})$ means a smaller order than $n^{-1/2}$ in probability, and by
$\mathcal{O}_{a.s.}$ (or ${\scriptstyle{\mathcal{O}}}_{a.s.}$) almost surely
$\mathcal{O}$ (or ${\scriptstyle{\mathcal{O}}}$). In addition, $\mathcal{U}%
_{p}$ denotes a sequence of random functions which are $\mathcal{O}_{p}$
uniformly defined in the domain.

For any vector $\mathbf{a}=\left(a_{1},\ldots,a_{n}\right) \in \mathcal{R}%
^{n}$, denote the norm $\left\Vert \mathbf{a}\right\Vert _{r}=(\left\vert
a_{1}\right\vert ^{r}+$ $\cdots $ $+\left\vert a_{n}\right\vert ^{r})^{1/r}$%
, $1\leq r<+\infty $, $\left\Vert \mathbf{a}\right\Vert_{\infty}=\max
\left(\left\vert a_{1}\right\vert ,\ldots,\left\vert a_{n}\right\vert
\right) $. For any matrix $\mathbf{A}=\left(a_{ij}\right) _{i=1,j=1}^{m,n}$,
denote its $L_{r}$ norm as $\left\Vert \mathbf{A}\right\Vert _{r}=\max_{%
\mathbf{a}\in \mathcal{R}^{n},\mathbf{a}\neq \mathbf{0}}\left\Vert \mathbf{Aa%
}\right\Vert _{r}$ $\left\Vert \mathbf{a}\right\Vert _{r}^{-1}$, for $%
r<+\infty $ and $\left\Vert \mathbf{A}\right\Vert _{r}=\max_{1\leq i\leq
m}\sum_{j=1}^{n}\left\vert a_{ij}\right\vert $, for $r=\infty $.

\subsection{Lemmas}
\label{subsection:lem}

Let $\mathbf{Y}_{i}=\left(Y_{i1},\ldots,Y_{iN}\right)^{\top}$, then the
spline estimator $\widehat{\eta}_{i}(x)$ in (\ref{EQ:eta-i-hat}) can be
represented as $\widehat{\eta}_{i}(x)=\mathbf{B}(x)^{\top}(\mathbf{B}^{\top}%
\mathbf{B})^{-1}\mathbf{B}^{\top}\mathbf{Y}_{i}$, where $\mathbf{B}$ is
given in (\ref{DEF:B}). Define the empirical inner product matrix of
B-spline basis $\left\{B_{\ell,p}(x) \right\} _{\ell=1}^{J_{s}+p}$ as
$\mathbf{V}_{n,p} =\left\{\left\langle
B_{\ell,p},B_{\ell^{\prime},p}\right\rangle _{N}\right\}
_{\ell,\ell^{\prime}=1}^{J_{s}+p}=N^{-1}\mathbf{B}^{\top}\mathbf{B}$,
and according to Lemma A.3 in \cite{CYT12}, for some constant $C_{p}>0 $
\begin{equation}
\left\Vert \mathbf{V}_{n,p}^{-1}\right\Vert_{\infty}\leq C_{p} J_{s}.
\label{EQ:Vhatp-invbound}
\end{equation}

According to model (\ref{DEF:model j/N}), $\mathbf{\eta }_{i}=\mathbf{m}+%
\mathbf{Z}_{i}$, where $\mathbf{\eta }_{i}=\left\{ \eta _{i}\left(
1/N\right) ,\ldots ,\eta _{i}\left( N/N\right) \right\} ^{\top }$, $\mathbf{Z%
}_{i}=\left\{ Z_{i}\left( 1/N\right) ,\ldots , Z_{i}\left( N/N\right)
\right\} ^{\top }$, $\mathbf{m}=\left\{ m\left(1/N\right) ,\ldots ,m\left(
N/N\right) \right\} ^{\top }$, then the approximation error $\widehat{\eta }%
_{i}(x)-\eta _{i}(x)$ can be decomposed into the following:
\begin{equation}
\widehat{\eta }_{i}(x)-\eta _{i}(x)=\widetilde{\eta }_{i}(x)-\eta _{i}(x)+%
\widetilde{\varepsilon }_{i}(x),  \label{DEF:etahat_project}
\end{equation}%
where $\widetilde{\varepsilon }_{i}(x) =N^{-1}\mathbf{B}(x)^{\mathbf{\top }}%
\mathbf{V}_{n,p}^{-1}\mathbf{B}^{\top }\mathbf{\varepsilon }_{i}$, and
\begin{align}
\widetilde{\eta }_{i}(x)& =N^{-1}\mathbf{B}(x)^{\top }\mathbf{V}_{n,p}^{-1}%
\mathbf{B}^{\top }\mathbf{\eta }_{i}=\widetilde{m}(x)+\widetilde{Z}_{i}(x),
\label{DEF:etai_tilde} \\
\widetilde{m}(x)& =N^{-1}\mathbf{B}(x)^{\top }\mathbf{V}_{n,p}^{-1}\mathbf{B}%
^{\top }\mathbf{m},~\widetilde{Z}_{i}(x)=N^{-1}\mathbf{B}(x)^{\mathbf{\top }}%
\mathbf{V}_{n,p}^{-1}\mathbf{B}^{\top }\mathbf{Z}_{i},  \label{DEF:mZ_tilde}
\end{align}%
where $\mathbf{\varepsilon }_{i}=(\sigma \left( 1/N\right)
\varepsilon_{i1},\ldots ,\sigma \left( N/N\right) \varepsilon _{iN})^{\top }
$. Thus, one has $\widehat{\eta }_{i}(x)-\eta _{i}(x)=\widetilde{Z}%
_{i}(x)-Z_{i}(x)+\widetilde{m}(x)-m(x)+\widetilde{\varepsilon }_{i}(x)$.
Therefore, by (\ref{DEF:Z-residuals}), (\ref{EQ:mhat}) and (\ref%
{DEF:etahat_project}), the approximation error of $\widehat{Z}_{i}(x)$ in (%
\ref{DEF:Z-residuals}) to $Z_{i}(x)$ can be represented by
\begin{equation}
\widehat{Z}_{i}(x)-Z_{i}(x)=\widetilde{Z}_{i}(x)-Z_{i}(x)+\widetilde{%
\varepsilon }_{i}(x)-\frac{1}{n}\sum_{i^{\prime }=1}^{n}\left\{ \widetilde{Z}%
_{i^{\prime }}(x)+\widetilde{\varepsilon }_{i^{\prime }}(x)\right\} .
\label{EQ:Zhati-Zi}
\end{equation}

\begin{lemma}
\label{LEM:max-etatilde-eta} Under Assumptions (A1)--(A6), as $N\rightarrow
\infty$, one has
\begin{align}
\max_{1\leq i\leq n}\left\Vert \widetilde{\eta}_{i}-\eta _{i}\right\Vert
_{\infty }&=\mathcal{O}_{a.s.} \{J_{s}^{-p^* }\left(n\log n\right)^{2/r_{1}}
\},  \label{EQ:etabiasbound} \\
\max_{1\leq i\leq n}\Vert\widetilde{Z}_{i}-Z_{i}\Vert_{\infty}=\mathcal{O}%
_{a.s.} \{J_{s}^{-p^* } (n\log n )^{2/r_{1}} \}, &~
\max_{1\leq i\leq n}\left\Vert Z_{i}\right\Vert_{\infty}=\mathcal{O}_{a.s.}
\{(n\log n)^{2/r_{1}} \}. \notag
\end{align}
\end{lemma} \medskip


\begin{lemma}
\label{LEM:bound_B_eps_ij} Under Assumptions (A1)--(A6), as $N\rightarrow
\infty $, one has
\begin{equation*}
\max_{1\leq i\leq n}\left\Vert \widetilde{\varepsilon }_{i}\right\Vert
_{\infty }=\mathcal{O}_{a.s}\{ J_{s}^{1/2}N^{-1/2}(\log N)^{1/2}\}.
\end{equation*}
\end{lemma} \medskip

The next lemma follows from Lemmas \ref{LEM:max-etatilde-eta}, \ref%
{LEM:bound_B_eps_ij}, (\ref{DEF:etahat_project}) and (\ref{EQ:Zhati-Zi}).

\begin{lemma}
\label{LEM:Zij'-Zij'hat} Under Assumptions (A1)--(A6), as $N\rightarrow
\infty $,
\begin{align}
\max_{1\leq i\leq n}\left\Vert \widehat{\eta}_{i}-\eta
_{i}\right\Vert_{\infty}&=\mathcal{O}_{p}\left\{J_{s}^{-p^* }\left(n\log
n\right)^{2/r_{1}}+J_{s}^{1/2}N^{-1/2}\left(\log N\right)^{1/2}\right\},
\notag \\
\max_{1\leq i\leq n}\Vert\widehat{Z}_{i}-Z_{i}\Vert_{\infty}&=\mathcal{O}%
_{p}\left\{J_{s}^{-p^* }\left(n\log
n\right)^{2/r_{1}}+J_{s}^{1/2}N^{-1/2}\left(\log N\right)^{1/2}\right\}.
\label{EQ:Zhat-Z}
\end{align}
\end{lemma} \medskip


\begin{lemma}
\label{LEM: strong-moment} Assumption (A5) holds under Assumptions (A4) and
(A5').
\end{lemma} \medskip


\begin{lemma}
\label{LEM: sup_Zi*Zitilde-Zi} Under Assumptions (A1)--(A6),%
\begin{equation*}
\sup_{h\in \left[ 0,h_{0}\right] }\left\vert \frac{1}{n(1-h)}\int_{0}^{1-h}
\sum_{i=1}^{n}Z_{i}(x+h)\left\{\widetilde{Z}_{i}(x)-Z_{i}(x)\right\}
dx\right\vert ={\scriptstyle{\mathcal{O}}}_{p}(n^{-1/2}).
\end{equation*}
\end{lemma} \medskip


\begin{lemma}
\label{LEM:max_B_Uijeps}Under Assumptions (A1)--(A6),
\begin{equation*}
\max_{1\leq i\leq n}\max_{1\leq \ell \leq J_{s}+p}\left\vert
N^{-1}\sum_{j=1}^{N}B_{\ell ,p}(j/N) \sigma \left( j/N\right)
U_{ij,\varepsilon }\right\vert =\mathcal{O}_{a.s.}(N^{-1/2}J_{s}^{-1/2}\log
^{1/2}N),
\end{equation*}%
where $U_{ij,\varepsilon }$, $1\leq i\leq n$, $1\leq j\leq N$, are {iid}
standard normal random variables.
\end{lemma} \medskip


\begin{lemma}
\label{LEM:max_kxi_eps-Uijeps} Under Assumptions (A1)--(A6),%
\begin{equation*}
\max_{1\leq k\leq k_{n} \atop 1\leq \ell \leq J_{s}+p} \left\vert \frac{1}{nN}%
\sum_{i=1}^{n}U_{ik,\xi }\left\{ \sum_{j=1}^{N}B_{\ell ,p}\left( \frac{j}{N}%
\right) \sigma \left( j/N\right) (\varepsilon _{ij}-U_{ij,\varepsilon
})\right\} \right\vert =\mathcal{O}_{a.s.}\left( n^{-1/2}N^{\beta
_{2}-1}\log ^{1/2}N\right) ,
\end{equation*}%
where $0<\beta _{2}<1/2$.
\end{lemma} \medskip


\begin{lemma}
\label{LEM:max_kxi-Ukxi*Uijeps} Under Assumptions (A1)--(A6), one has
\begin{align*}
& \max_{1\leq k\leq k_{n} \atop 1\leq \ell \leq J_{s}+p}\left\vert
(nN)^{-1}\sum_{i=1}^{n}\left( \xi _{ik}-U_{ik,\xi }\right)
\sum_{j=1}^{N}B_{\ell ,p}\left( \frac{j}{N}\right) \sigma \left( \frac{j}{N}%
\right) U_{ij,\varepsilon }\right\vert \\
& =\mathcal{O}_{a.s.}(n^{\beta _{1}-1/2}N^{-1/2}J_{s}^{-1/2}\log ^{1/2}N),
\textrm{~ for ~} 0<\beta _{1}<1/2.
\end{align*}%
\end{lemma} \medskip


\begin{lemma}
\label{LEM:max-kxi_ik-U_ikkxi} Under Assumptions (A2)--(A6),%
\begin{equation*}
\max_{1\leq k\leq k_{n} \atop 1\leq \ell \leq J_{s}+p}\left\vert \frac{1}{nN}%
\sum_{i=1}^{n}\left( \xi _{ik}-U_{ik,\xi }\right) \left\{
\sum_{j=1}^{N}B_{\ell ,p}\left( \frac{j}{N}\right) \sigma \left( \frac{j}{N}%
\right) (\varepsilon _{ij}-U_{ij,\varepsilon })\right\} \right\vert =%
\mathcal{O}_{a.s.}\left( n^{\beta _{1}}N^{\beta _{2}-1}\right) .
\end{equation*}
\end{lemma} \medskip


\begin{lemma}
\label{LEM: sup_meanZi_epsi_tilde} Under Assumptions (A2)--(A6),
\begin{equation*}
\sup_{h\in \left[ 0,h_{0}\right] }\sup_{x\in \left[ 0,1\right] }\left\vert
\frac{1}{n}\sum_{i=1}^{n}Z_{i}(x+h)\widetilde{\varepsilon }%
_{i}(x)\right\vert ={\scriptstyle{\mathcal{O}}}_{p}(n^{-1/2}).
\end{equation*}
\end{lemma} \medskip


\begin{lemma}
\label{LEM:supsup_meanZimeanZi'} Under Assumptions (A2)--(A6), one has%
\begin{equation*}
\sup_{h\in \left[ 0,h_{0}\right] }\sup_{x\in \left[ 0,1\right] }\left\vert
\frac{1}{n}\sum\limits_{i=1}^{n}Z_{i}(x+h)\frac{1}{n}\sum\limits_{i^{\prime
}=1}^{n}\widetilde{Z}_{i^{\prime}}(x)\right\vert ={\scriptstyle{\mathcal{O}}}%
_{p}(n^{-1/2}).
\end{equation*}
\end{lemma} \medskip


\begin{lemma}
\label{LEM:maxGaussian} Let $W_{i}\sim N\left( 0,\sigma _{i}^{2}\right)
,\sigma _{i}>0,i=1,\ldots ,n$, then for   $a>2$%
\begin{equation}
\Pr \left( \max_{1\leq i\leq n}\left\vert W_{i}/\sigma _{i}\right\vert >a%
\sqrt{\log n}\right) <2n^{1-a^{2}/2}.
\label{EQ:maxGaussiantailprob}
\end{equation}%
Hence, $\left( \max_{1\leq i\leq n}\left\vert W_{i}\right\vert \right)
/\left( \max_{1\leq i\leq n}\sigma _{i}\right) \leq \max_{1\leq i\leq
n}\left\vert W_{i}/\sigma _{i}\right\vert ={{\mathcal{O}}}_{a.s.}\left(
\sqrt{\log n}\right)$.
\end{lemma}

\subsection{Proof of Proposition \protect\ref{Pro: Uniform Ctilde-C}}

Let $\mathcal{F}_{t}=\sigma \left( \bar{\xi}_{\cdot 11},\bar{\xi}_{\cdot 12},\ldots ,\bar{\xi%
}_{\cdot 1t},\bar{\xi}_{\cdot 22},\ldots ,\bar{\xi}_{\cdot t-1,t},\bar{\xi}%
_{\cdot tt}\right) $, so that $\mathcal{F}_{2}\subseteq \mathcal{F}%
_{3}\subseteq \mathcal{F}_{4}\subseteq \mathcal{\cdots }$ is an increasing
sequence of $\sigma $-fields. Denote
\begin{align*}
S_{t}(h) =&\sqrt{n}\Delta (\cdot )=\sqrt{n}\textstyle\sum\limits_{1\leq
k\neq k^{\prime }\leq t}\bar{\xi}_{\cdot kk^{\prime }}\frac{1}{1-h}%
\int_{0}^{1-h}\phi _{k}(x)\phi _{k^{\prime }}(x+h)dx \\
&+\sqrt{n}\textstyle\sum\limits_{1\leq k\leq t}\left( \bar{\xi}_{\cdot
kk}-1\right) \frac{1}{1-h}\int_{0}^{1-h}\phi _{k}(x)\phi _{k}(x+h)dx,
\end{align*}%
for $t=1,\ldots ,k_{n}$, where $k_{n}$ satisfies Assumption (A4). We show
that $S_{t}(h)$ is a martingale process in $h\in \left[ 0,h_{0}\right] $.

Define $D_{t}(h)=S_{t}(h)-S_{t-1}(h)$, thus,
\begin{equation*}
D_{t}(h)=\frac{\sqrt{n}}{1-h}\left\{ \textstyle\sum\limits_{k=1}^{t-1}\bar{%
\xi}_{\cdot kt}\int_{0}^{1-h}\left\{ \phi _{k}(x)\phi _{t}(x+h)+\phi
_{t}(x)\phi _{k}(x+h)\right\} dx+\left( \bar{\xi}_{\cdot tt}-1\right)
\int_{0}^{1-h}\phi _{t}(x)\phi _{t}(x+h)dx\right\} ,
\end{equation*}%
which is $\mathcal{F}_{t}$-measurable. While notice that for any $t$,
\begin{align*}
& \mathrm{E}\left( \left. D_{t}(h)\right\vert \mathcal{F}_{t-1}\right) =%
\frac{\sqrt{n}}{1-h}\mathrm{E}\left\{ \textstyle\sum\limits_{k=1}^{t-1}\bar{%
\xi}_{\cdot kt}\int_{0}^{1-h}\left\{ \phi _{k}(x)\phi _{t}(x+h)+\phi
_{t}(x)\phi _{k}(x+h)\right\} dx\right. \\
& \quad \left. \left. +\left( \bar{\xi}_{\cdot tt}-1\right)
\int_{0}^{1-h}\phi _{t}(x)\phi _{t}(x+h)dx\right\vert \mathcal{F}%
_{t-1}\right\} \\
& =\frac{\sqrt{n}}{1-h}\mathrm{E}\left\{ \left. \frac{1}{n}\textstyle%
\sum\limits_{i=1}^{n}\xi _{it}\textstyle\sum\limits_{k=1}^{t-1}\xi
_{ik}\int_{0}^{1-h}\left\{ \phi _{k}(x)\phi _{t}(x+h)+\phi _{t}(x)\phi
_{k}(x+h)\right\} dx\right\vert \mathcal{F}_{t-1}\right\} \\
& \quad +\sqrt{n}\mathrm{E}\left\{ \left. \left( \bar{\xi}_{\cdot
tt}-1\right) \int_{0}^{1-h}\phi _{t}(x)\phi _{t}(x+h)dx\right\vert \mathcal{F%
}_{t-1}\right\} =0,
\end{align*}%
which implies that $\left\{ D_{t}(h),t=2,3,\ldots \right\} $ is a martingale
difference process with respect to $\left\{ \mathcal{F}_{t-1},t=2,3,\ldots
\right\} $.

Next denote
\begin{equation}
\mathrm{E}\left( \left. D_{t}^{2}(h)\right\vert \mathcal{F}_{t-1}\right) =%
\mathrm{V}_{t}^{\left( 1\right) }(h)+\mathrm{V}_{t}^{\left( 2\right) }(h)+%
\mathrm{V}_{t}^{\left( 3\right) }(h),  \label{EQ:EDt^2}
\end{equation}%
in which
\begin{align*}
\mathrm{V}_{t}^{\left( 1\right) }(h)=& n\mathrm{E}\left[ \left. \left\{
n^{-1}\textstyle\sum\limits_{i=1}^{n}\xi _{it}\textstyle\sum%
\limits_{k=1}^{t-1}\xi _{ik}\frac{1}{1-h}\int_{0}^{1-h}\left\{ \phi
_{k}(x)\phi _{t}(x+h)+\phi _{t}(x)\phi _{k}(x+h)\right\} dx\right\}
^{2}\right\vert \mathcal{F}_{t-1}\right] , \\
\mathrm{V}_{t}^{\left( 2\right) }(h)=& n\mathrm{E}\left[ \left. \left\{
\left( \bar{\xi}_{\cdot tt}-1\right) \frac{1}{1-h}\int_{0}^{1-h}\phi
_{t}(x)\phi _{t}(x+h)dx\right\} ^{2}\right\vert \mathcal{F}_{t-1}\right] , \\
\mathrm{V}_{t}^{\left( 3\right) }(h)=& 2n\mathrm{E}\left[ \left\{ n^{-1}%
\textstyle\sum\limits_{i=1}^{n}\xi _{it}\textstyle\sum\limits_{k=1}^{t-1}\xi
_{ik}\frac{1}{1-h}\int_{0}^{1-h}\left\{ \phi _{k}(x)\phi _{t}(x+h)+\phi
_{t}(x)\phi _{k}(x+h)\right\} dx\right\} \right. \\
& \times \left. \left. \left( \bar{\xi}_{\cdot tt}-1\right) \frac{1}{1-h}%
\int_{0}^{1-h}\phi _{t}(x)\phi _{t}(x+h)dx\right\vert \mathcal{F}_{t-1}%
\right] .
\end{align*}%
Moreover, one can show that
\begin{align*}
\mathrm{V}_{t}^{\left( 1\right) }(h)& =\mathrm{E}\left[ \left. \left\{ \frac{%
1}{n}\textstyle\sum\limits_{i=1}^{n}\xi _{it}\textstyle\sum%
\limits_{k=1}^{t-1}\xi _{ik}\frac{1}{1-h}\int_{0}^{1-h}\left\{ \phi
_{k}(x)\phi _{t}(x+h)+\phi _{t}(x)\phi _{k}(x+h)\right\} dx\right\}
^{2}\right\vert \mathcal{F}_{t-1}\right] \\
& =\mathrm{E}\textstyle\sum\limits_{k=1}^{t-1}\left[ \left. \frac{1}{n}%
\textstyle\sum\limits_{i=1}^{n}\xi _{it}^{2}\xi _{ik}^{2}\left( \frac{1}{1-h}%
\int_{0}^{1-h}\left\{ \phi _{k}(x)\phi _{t}(x+h)+\phi _{t}(x)\phi
_{k}(x+h)\right\} dx\right) ^{2}\right\vert \mathcal{F}_{t-1}\right] \\
& =\mathrm{E}\xi _{1t}^{2}\textstyle\sum\limits_{k=1}^{t-1}\bar{\xi}_{\cdot
kk}\left[ \frac{1}{1-h}\int_{0}^{1-h}\left\{ \phi _{k}(x)\phi _{t}(x+h)+\phi
_{t}(x)\phi _{k}(x+h)\right\} dx\right] ^{2},
\end{align*}%
therefore, one has when $n\rightarrow \infty $,
\begin{align*}
\textstyle\sum\limits_{t=2}^{k_{n}}& \mathrm{V}_{t}^{\left( 1\right)
}(h)\rightarrow \textstyle\sum\limits_{k\neq k^{\prime }}^{\infty }\left\{
\frac{1}{1-h}\int_{0}^{1-h}\phi _{k}(x)\phi _{k^{\prime }}(x+h)dx\right\}
^{2} \\
& +\textstyle\sum\limits_{k\neq k^{\prime }}^{\infty }\left\{ \frac{1}{1-h}%
\int_{0}^{1-h}\phi _{k}(x)\phi _{k^{\prime }}(x+h)dx\right\} \left\{ \frac{1%
}{1-h}\int_{0}^{1-h}\phi _{k}(x+h)\phi _{k^{\prime }}(x)dx\right\} <\infty .
\end{align*}%
Note that $\mathrm{V}_{t}^{\left( 2\right) }(h)=\left( \mathrm{E}\xi
_{1t}^{4}-1\right) \left\{ (1-h)^{-1}\int_{0}^{1-h}\phi _{t}(x)\phi
_{t}(x+h)dx\right\} ^{2}<\infty $, so one has that $\textstyle%
\sum_{t=2}^{k_{n}}\mathrm{V}_{t}^{\left( 2\right) }\left( h\right)
\rightarrow \textstyle\sum_{k=1}^{\infty }\left( \mathrm{E}\xi
_{1k}^{4}-1\right) \left\{ (1-h)^{-1}\int_{0}^{1-h}\phi _{k}(x)\phi
_{k}(x+h)dx\right\} ^{2}<\infty $. Similarly,
\begin{align*}
\mathrm{V}_{t}^{\left( 3\right) }(h)& =2n\mathrm{E}\left[ \left\{ \textstyle%
\sum\limits_{k=1}^{t-1}\frac{1}{n}\textstyle\sum\limits_{i=1}^{n}\xi
_{it}\xi _{ik}\frac{1}{1-h}\int_{0}^{1-h}\left\{ \phi _{k}(x)\phi
_{t}(x+h)+\phi _{t}(x)\phi _{k}(x+h)\right\} dx\right\} \right. \\
& \times \left. \left( \frac{1}{n}\textstyle\sum\limits_{i=1}^{n}\xi
_{it}^{2}-1\right) \frac{1}{1-h}\int_{0}^{1-h}\phi _{t}(x)\phi _{t}(x+h)dx%
\mathcal{F}_{t-1}\right] \\
& =2\left( \mathrm{E}\xi _{1t}^{3}-1\right) \mathrm{E}\left[ \textstyle%
\sum\limits_{k=1}^{t-1}\bar{\xi}_{\cdot k}\frac{1}{1-h}\int_{0}^{1-h}\left\{
\phi _{k}(x)\phi _{t}(x+h)+\phi _{t}(x)\phi _{k}(x+h)\right\} dx\right. \\
& \quad \times \left. \frac{1}{1-h}\int_{0}^{1-h}\phi _{t}(x)\phi _{t}(x+h)dx%
\mathcal{F}_{t-1}\right] ,
\end{align*}%
where $\bar{\xi}_{\cdot k}=n^{-1}\textstyle\sum_{i=1}^{n}\xi _{ik}$. Next,
notice that
\begin{equation*}
\sup_{h\in \left[ 0,h_{0}\right] }\textstyle\sum\limits_{k=1}^{\infty }%
\textstyle\sum\limits_{k^{\prime }=1}^{\infty }\frac{1}{1-h}%
\int_{0}^{1-h}\left\{ \phi _{k}(x)\phi _{t}(x+h)+\phi _{t}(x)\phi
_{k}(x+h)\right\} dx<\infty .
\end{equation*}%
Therefore, one has%
\begin{align*}
\textstyle\sum_{t=2}^{k_{n}}& \mathrm{V}_{t}^{\left( 3\right)
}(h)\rightarrow 2\textstyle\sum_{t=2}^{k_{n}}\left( \mathrm{E}\xi
_{1t}^{3}-1\right) \textstyle\sum_{k=1}^{\infty }\mathrm{E}\left( \left.
\bar{\xi}_{\cdot k}\right\vert \mathcal{F}_{t-1}\right) \times \frac{1}{%
(1-h)^{2}} \\
& \times \int_{0}^{1-h}\left\{ \phi _{k}(x)\phi _{t}(x+h)+\phi _{t}(x)\phi
_{k}(x+h)\right\} dx\times \int_{0}^{1-h}\phi _{t}(x)\phi
_{t}(x+h)dx\rightarrow _{p}0,
\end{align*}%
as $n\rightarrow \infty $.

According to (\ref{EQ:EDt^2}), as $n\rightarrow \infty $, one has
\begin{align*}
\textstyle\sum_{t=2}^{k_{n}} &\mathrm{E}\left( \left. D_{t}^{2}\left(
h\right) \right\vert \mathcal{F}_{t-1}\right) \rightarrow_{p} \textstyle%
\sum\limits_{k\neq k^{\prime }}^{\infty }\left\{ \frac{1}{1-h}
\int_{0}^{1-h}\phi _{k}(x)\phi _{k^{\prime }}(x+h)dx\right\} ^{2} \\
&+\textstyle\sum\limits_{k\neq k^{\prime }}^{\infty }\left\{ \frac{1}{1-h}
\int_{0}^{1-h}\phi _{k}(x)\phi _{k^{\prime }}(x+h)dx\right\} \left\{ \frac{1%
}{1-h}\int_{0}^{1-h}\phi _{k}(x+h)\phi _{k^{\prime }}(x)dx\right\} \\
&+\left( \mathrm{E}\xi _{1t}^{4}-1\right) \left\{ \frac{1}{1-h}
\int_{0}^{1-h}\phi _{t}(x)\phi _{t}(x+h)dx\right\} ^{2}.
\end{align*}
Denote by $\mathrm{E}\left( \left. D_{t}^{3}(h) \right\vert \mathcal{F}%
_{t-1}\right) =d_{t}^{\left( 1\right) }(h) +3d_{t}^{\left( 2\right) }(h)
+3d_{t}^{\left( 3\right) }(h) +d_{t}^{\left( 4\right) }(h)$, where
\begin{align*}
d_{t}^{\left( 1\right) }(h) =&n^{3/2}\mathrm{E}\left[ \left. \left\{ %
\textstyle\sum\limits_{k=1}^{t-1}\bar{\xi}_{\cdot kt}\frac{1}{1-h}%
\int_{0}^{1-h}\left\{ \phi _{k}(x)\phi _{t}(x+h)+\phi _{t}(x)\phi
_{k}(x+h)\right\} dx\right\} ^{3}\right\vert \mathcal{F}_{t-1}\right], \\
d_{t}^{\left( 2\right) }(h) =&n^{3/2}\mathrm{E}\left[ \left\{ \textstyle%
\sum\limits_{k=1}^{t-1}\bar{\xi}_{\cdot kt}\frac{1}{1-h}\int_{0}^{1-h}\left%
\{ \phi _{k}(x)\phi _{t}(x+h)+\phi _{t}(x)\phi _{k}(x+h)\right\} dx\right\}
^{2}\right. \\
&\times \left. \left. \left( \bar{\xi}_{\cdot tt}-1\right) \frac{1}{1-h}%
\int_{0}^{1-h}\phi _{t}(x)\phi _{t}(x+h)dx\right\vert \mathcal{F}_{t-1}%
\right], \\
d_{t}^{\left( 3\right) }(h) =&n^{3/2}\mathrm{E}\left[ \left\{ \textstyle%
\sum\limits_{k=1}^{t-1}\bar{\xi}_{\cdot kt}\frac{1}{1-h}\int_{0}^{1-h}\left%
\{ \phi _{k}(x)\phi _{t}(x+h)+\phi _{t}(x)\phi _{k}(x+h)\right\} dx\right\}
\right. \\
&\times \left. \left. \left( \bar{\xi}_{\cdot tt}-1\right) ^{2}\left\{ \frac{%
1}{1-h}\int_{0}^{1-h}\phi _{t}(x)\phi _{t}(x+h)dx\right\} ^{2}\right\vert
\mathcal{F}_{t-1}\right], \\
d_{t}^{\left( 4\right) }(h) =&n^{3/2}\mathrm{E}\left[ \left. \left\{ \left(
\bar{\xi}_{\cdot tt}-1\right) \frac{1}{1-h}\int_{0}^{1-h}\phi _{t}(x)\phi
_{t}(x+h)dx\right\} ^{3}\right\vert \mathcal{F}_{t-1}\right].
\end{align*}%
Applying similar arguments in Lemma 6 of \cite{CWLY16}, one has $\textstyle%
\sum_{t=2}^{k_{n}}\mathrm{E}\{d_{t}^{\left( i\right) }(h) \vert \mathcal{F}%
_{t-1}\} \rightarrow_{p} 0$, for $i=1,2,3,4$. Hence, for any $\epsilon >0$, $%
\sup_{h\in \left[ 0,h_{0}\right] }\textstyle\sum_{t=2}^{k_{n}}\mathrm{E}%
\left\{ \left. D_{t}^{3}(h) I\left( D_{t}^{2}(h) >\epsilon \right)
\right\vert \mathcal{F}_{t-1}\right\} \rightarrow_{p} 0$.

By the uniform central limit theorem, one has $\sqrt{n}\Delta (\cdot
)=S_{t}(h) \rightarrow _{D}\zeta (\cdot )$, as $n\rightarrow \infty $, where
$\zeta (h)$ is a Gaussian process such that $\mathrm{E}\zeta (h)=0$,
\begin{align*}
\Xi (h) &=\mathrm{E}\zeta ^{2}(h)=\sum_{k=1}^{\infty }\left( \mathrm{E}\xi
_{1k}^{4}-1\right) \left( \frac{1}{1-h}\int_{0}^{1-h}\phi _{k}(x)\phi
_{k}(x+h)dx\right) ^{2} \\
&+\sum_{k<k^{\prime }}^{\infty }\left\{ \frac{1}{1-h}\left(
\int_{0}^{1-h}\phi _{k}(x)\phi _{k^{\prime }}(x+h)dx+\int_{0}^{1-h}\phi
_{k^{\prime }}(x)\phi _{k}\left( x+h\right) dx\right) \right\} ^{2},
\end{align*}%
and covariance function
\begin{align*}
\Omega \left( h,h^{\prime }\right) =& \mathrm{Cov}\left( \zeta (h),\zeta
\left( h^{\prime }\right) \right) =\frac{1}{1-h}\frac{1}{1-h^{\prime }} \\
& \times \Bigg\{\int_{0}^{1-h}\int_{0}^{1-h^{\prime }}\sum_{k,k^{\prime
}=1}^{\infty }\phi _{k}(x)\phi _{k}\left( x^{\prime }\right) \phi
_{k^{\prime }}(x+h)\phi _{k^{\prime }}\left( x^{\prime }+h^{\prime }\right)
dxdx^{\prime } \\
& +\int_{0}^{1-h}\int_{0}^{1-h^{\prime }}\sum_{k,k^{\prime }=1}^{\infty
}\phi _{k}(x)\phi _{k}\left( x^{\prime }+h^{\prime }\right) \phi _{k^{\prime
}}\left( x+h\right) \phi _{k^{\prime }}\left( x^{\prime }\right)
dxdx^{\prime } \\
& +\int_{0}^{1-h}\int_{0}^{1-h^{\prime }}\sum_{k=1}^{\infty }\left( \mathbb{E
}\xi _{1k}^{4}-3\right) \phi _{k}(x)\phi _{k}(x+h)\phi _{k}\left( x^{\prime
}\right) \phi _{k}\left( x^{\prime }+h^{\prime }\right) dxdx^{\prime }\Bigg\},
\end{align*}%
for any $h,h^{\prime }\in \left[ 0,h_{0}\right] $. The proposition is
proved.

\subsection{Proof of Proposition \protect\ref{PROP:Uniform Chat-Ctilde}}

We decompose the difference between $\widehat{C}(h)$ and $\widetilde{C}(h)$ into the
following three terms:
$\widehat{C}(h)-\widetilde{C}(h)=\mathrm{I}(h) +\mathrm{II}(h) +\mathrm{III}%
(h)$, where
\begin{align}
\mathrm{I}(h) &=\frac{1}{n(1-h)}\int_{0}^{1-h} \sum_{i=1}^{n}\left\{\widehat{%
Z}_{i}(x)-Z_{i}(x)\right\} \left\{\widehat{Z}_{i}(x+h)-Z_{i}(x+h)\right\} dx,
\label{EQ:I and II and III} \\
\mathrm{II}(h) &=\frac{1}{n(1-h)}\int_{0}^{1-h}
\sum_{i=1}^{n}Z_{i}(x+h)\left\{\widehat{Z}_{i}(x)-Z_{i}(x)\right\} dx,
\notag \\
\mathrm{III}(h) &=\frac{1}{n(1-h)}\int_{0}^{1-h}
\sum_{i=1}^{n}Z_{i}(x)\left\{\widehat{Z}_{i}(x+h)-Z_{i}(x+h)\right\} dx.
\notag
\end{align}
Note that by (\ref{EQ:Zhati-Zi}), $\sup_{h\in \left[ 0,h_{0}\right]
}\left\vert \mathrm{I}(h) \right\vert \leq \max_{1\leq i\leq n}\Vert
\widehat{Z}_{i}-Z_{i}\Vert_{\infty}^{2}$. According to (\ref{EQ:Zhat-Z}),
\begin{equation*}
\max_{1\leq i\leq n}\Vert\widehat{Z}_{i}-Z_{i}\Vert_{\infty}=\mathcal{O}%
_{p}\left\{J_{s}^{-p^* }n\log n+J_{s}^{1/2}N^{-1/2}\left(\log
N\right)^{1/2}\right\}.
\end{equation*}

By (\ref{EQ:Zhati-Zi}), one has
\begin{align*}
\mathrm{II}(h) =&\frac{1}{n(1-h)}\int_{0}^{1-h}
\sum_{i=1}^{n}Z_{i}(x+h)\left\{\widehat{Z}_{i}(x)-Z_{i}(x)\right\} dx \\
=&\frac{1}{n(1-h)}\left[\int_{0}^{1-h}\sum_{i=1}^{n}Z_{i}(x+h)\left\{
\widetilde{Z}_{i}(x)-Z_{i}(x)\right\} dx
+\int_{0}^{1-h}\sum_{i=1}^{n}Z_{i}(x+h)\widetilde{\varepsilon}_{i}(x)dx %
\right] \\
&-\frac{1}{n^2(1-h)}\left[\int_{0}^{1-h}\sum_{i=1}^{n}Z_{i}(x+h)\sum_{i^{%
\prime}=1}^{n}\widetilde{Z}_{i^{\prime}}(x)dx
+\int_{0}^{1-h}\sum_{i=1}^{n}Z_{i}(x+h)\sum_{i^{\prime}=1}^{n}\widetilde{%
\varepsilon}_{i^{\prime}}(x) dx \right].
\end{align*}
Similar to the proof of Lemma \ref{LEM: sup_meanZi_epsi_tilde}, it is easy
to see
\begin{equation*}
\sup\limits_{h\in \left[ 0,h_{0}\right] }\frac{1}{n^2(1-h)} \left\vert
\int_{0}^{1-h}\sum_{i=1}^{n}Z_{i}(x+h)\sum_{i^{\prime}=1}^{n}\widetilde{%
\varepsilon}_{i^{\prime}}(x) dx\right\vert={\scriptstyle{\mathcal{O}}}%
_{p}(n^{-1/2}).
\end{equation*}%
Consequently, by Lemmas \ref{LEM: sup_Zi*Zitilde-Zi}, \ref{LEM:
sup_meanZi_epsi_tilde} and \ref{LEM:supsup_meanZimeanZi'}, one has
\begin{equation*}
\sup\limits_{h\in \left[ 0,h_{0}\right] }\left\vert \mathrm{II}(h)
\right\vert =\sup_{h\in \left[ 0,h_{0}\right] }\frac{1}{n(1-h)} \left\vert
\int_{0}^{1-h}\sum_{i=1}^{n}Z_{i}(x+h)\left\{ \widehat{Z}_{i}(x)-Z_{i}(x)%
\right\} dx\right\vert ={\scriptstyle{\mathcal{O}}}_{p}(n^{-1/2}).
\end{equation*}

Similarly, one can show that $\sup_{h\in \left[ 0,h_{0}\right] }\left\vert
\mathrm{III}(h) \right\vert =\sup_{h\in \left[ 0,h_{0}\right] }\left\vert
\mathrm{II}(h) \right\vert $. Consequently,
\begin{equation*}
\sup_{h\in \left[ 0,h_{0}\right] } \vert\widehat{C}(h)-\widetilde{C}
(h)\vert =\sup_{h\in \left[ 0,h_{0}\right] }\vert \mathrm{I} (h) +\mathrm{II}%
(h) +\mathrm{III}(h)\vert ={\scriptstyle{\mathcal{O}}}_{p}(n^{-1/2}).
\end{equation*}

\setcounter{chapter}{7} \renewcommand{\theequation}{B.\arabic{equation}} %
\renewcommand{\thesection}{B.\arabic{section}} \renewcommand{%
\thesubsection}{B.\arabic{subsection}} \renewcommand{\thetheorem}{B.%
\arabic{theorem}} \renewcommand{\thelemma}{B.\arabic{lemma}} %
\renewcommand{\theproposition}{B.\arabic{proposition}} \renewcommand{%
\thecorollary}{B.\arabic{corollary}} \renewcommand{\thefigure}{B.%
\arabic{figure}} \renewcommand{\thetable}{B.\arabic{table}} %
\setcounter{table}{0} \setcounter{figure}{0} \setcounter{equation}{0} %
\setcounter{theorem}{0} \setcounter{lemma}{0} \setcounter{proposition}{0} %
\setcounter{section}{0} \setcounter{subsection}{0}

\section*{B. Proofs of Technical Lemmas}
\label{APP:Proofs}

In this section, we provide the proofs of technical lemmas introduced in Appendix A. \medskip

\textsc{Proof of Lemma \ref{LEM:max-etatilde-eta}.} For any $k=1,2,\ldots$,
let $\mathbf{\phi}_{k}=\left(\phi _{k}(1/N),\ldots ,\phi _{k}\left(
N/N\right) \right)^{\top}$, and denote $\widetilde{\phi}_{k}(x) =N^{-1}%
\mathbf{B}(x)^{\top}\mathbf{V}_{n,p}^{-1}\mathbf{B}^{\top}\mathbf{\phi }_{k}$%
. According to (\ref{DEF:etai_tilde}), $\widetilde{\eta}_{i}(x)=\widetilde{m}%
(x)+\sum_{k=1}^{\infty }\xi _{ik}\widetilde{\phi}_{k}(x) $, therefore,
\begin{equation*}
\widetilde{\eta}_{i}(x)-\eta _{i}(x)=\widetilde{m}(x) -m(x)
+\sum_{k=1}^{\infty }\xi _{ik}\left\{\widetilde{\phi}_{k}(x) -\phi _{k}(x)
\right\} .
\end{equation*}
By Lemma A.4 of \ \cite{CYT12}, there exists a constant $C_{q,\mu }>0$, such
that
\begin{equation*}
\Vert \widetilde{m}-m\Vert_{\infty} \leq C_{q,\mu }\left\Vert m\right\Vert
_{q,\mu }J_{s}^{-p^* },  ~
\Vert \widetilde{\phi}_k-\phi_k\Vert_{\infty}\leq C_{q,\mu }\left\Vert
\phi_k\right\Vert _{q,\mu }J_{s}^{-p^* }, ~ k\geq 1.
\end{equation*}
Thus, by Assumption (A4), one obtains
\begin{equation*}
\left\Vert \widetilde{\eta}_{i}-\eta _{i}\right\Vert_{\infty} \leq
\left\Vert \widetilde{m}-m\right\Vert_{\infty}+\sum_{k=1}^{\infty }\vert \xi
_{ik}\vert \Vert \widetilde{\phi}_{k}-\phi _{k}\Vert_{\infty} \leq C_{q,\mu
}W_{i}J_{s}^{-p^*},
\end{equation*}
{where $W_{i}=\left\Vert m\right\Vert _{q,\mu }+\sum_{k=1}^{\infty
}\left\vert \xi _{ik}\right\vert \left\Vert \phi _{k}\right\Vert _{q,\mu }$,
$i=1,\ldots,n,$ are iid nonnegative random variables with finite absolute
moment.} According to Assumption (A6), one has
\begin{equation*}
\Pr \left\{\max_{1\leq i\leq n}W_{i}>\left(n\log n\right)^{2/r_{1}}\right\}
\leq n\frac{\mathrm{E}W_{i}^{r_{1}}}{\left(n\log n\right)^{2}}=\mathrm{E}%
W_{i}^{r_{1}}n^{-1}\left(\log n\right)^{-2},
\end{equation*}
thus, $\sum\limits_{n=1}^{\infty }\Pr \left\{\max_{1\leq i\leq n}W_{i}>n\log
n\right\} \leq \mathrm{E}W_{i}^{r_{1}}\sum\limits_{n=1}^{\infty
}n^{-1}\left(\log n\right)^{-2}<+\infty $, so $\max_{1\leq i\leq n}W_{i}=%
\mathcal{O}_{a.s.}\left\{\left(n\log n\right)^{2/r_{1}}\right\}$ and (\ref%
{EQ:etabiasbound}) is proved. Similarly, one obtains that $\max_{1\leq i\leq
n}\left\Vert Z_{i}\right\Vert_{\infty}=\mathcal{O}_{a.s.}\left\{(n\log
n)^{2/r_{1}}\right\}$ and $\max_{1\leq i\leq n}\Vert \widetilde{Z}%
-Z_{i}\Vert_{\infty}=\mathcal{O}_{a.s.}\left\{J_{s}^{-p^*}(n\log
n)^{2/r_{1}}\right\}$. Lemma \ref{LEM:max-etatilde-eta} holds consequently. $%
\blacksquare $ \medskip

\textsc{Proof of Lemma \ref{LEM:bound_B_eps_ij}.} Note that $\left\{
\varepsilon _{ij}\right\} $ are iid variables with $\mathrm{E}\left(
\varepsilon _{ij}^{2}\right) =1$ and Lemma 2 of \cite{W12} implies that $%
\left\Vert B_{\ell ,p}\right\Vert _{2,N}^{2}\asymp J_{s}^{-1}$ uniformly for
all $1-p\leq \ell \leq J_{s}$. Thus, one has uniformly for all $1-p\leq \ell
\leq J_{s}$,
\begin{equation*}
\mathrm{E}\left\{ \frac{1}{N}\sum_{j=1}^{N}B_{\ell ,p}\left( \frac{j}{N}%
\right) \sigma \left( \frac{j}{N}\right) \varepsilon _{ij}\right\} ^{2}=%
\frac{1}{N^{2}}\sum_{j=1}^{N}B_{\ell ,p}^{2}\left( \frac{j}{N}\right) \sigma
^{2}\left( \frac{j}{N}\right) =\frac{1}{N}\left\Vert B_{\ell ,p}\sigma
\right\Vert _{2,N}^{2}\asymp J_{s}^{-1}N^{-1}.
\end{equation*}%
Applying Bernstein inequality of Theorem 1.2 of \cite{B98},
similar to the proof of Lemma \ref{LEM:max_B_Uijeps}, one has $\max_{1\leq
i\leq n}\left\Vert N^{-1}\mathbf{B}^{T}\mathbf{\varepsilon }_{i}\right\Vert
_{\infty }=\mathcal{O}_{a.s}\{J_{s}^{-1/2}N^{-1/2}(\log N)^{1/2}\}$.
Therefore, by recalling  (\ref{EQ:Vhatp-invbound}) and
$\widetilde{\varepsilon }_{i}(x)$ in (\ref{DEF:etahat_project}),
one obtains
\begin{equation*}
\max_{1\leq i\leq n}\left\Vert \widetilde{\varepsilon }_{i}\right\Vert
_{\infty }=\max_{1\leq i\leq n}\left\Vert N^{-1}\mathbf{B}(x)^{\mathbf{\top }%
}\mathbf{V}_{n,p}^{-1}\mathbf{B}^{T}\mathbf{\varepsilon }_{i}\right\Vert
_{\infty }= \mathcal{O}_{a.s}\left\{ J_{s}^{1/2}N^{-1/2}(\log
N)^{1/2}\right\} .
\end{equation*}%
The lemma follows. $\blacksquare $ \medskip

\textsc{Proof of Lemma \ref{LEM: strong-moment}. } Under Assumption (A5'), $%
\mathrm{E}\left\vert \xi _{ik}\right\vert ^{r_{1}}<+\infty $, $%
r_{1}>4+2\omega $, $\mathrm{E}\left\vert \varepsilon _{ij}\right\vert
^{r_{2}}<+\infty $, $r_{2}>4+2\theta $, where $\omega$ is defined in
Assumption (A4) and $\theta$ is defined in Assumption (A3), so there exists
some $\beta _{1},\beta _{2}\in \left( 0,1/2\right) $, such that $%
r_{1}>\left( 2+\omega \right) /\beta _{1}$, $r_{2}>\left( 2+\theta \right)
/\beta _{2}$.

Let $H(x)=x^{r_{1}}$. Theorem 2.6.7 of \cite{CR81} entails that there exist constants
$c_{1k}$ and $a_{k}$ depending on the distribution of $\xi_{ik}$, such that
for $x_{n}=n^{\beta _{1}}$, $n/H\left( a_{k}x_{n}\right)
=a_{k}^{-r_{1}}n^{1-r_{1}\beta _{1}}$\ and iid $\mathcal{N}(0,1)$ variables $%
U_{ik,\xi }$,
\begin{equation*}
\Pr \left\{ \max_{1\leq t\leq n}\left\vert \sum_{i=1}^{t}\xi
_{ik}-\sum_{i=1}^{t}U_{ik,\xi }\right\vert >n^{\beta _{1}}\right\}
<c_{1k}a_{k}^{-r_{1}}n^{1-r_{1}\beta _{1}},
\end{equation*}%
by noticing that $r_{1}> \left( 2+\omega \right) /\beta _{1}$, $\gamma
_{1}=r_{1}\beta _{1}-1-\omega >1$. Since Assumption (A5') ensures that the
number of distinct distributions for $\xi _{ik}$ is finite, there is a
common $c_{1}>0$, such that $\max_{1\leq k\leq k_{n}}\Pr \left\{ \max_{1\leq
t\leq n}\left\vert \sum_{i=1}^{t}\xi _{ik}-\sum_{i=1}^{t}U_{ik,\xi
}\right\vert >n^{\beta _{1}}\right\} <c_{1}n^{-\gamma _{1}}$, and
consequently, there is a $C_{1}>0$ such that%
\begin{equation*}
\Pr \left\{ \max_{1\leq k\leq k_{n}\atop 1\leq t\leq n}\left\vert
\sum_{i=1}^{t}\xi _{ik}-\sum_{i=1}^{t}U_{ik,\xi }\right\vert >n^{\beta
_{1}}\right\} <k_{n}c_{1}n^{1-r_{1}\beta_1}\leq C_{1}n^{ -\gamma _{1}}.
\end{equation*}

Likewise, under Assumption (A5), taking $H(x)=x^{r_{2}}$, Theorem 2.6.7 of \cite{CR81}
implies that there exists constants $c_{2}$ and $b$ depending on the
distribution of $\varepsilon _{ij}$, such that for $x_{N}=N^{\beta _{2}}$, $%
N/H(ax_{N})=b^{-r_{2}}c_{2}^{-r_{2}}N^{1-r_{2}\beta _{2}}$ and iid standard
normal random variables $U_{ij,\varepsilon }$ such that
\begin{equation*}
\max_{1\leq i\leq n}\Pr \left\{ \max_{1\leq t\leq N}\left\vert
\sum_{j=1}^{t}\varepsilon _{it}-\sum_{j=1}^{t}U_{it,\varepsilon }\right\vert
>N^{\beta _{2}}\right\} <c_{2}b^{-r_{2}}N^{1-\gamma _{2}\beta _{2}},
\end{equation*}%
and consequently there is a $C_{2}>0$ such that
\begin{equation*}
\Pr \left\{ \max_{1\leq i\leq n \atop 1\leq t\leq N}\left\vert
\sum_{j=1}^{t}\varepsilon _{it}-\sum_{j=1}^{t}U_{it,\varepsilon }\right\vert
>N^{\beta _{2}}\right\} <c_{2}b^{-r_{2}}n\times N^{1-\gamma _{2}\beta
_{2}}\leq C_{2}N^{\theta +1-\gamma _{2}\beta _{2}}.
\end{equation*}
Since $r_{2}\beta _{2}>\left( 2+\theta \right) $, there is $\gamma
_{2}=r_{2}\beta _{2}-1-\theta >1$ and Assumption (A5) follows. The lemma
holds consequently. $\blacksquare $ \medskip

\textsc{Proof of Lemma \ref{LEM: sup_Zi*Zitilde-Zi}.} According to Lemma \ref%
{LEM:max-etatilde-eta}, one has
\begin{align*}
& \sup_{h\in \left[ 0,h_{0}\right] } \left\vert \frac{1}{n(1-h)}%
\int_{0}^{1-h}\sum_{i=1}^{n}Z_{i}(x+h)\left\{ \widetilde{Z}%
_{i}(x)-Z_{i}(x)\right\} dx\right\vert \leq \max_{1\leq i\leq n} \sup_{h\in \left[ 0,h_{0}\right]
}\sup_{x\in \left[ 0,1\right] }\left\vert
Z_{i}(x+h)\right\vert \\
&\quad \times \max_{1\leq i\leq n}\Vert \widetilde{Z}_{i}-Z_{i}\Vert
_{\infty } =\mathcal{O}_{a.s.}\left\{ J_{s}^{-p^{\ast }}\left( n\log
n\right) ^{4/r_{1}}\right\} ={\scriptstyle{\mathcal{O}}}_{p}(n^{-1/2}).
\end{align*}%
The proof is completed. $\blacksquare $ \medskip

\textsc{Proof of Lemma \ref{LEM:max_B_Uijeps}.} We apply Lemma \ref%
{LEM:maxGaussian} to obtain the uniform bound for the zero mean Gaussian
variables $N^{-1}\sum_{j=1}^{N}B_{\ell ,p}(j/N) \sigma \left( j/N\right)
U_{ij,\varepsilon }$, $1\leq i\leq n,1\leq \ell \leq J_{s}+p$ with variance $%
N^{-1}\left\Vert B_{\ell ,p} \sigma \right\Vert _{2,N}^{2}\leq
CN^{-1}J_{s}^{-1}$. It follows from Lemma \ref{LEM:maxGaussian} that
\begin{align}
\max_{1\leq i\leq n \atop 1\leq \ell \leq J_{s}+p}\left\vert
N^{-1}\sum_{j=1}^{N}B_{\ell ,p}(j/N)\sigma \left( j/N\right)
U_{ij,\varepsilon }\right\vert & =\mathcal{O}_{a.s.}\left\{
N^{-1/2}J_{s}^{-1/2}\log ^{1/2}\left( J_{s}+p\right) n\right\}  \notag \\
& =\mathcal{O}_{a.s.}\left( N^{-1/2}J_{s}^{-1/2}\log ^{1/2}N\right) ,
\label{EQ:max_.Uij_eps}
\end{align}%
where the last step follows from Assumptions (A4) and (A3) on the order of $%
J_{s}$ and $n$ relative to $N$. The lemma is proved. $\blacksquare $ \medskip


\textsc{Proof of Lemma \ref{LEM:max_kxi_eps-Uijeps}.} Applying Lemma \ref%
{LEM: strong-moment}, one has
\begin{equation*}
\max_{1\leq i\leq n \atop 1\leq j\leq N}\left\vert
N^{-1}\sum_{t=1}^{j}\left( \varepsilon _{it}-U_{it,\varepsilon }\right)
\right\vert =\mathcal{O}_{a.s.}(N^{\beta _{2} -1}).
\label{EQ:max_strong_eps_approx}
\end{equation*}

Next, we denote the following sequences:
\begin{align*}
\mathrm{Q}_{1,lk}=& (nN)^{-1}\sum_{i=1}^{n}U_{ik,\xi }\left[
\sum_{j=1}^{N-1}\left\{ B_{\ell ,p}\left( \frac{j}{N}\right) \sigma \left(
\frac{j}{N}\right) -B_{\ell ,p}\left( \frac{j+1}{N}\right) \sigma \left(
\frac{j+1}{N}\right) \right\} \sum_{t=1}^{j}\left( \varepsilon
_{it}-U_{it,\varepsilon }\right) \right] , \\
\mathrm{Q}_{2,lk}=& (nN)^{-1}\sum_{i=1}^{n}U_{ik,\xi }\left\{ B_{\ell
,p}(1)\sigma \left( 1\right) \sum_{t=1}^{N}\left( \varepsilon
_{it}-U_{it,\varepsilon }\right) \right\} .
\end{align*}%
Further denote the $\sigma $-field $\mathcal{F}_{\varepsilon }=\sigma
\left\{ \varepsilon _{ij},i,j=1,2,\ldots \right\} $, then for $1\leq k\leq
k_{n},1\leq \ell \leq J_{s}+p$, $\left. \mathrm{Q}_{1,lk}\right\vert
\mathcal{F}_{\varepsilon }=_{D}N\left( 0,\sigma _{lk}^{2}\right) $, in which
\begin{equation*}
\sigma _{lk}^{2}=n^{-2}\sum_{i=1}^{n}\left[ \sum_{j=1}^{N-1}\left\{ B_{\ell
,p}\left( \frac{j}{N}\right) \sigma \left( \frac{j}{N}\right) -B_{\ell
,p}\left( \frac{j+1}{N}\right) \sigma \left( \frac{j+1}{N}\right) \right\}
N^{-1}\sum_{t=1}^{j}\left( \varepsilon _{it}-U_{it,\varepsilon }\right) %
\right] ^{2}.
\end{equation*}

Applying (\ref{EQ:maxGaussiantailprob}), for any $a>2$%
\begin{equation*}
\Pr \left[ \left. \max_{1\leq k\leq k_{n}\atop 1\leq \ell \leq
J_{s}+p}\left\vert \mathrm{Q}_{1,lk}\right\vert /\sigma _{lk}>a\left\{ \log
k_{n}\left( J_{s}+p\right) \right\} ^{1/2}\right\vert \mathcal{F}%
_{\varepsilon }\right] <2\left\{ k_{n}\left( J_{s}+p\right) \right\}
^{1-a^{2}/2},
\end{equation*}%
and hence%
\begin{equation*}
\Pr \left[ \max_{1\leq k\leq k_{n} \atop 1\leq \ell \leq J_{s}+p}\left\vert
\mathrm{Q}_{1,lk}\right\vert /\sigma _{lk}>a\left\{ \log k_{n}\left(
J_{s}+p\right) \right\} ^{1/2}\right] <2\left\{ k_{n}\left( J_{s}+p\right)
\right\} ^{1-a^{2}/2}.
\end{equation*}%
Taking large enough $a$, while noting Assumptions (A6) and (A4) on the order
of $J_{s}$ and $k_{n}$ relative to $N$, one concludes with Borel-Cantelli
Lemma that%
\begin{equation}
\max_{1\leq k\leq k_{n}\atop 1\leq \ell \leq J_{s}+p}\left\vert \mathrm{Q}%
_{1,lk}\right\vert /\sigma _{lk}=\mathcal{O}_{a.s.}\left( \left\{ \log
k_{n}\left( J_{s}+p\right) \right\} ^{1/2}\right) =\mathcal{O}_{a.s.}(\log
^{1/2}N).  \label{EQ:standardizednormbd}
\end{equation}%
Next, the B spline basis satisfies%
\begin{equation*}
\left\vert B_{\ell ,p}\left( \frac{j}{N}\right) -B_{\ell ,p}\left( \frac{j+1%
}{N}\right) \right\vert \leq N^{-1}\left\Vert B_{\ell ,p}\right\Vert
_{0,1}\leq CJ_{s}N^{-1}
\end{equation*}%
uniformly over $1\leq j\leq N$ and $1\leq \ell \leq J_{s}+p$, while
Assumptions (A2) and (A6) imply that $J_{s}N^{-1}\sim N^{\gamma
}d_{N}N^{-1}\sim N^{\gamma -1}d_{N}\gg N^{-\nu }$, hence
\begin{equation*}
\left\vert \sigma \left( \frac{j}{N}\right) -\sigma \left( \frac{j+1}{N}%
\right) \right\vert \leq N^{-\nu }\left\Vert \sigma \right\Vert _{0,\nu
}\leq CJ_{s}N^{-1}
\end{equation*}%
uniformly over $1\leq j\leq N$. it then follows that
\begin{align}
\sigma _{lk}^{2}& \leq n^{-1}\left\{ \max_{1\leq i\leq n \atop 1\leq j\leq
N}\left\vert N^{-1}\sum_{t=1}^{j}\left( \varepsilon _{it}-U_{it,\varepsilon
}\right) \right\vert \right\} ^{2}\left\{ CNJ_{s}^{-1}\times \left\Vert
B_{\ell ,p}\right\Vert _{0,1}N^{-1}\right\} ^{2}  \label{EQ:sigmalkbds} \\
& \leq Cn^{-1}\left\{ \max_{1\leq i\leq n\atop 1\leq j\leq N}\left\vert
N^{-1}\sum_{t=1}^{j}\left( \varepsilon _{it}-U_{it,\varepsilon }\right)
\right\vert \right\} ^{2}=\mathcal{U}_{a.s.}(n^{-1}N^{2\beta _{2}-2}).
\notag
\end{align}%
Putting together the bounds in (\ref{EQ:standardizednormbd}) and (\ref%
{EQ:sigmalkbds}), one obtains that%
\begin{equation*}
\max_{1\leq k\leq k_{n}\atop 1\leq \ell \leq J_{s}+p}\left\vert \mathrm{Q}%
_{1,lk}\right\vert =\mathcal{O}_{a.s.}(n^{-1/2}N^{\beta _{2}-1}\log ^{1/2}N).
\end{equation*}%
Similarly, $\max_{1\leq k\leq k_{n}}\max_{1\leq \ell \leq J_{s}+p}\left\vert
\mathrm{Q}_{2,lk}\right\vert =\mathcal{O}_{a.s.}(n^{-1/2}N^{\beta
_{2}-1}\log ^{1/2}N)$. Finally, the Lemma is proved by noticing that
\begin{align*}
& \left\vert n^{-1}\sum_{i=1}^{n}U_{ik,\xi }\left\{
N^{-1}\sum_{j=1}^{N}B_{\ell ,p}\left( \frac{j}{N}\right) \sigma \left( \frac{%
j}{N}\right) (\varepsilon _{ij}-U_{ij,\varepsilon })\right\} \right\vert \\
& =\left\vert (nN)^{-1}\sum_{i=1}^{n}U_{ik,\xi }\left[ \sum_{j=1}^{N-1}\left%
\{ B_{\ell ,p}\left( \frac{j}{N}\right) \sigma \left( \frac{j}{N}\right)
-B_{\ell ,p}\left( \frac{j+1}{N}\right) \sigma \left( \frac{j+1}{N}\right)
\right\} \sum_{t=1}^{j}\left( \varepsilon _{it}-U_{it,\varepsilon }\right) %
\right] \right. \\
& \quad \left. +(nN)^{-1}\sum_{i=1}^{n}U_{ik,\xi }\left\{ B_{\ell ,p}\left(
1\right) \sigma \left( 1\right) \sum_{t=1}^{N}\left( \varepsilon
_{it}-U_{it,\varepsilon }\right) \right\} \right\vert \leq \left\vert
\mathrm{Q}_{1,lk}\right\vert +\left\vert \mathrm{Q}_{2,lk}\right\vert . ~\blacksquare
\end{align*}


\textsc{Proof of Lemma \ref{LEM:max_kxi-Ukxi*Uijeps}.} According to Lemma
A.5 in \cite{CYT12}, under Assumptions (A4)--(A5),
$\max_{1\leq k\leq k_{n}}\max_{1\leq t\leq n}\left\vert \sum_{i=1}^{t}\left(
\xi _{ik}-U_{ik,\xi }\right) \right\vert =\mathcal{O}_{a.s.}(n^{\beta _{1}})$.
Next, denote
\begin{align*}
\mathrm{Q}_{3,lk}& =\left( nN\right)
^{-1}\sum_{t=1}^{n-1}\sum_{i=1}^{t}\left( \xi _{ik}-U_{ik,\xi }\right)
\sum_{j=1}^{N}B_{\ell ,p}\left( \frac{j}{N}\right) \sigma \left( \frac{j}{N}%
\right) U_{tj,\varepsilon }, \\
\mathrm{Q}_{4,lk}& =-\left( nN\right)
^{-1}\sum_{t=1}^{n-1}\sum_{i=1}^{t}\left( \xi _{ik}-U_{ik,\xi }\right)
\sum_{j=1}^{N}B_{\ell ,p}\left( \frac{j}{N}\right) \sigma \left( \frac{j}{N}%
\right) U_{\left( t+1\right) j,\varepsilon }, \\
\mathrm{Q}_{5,lk}& =(nN)^{-1}\sum_{t=1}^{n}\left( \xi _{tk}-U_{tk,\xi
}\right) \sum_{j=1}^{N}B_{\ell ,p}\left( \frac{j}{N}\right) \sigma \left(
\frac{j}{N}\right) U_{nj,\varepsilon }.
\end{align*}%
Denote the $\sigma $-field $\mathcal{F}_{\xi }=\sigma \left\{ \xi
_{ij},~i,j=1,2,\ldots \right\} $, then for $1\leq k\leq k_{n},1\leq \ell
\leq J_{s}+p$, one has $\left. \mathrm{Q}_{3,lk}\right\vert \mathcal{F}_{\xi
}=_{D}N\left( 0,\sigma _{lk,3}^{2}\right) $, where
\begin{equation*}
\sigma_{lk,3}^{2}=2(nN)^{-2}\sum_{t=1}^{n-1}\left\{ \sum_{i=1}^{t}\left( \xi
_{ik}-U_{ik,\xi }\right) \right\} ^{2}\sum_{j=1}^{N}B_{\ell ,p}^{2}\left(
\frac{j}{N}\right) \sigma ^{2}\left( \frac{j}{N}\right).
\end{equation*}

Similar to Lemma \ref{LEM:max_kxi_eps-Uijeps}, applying (\ref%
{EQ:maxGaussiantailprob}), for any $a>2$
\begin{equation*}
\Pr \left[ \left. \max_{1\leq k\leq k_{n}\atop 1\leq \ell \leq
J_{s}+p}\left\vert \mathrm{Q}_{3,lk}\right\vert /\sigma _{lk,3}>a\left\{
\log k_{n}\left( J_{s}+p\right) \right\} ^{1/2}\right\vert \mathcal{F}%
_{\varepsilon }\right] \leq 2\left\{ k_{n}\left( J_{s}+p\right) \right\}
^{1-a^{2}/2},
\end{equation*}%
and hence%
\begin{equation*}
\Pr \left[ \max_{1\leq k\leq k_{n} \atop 1\leq \ell \leq J_{s}+p}\left\vert
\mathrm{Q}_{3,lk}\right\vert /\sigma _{lk,3}>a\left\{ \log k_{n}\left(
J_{s}+p\right) \right\} ^{1/2}\right] \leq 2\left\{ k_{n}\left(
J_{s}+p\right) \right\} ^{1-a^{2}/2}.
\end{equation*}%
Taking large enough $a$, according to Assumptions (A4) and (A6), one
concludes with Borel-Cantelli Lemma that%
\begin{equation}
\max_{1\leq k\leq k_{n}\atop 1\leq \ell \leq J_{s}+p}\left\vert \mathrm{Q}%
_{3,lk}\right\vert /\sigma _{lk,3}=\mathcal{O}_{a.s.}\left( \left\{ \log
k_{n}\left( J_{s}+p\right) \right\} ^{1/2}\right) =\mathcal{O}_{a.s.}(\log
^{1/2}N).  \label{EQ:Q3,lk/sigma3}
\end{equation}%
Noticing that $\max_{1\leq \ell \leq J_{s}+p}\left\Vert B_{\ell ,p}\sigma
\right\Vert _{2,N}^{2}=\mathcal{O}\left( J_{s}^{-1}\right) $, one has
\begin{align}
& \sigma _{lk,3}^{2}=2n^{-2}\left\{ \sum_{t=1}^{n-1}\sum_{i=1}^{t}\left( \xi
_{ik}-U_{ik,\xi }\right) \right\} ^{2}N^{-2}\sum_{j=1}^{N}B_{\ell
,p}^{2}\left( \frac{j}{N}\right) \sigma ^{2}\left( \frac{j}{N}\right)  \notag
\\
& \leq 2N^{-1}\left\Vert B_{\ell ,p}\sigma \right\Vert _{2,N}^{2}\left\{
\left( n-1\right) \times \max_{1\leq t\leq n}\left\vert
n^{-1}\sum_{t=1}^{n-1}\left( \xi _{tk}-U_{tk,\xi }\right) \right\vert
\right\} ^{2}\leq cJ_{s}^{-1}N^{-1}n^{2\beta _{1}-1}.
\label{EQ: sigma3-bound}
\end{align}

Putting together the bounds in (\ref{EQ:Q3,lk/sigma3}) and (\ref{EQ:
sigma3-bound}), one obtains that%
\begin{equation*}
\max_{1\leq k\leq k_{n}}\max_{1\leq \ell\leq J_{s}+p}\left\vert \mathrm{Q}%
_{3,lk}\right\vert =\mathcal{O}_{a.s.} \left(n^{\beta
_{1}-1/2}J_{s}^{-1/2}N^{-1/2}\log ^{1/2}N\right).
\end{equation*}

Thus, one can show the following similarly,%
\begin{equation*}
\max_{1\leq k\leq k_{n}}\max_{1\leq \ell \leq J_{s}+p}\left\vert \mathrm{Q}%
_{4,lk}\right\vert =\mathcal{O}_{a.s.}\left( n^{\beta
_{1}-1/2}N^{-1/2}J_{s}^{-1/2}\log ^{1/2}N\right) .
\end{equation*}%
\begin{equation*}
\max_{1\leq k\leq k_{n}}\max_{1\leq \ell \leq J_{s}+p}\left\vert \mathrm{Q}%
_{5,lk}\right\vert =\mathcal{O}_{a.s.}\left( n^{\beta
_{1}-1/2}N^{-1/2}J_{s}^{-1/2}\log ^{1/2}N\right) .
\end{equation*}%
Therefore, the lemma holds by noticing that%
\begin{align*}
\max_{1\leq k\leq k_{n}}& \max_{1\leq \ell \leq J_{s}+p}\left\vert
n^{-1}\sum_{i=1}^{n}\left( \xi _{ik}-U_{ik,\xi }\right)
N^{-1}\sum_{j=1}^{N}B_{\ell ,p}\left( \frac{j}{N}\right) \sigma \left( \frac{%
j}{N}\right) U_{ij,\varepsilon }\right\vert \\
& \leq \max_{1\leq k\leq k_{n}}\max_{1\leq \ell \leq J_{s}+p}\left\vert
\mathrm{Q}_{3,lk}\right\vert +\max_{1\leq k\leq k_{n}}\max_{1\leq \ell \leq
J_{s}+p}\left\vert \mathrm{Q}_{4,lk}\right\vert +\max_{1\leq k\leq
k_{n}}\max_{1\leq \ell \leq J_{s}+p}\left\vert \mathrm{Q}_{5,lk}\right\vert. ~
\blacksquare
\end{align*}%


\textsc{Proof of Lemma \ref{LEM:max-kxi_ik-U_ikkxi}. }Simple algebra
provides that
\begin{align*}
\sum_{i=1}^{n}& \left( \xi _{ik}-U_{ik,\xi }\right) \left\{
\sum_{j=1}^{N}B_{\ell ,p}\left( \frac{j}{N}\right) \sigma \left( \frac{j}{N}%
\right) (\varepsilon _{ij}-U_{ij,\varepsilon })\right\} \\
& =\sum_{t=1}^{n-1}\sum_{i=1}^{t}\left( \xi _{ik}-U_{ik,\xi }\right) \left\{
\sum_{j=1}^{N}B_{\ell ,p}\left( \frac{j}{N}\right) \sigma \left( \frac{j}{N}%
\right) (\varepsilon _{tj}-U_{tj,\varepsilon })\right\} \\
& \quad -\sum_{t=1}^{n-1}\sum_{i=1}^{t}\left( \xi _{ik}-U_{ik,\xi }\right)
\left\{ \sum_{j=1}^{N}B_{\ell ,p}\left( \frac{j}{N}\right) \sigma \left(
\frac{j}{N}\right) (\varepsilon _{\left( t+1\right) j}-U_{\left( t+1\right)
j,\varepsilon })\right\} \\
& \quad -\sum_{i=1}^{n}\left( \xi _{ik}-U_{ik,\xi }\right) \left\{
\sum_{j=1}^{N}B_{\ell ,p}\left( \frac{j}{N}\right) \sigma \left( \frac{j}{N}%
\right) (\varepsilon _{nj}-U_{nj,\varepsilon })\right\} .
\end{align*}%
By noticing that
\begin{align*}
\max_{1\leq t\leq n}\max_{1\leq k\leq k_{n}}\left\vert
n^{-1}\sum_{i=1}^{t}\left( \xi _{ik}-U_{ik,\xi }\right) \right\vert & ={{%
\mathcal{O}}}_{a.s.}\left( n^{\beta _{1}-1}\right) , \\
\max_{1\leq t\leq n}\max_{1\leq \ell \leq J_{s}+p}\left\vert
N^{-1}\sum_{j=1}^{N}B_{\ell ,p}\left( \frac{j}{N}\right) \sigma \left( \frac{%
j}{N}\right) (\varepsilon _{tj}-U_{tj,\varepsilon })\right\vert & =\mathcal{O%
}_{a.s.}(N^{\beta _{2}-1}),
\end{align*}%
one has
\begin{align*}
& \max_{1\leq k\leq k_{n}}\max_{1\leq \ell \leq J_{s}+p}\left\vert
n^{-1}\sum_{i=1}^{n}\left( \xi _{ik}-U_{ik,\xi }\right) \left\{
N^{-1}\sum_{j=1}^{N}B_{\ell ,p}\left( \frac{j}{N}\right) \sigma \left( \frac{%
j}{N}\right) (\varepsilon _{ij}-U_{ij,\varepsilon })\right\} \right\vert \\
& \leq n\max_{1\leq k\leq k_{n}}\left\vert n^{-1}\sum_{i=1}^{n}\left( \xi
_{ik}-U_{ik,\xi }\right) \right\vert \max_{1\leq \ell \leq
J_{s}+p}\max_{1\leq i\leq n}\left\vert N^{-1}\sum_{j=1}^{N}B_{\ell ,p}\left(
\frac{j}{N}\right) \sigma \left( \frac{j}{N}\right) (\varepsilon
_{ij}-U_{ij,\varepsilon })\right\vert \\
& =\mathcal{O}_{a.s.}(n^{\beta _{1}}N^{\beta _{2}-1}).
\end{align*}%
The lemma holds. $\blacksquare $ \medskip


\textsc{Proof of Lemma \ref{LEM: sup_meanZi_epsi_tilde}.} By the definition
of $\widetilde{\varepsilon }_{i}(x)$ in (\ref{DEF:etahat_project}), one has%
\begin{equation*}
Z_{i}(x+h)\widetilde{\varepsilon }_{i}(x)=N^{-1}\mathbf{B}(x)^{\top }\mathbf{%
V}_{n,p}^{-1}\mathbf{B}^{\top }\left\{ Z_{i}(x+h)\sigma \left( \frac{j}{N}%
\right) \varepsilon _{ij}\right\} _{j=1}^{N},
\end{equation*}%
which implies that
\begin{align*}
\sum_{i=1}^{n}& Z_{i}(x+h)\widetilde{\varepsilon }_{i}(x)=\mathbf{B}%
(x)^{\top }\mathbf{V}_{n,p}^{-1}\left\{ \sum_{k=1}^{\infty }\phi _{k}(x+h)%
\frac{1}{N}\sum_{i=1}^{n}\sum_{j=1}^{N}B_{\ell ,p}\left( \frac{j}{N}\right)
\sigma \left( \frac{j}{N}\right) \xi _{ik}\varepsilon _{ij}\right\}  \\
=& \mathbf{B}(x)^{\top }\mathbf{V}_{n,p}^{-1}\left\{ \left(
\sum_{k=1}^{k_{n}}+\sum_{k=k_{n}+1}^{\infty }\right) \phi _{k}(x+h)\frac{1}{N%
}\sum_{i=1}^{n}\sum_{j=1}^{N}B_{\ell ,p}\left( \frac{j}{N}\right) \sigma
\left( \frac{j}{N}\right) \xi _{ik}\varepsilon _{ij}\right\} .
\end{align*}%
First, uniformly for $1\leq \ell \leq J_{s}+p,h\in \left[ 0,h_{0}\right] $,
one has
\begin{equation*}
\left\vert \frac{1}{nN}\sum_{k=k_{n}+1}^{\infty }\phi
_{k}(x+h)\sum_{i=1}^{n}\sum_{j=1}^{N}B_{\ell ,p}\left( \frac{j}{N}\right)
\sigma \left( \frac{j}{N}\right) \xi _{ik}\varepsilon _{ij}\right\vert \leq
\mathrm{D},
\end{equation*}%
where $\mathrm{D}=\max_{1\leq \ell \leq J_{s}+p}\sum_{k=k_{n}+1}^{\infty
}\left\Vert \phi _{k}\right\Vert _{\infty }\left( nN\right)
^{-1}\sum_{i=1}^{n}\sum_{j=1}^{N}B_{\ell ,p}(j/N)\sigma \left( \frac{j}{N}%
\right) \left\vert \xi _{ik}\right\vert \left\vert \varepsilon
_{ij}\right\vert $, and
\begin{equation*}
\mathrm{E}\mathrm{D}\leq \max_{1\leq \ell \leq
J_{s}+p}\sum_{k=k_{n}+1}^{\infty }\left\Vert \phi _{k}\right\Vert _{\infty
}\left( nN\right) ^{-1}\sum_{i=1}^{n}\sum_{j=1}^{N}B_{\ell ,p}\left( \frac{j%
}{N}\right) \sigma \left( \frac{j}{N}\right) \mathrm{E}\left\vert \xi
_{ik}\right\vert \mathrm{E}\left\vert \varepsilon _{ij}\right\vert \leq
cJ_{s}^{-1}\sum_{k=k_{n}+1}^{\infty }\left\Vert \phi _{k}\right\Vert
_{\infty }.
\end{equation*}%
As Assumption (A4) guarantees that $\sum_{k=k_{n}+1}^{\infty }\left\Vert
\phi _{k}\right\Vert _{\infty }\ll n^{-1/2}$, while $\left\Vert \mathbf{V}%
_{n,p}^{-1}\right\Vert \leq CJ_{s}$ for large $N$, one has
\begin{equation}
\sup_{h\in \left[ 0,h_{0}\right] }\sup_{x\in \left[ 0,1\right] }\left\vert
\mathbf{B}(x)^{\top }\mathbf{V}_{n,p}^{-1}\left\{ \sum_{k=k_{n}+1}^{\infty
}\phi _{k}(x+h)\frac{1}{nN}\sum_{i=1}^{n}\sum_{j=1}^{N}B_{\ell ,p}\left(
\frac{j}{N}\right) \sigma \left( \frac{j}{N}\right) \xi _{ik}\varepsilon
_{ij}\right\} \right\vert ={\scriptstyle{\mathcal{O}}}_{p}(n^{-1/2}).
\label{EQ:max_inf_Zi_epsi_tilde}
\end{equation}

Next, one bounds the sum $\sum_{k=1}^{k_{n}} $. According to Assumption
(A4), (\ref{EQ:max_strong_eps_approx}) and Lemma \ref{LEM:max_kxi_eps-Uijeps}%
, one has
\begin{align*}
& \sup_{h\in \left[ 0,h_{0}\right] }\sup_{x\in \left[ 0,1\right] }\left\vert
\mathbf{B}(x)^{\top }\mathbf{V}_{n,p}^{-1}\left\{ \sum_{k=1}^{k_{n}}\phi
_{k}(x+h)\frac{1}{nN}\sum_{i=1}^{n}\xi _{ik}\sum_{j=1}^{N}B_{\ell ,p}\left(
\frac{j}{N}\right) \sigma \left( \frac{j}{N}\right) \left( \varepsilon
_{ij}-U_{ij,\varepsilon }\right) \right\} \right\vert \\
& ={\scriptstyle{\mathcal{O}}}_{p}(J_{s}n^{-1/2}N^{\beta _{2}-1}).
\end{align*}%
In addition, by Lemma \ref{LEM:max_kxi-Ukxi*Uijeps}, one obtains that
\begin{align*}
& \sup_{h\in \left[ 0,h_{0}\right] }\sup_{x\in \left[ 0,1\right] }\left\vert
\mathbf{B}(x)^{\top }\mathbf{V}_{n,p}^{-1}\left\{ \sum_{k=1}^{k_{n}}\phi
_{k}(x+h)\frac{1}{nN}\sum_{i=1}^{n}\left( \xi _{ik}-U_{ik,\xi }\right)
\sum_{j=1}^{N}B_{\ell ,p}\left( \frac{j}{N}\right) \sigma \left( \frac{j}{N}%
\right) U_{ij,\varepsilon }\right\} \right\vert \\
& ={\scriptstyle{\mathcal{O}}}_{p}(n^{\beta _{1}-1/2}N^{-1/2}J_{s}^{1/2}).
\end{align*}%
Note that $U_{ik,\xi }$ and $U_{ij,\varepsilon }$ are independent standard
normal random variables. Similar to (\ref{EQ:max_.Uij_eps}), one obtains
that
\begin{equation*}
\max_{1\leq \ell \leq J_{s}+p}\left\vert \frac{1}{n}\sum_{i=1}^{n}U_{ik,\xi
}\left\{ \frac{1}{N}\sum_{j=1}^{N}B_{\ell ,p}\left( \frac{j}{N}\right)
\sigma \left( \frac{j}{N}\right) U_{ij,\varepsilon }\right\} \right\vert =%
\mathcal{O}_{a.s.}\left( N^{-1/2}J_{s}^{-1/2}n^{-1/2}\sqrt{\log N}\right) .
\end{equation*}%
It is easy to see that
\begin{align*}
& \left\vert \mathbf{B}(x)^{\top }\mathbf{V}_{n,p}^{-1}\left\{%
\sum_{k=1}^{k_{n}}\phi _{k}(x+h)\frac{1}{nN}\sum_{i=1}^{n}\sum_{j=1}^{N}B_{%
\ell ,p}\left( \frac{j}{N}\right) \sigma \left( \frac{j}{N}\right) \xi
_{ik}\varepsilon _{ij}\right\} \right\vert \\
& \leq \left\vert \mathbf{B}(x)^{\top }\mathbf{V}_{n,p}^{-1}\left\{
\sum_{k=1}^{k_{n}}\phi _{k}(x+h)\frac{1}{nN}\sum_{i=1}^{n}\left( \xi
_{ik}-U_{ik,\xi }\right) \sum_{j=1}^{N}B_{\ell ,p}\left( \frac{j}{N}\right)
\sigma \left( \frac{j}{N}\right) \left( \varepsilon _{ij}-U_{ij,\varepsilon
}\right) \right\} \right\vert \\
& \quad +\left\vert \mathbf{B}(x)^{\top }\mathbf{V}_{n,p}^{-1}\left\{
\sum_{k=1}^{k_{n}}\phi _{k}(x+h)\frac{1}{nN}\sum_{i=1}^{n}U_{ik,\xi
}\sum_{j=1}^{N}B_{\ell ,p}\left( \frac{j}{N}\right) \sigma \left( \frac{j}{N}%
\right) \left( \varepsilon _{ij}-U_{ij,\varepsilon }\right) \right\}
\right\vert \\
& \quad +\left\vert \mathbf{B}(x)^{\top }\mathbf{V}_{n,p}^{-1}\left\{
\sum_{k=1}^{k_{n}}\phi _{k}(x+h)\frac{1}{n}\sum_{i=1}^{n}\left( \xi
_{ik}-U_{ik,\xi }\right) N^{-1}\sum_{j=1}^{N}B_{\ell ,p}\left( \frac{j}{N}%
\right) \sigma \left( \frac{j}{N}\right) U_{ij,\varepsilon }\right\}
\right\vert \\
& \quad +\left\vert \mathbf{B}(x)^{\top }\mathbf{V}_{n,p}^{-1}\left[
\sum_{k=1}^{k_{n}}\phi _{k}(x+h)\frac{1}{n}\sum_{i=1}^{n}U_{ik,\xi }\left\{
N^{-1}\sum_{j=1}^{N}B_{\ell ,p}\left( \frac{j}{N}\right) \sigma \left( \frac{%
j}{N}\right) U_{ij,\varepsilon }\right\} \right] \right\vert .
\end{align*}%
Therefore, combining Lemmas \ref{LEM:max_B_Uijeps}--\ref%
{LEM:max-kxi_ik-U_ikkxi}, one has
\begin{align*}
& \sup_{h\in \left[ 0,h_{0}\right] }\sup_{x\in \left[ 0,1\right] }\left\vert
\mathbf{B}(x)^{\top }\mathbf{V}_{n,p}^{-1}\left\{ \sum_{k=k_{n}+1}^{\infty
}\phi _{k}(x+h)\frac{1}{nN}\sum_{i=1}^{n}\sum_{j=1}^{N}B_{\ell ,p}\left(
\frac{j}{N}\right) \sigma \left( \frac{j}{N}\right) \xi _{ik}\varepsilon
_{ij}\right\} \right\vert \\
& =\mathcal{O}_{a.s.}\left( J_{s}n^{-1/2}N^{\beta _{2}-1}\log ^{1/2}N\right)
+\mathcal{O}_{a.s.}\left( n^{\beta _{1}-1/2}N^{-1/2}J_{s}^{1/2}\log
^{1/2}N\right) \\
& \quad +\mathcal{O}_{a.s.}\left( J_{s}n^{\beta _{1}}N^{\beta _{2}-1}\right) +%
\mathcal{O}_{a.s.}\left( n^{-1/2}N^{-1/2}J_{s}^{1/2}\log ^{1/2}N\right) .
\end{align*}%
Hence, the proof is completed by noticing that
\begin{align*}
& \sup_{h\in \left[ 0,h_{0}\right] }\sup_{x\in \left[ 0,1\right] }\left\vert
\frac{1}{n}\sum_{i=1}^{n}Z_{i}(x+h)\widetilde{\varepsilon }_{i}(x)\right\vert
\\
& =\sup_{h\in \left[ 0,h_{0}\right] }\sup_{x\in \left[ 0,1\right]
}\left\vert \mathbf{B}(x)^{\top }\mathbf{V}_{n,p}^{-1}\left\{
\sum_{k=1}^{\infty }\phi _{k}(x+h)\frac{1}{nN}\sum_{i=1}^{n}%
\sum_{j=1}^{N}B_{\ell ,p}\left( \frac{j}{N}\right) \sigma \left( \frac{j}{N}%
\right) \xi _{ik}\varepsilon _{ij}\right\} \right\vert \\
& ={\scriptstyle{\mathcal{O}}}_{p}(n^{-1/2})+\mathcal{O}_{a.s.}\left(
J_{s}n^{-1/2}N^{\beta _{2}-1}\log ^{1/2}N\right) +\mathcal{O}_{a.s.}\left(
n^{\beta _{1}-1/2}N^{-1/2}J_{s}^{1/2}\log ^{1/2}N\right) \\
& \quad +\mathcal{O}_{a.s.}\left( J_{s}n^{\beta _{1}}N^{\beta _{2}-1}\right)
+\mathcal{O}_{a.s.}\left( n^{-1/2}N^{-1/2}J_{s}^{1/2}\log ^{1/2}N\right) ={%
\scriptstyle{\mathcal{O}}}_{p}(n^{-1/2}). ~\blacksquare
\end{align*}%


\textsc{Proof of Lemma \ref{LEM:supsup_meanZimeanZi'}.} Notice that
\begin{align}
Z_{i}(x+h)& n^{-1}\sum_{i^{\prime }=1}^{n}\widetilde{Z}_{i^{\prime
}}(x)=N^{-1}\mathbf{B}(x)^{\top }\mathbf{V}_{n,p}^{-1}\mathbf{B}^{\top
}\left\{ Z_{i}(x+h)n^{-1}\sum_{i^{\prime }=1}^{n}\sum_{k=1}^{\infty }\xi
_{i^{\prime }k}\phi _{k}\left( j/N\right) \right\} _{j=1}^{N}  \notag \\
& =\frac{1}{N}\mathbf{B}(x)^{\top }\mathbf{V}_{n,p}^{-1}\mathbf{B}^{\top
}\left\{ \sum_{k=1}^{\infty }\xi _{ik}\phi _{k}(x+h)\frac{1}{n}%
\sum_{i^{\prime }=1}^{n}\sum_{k^{\prime }=1}^{\infty }\xi _{i^{\prime
}k^{\prime }}\phi _{k^{\prime }}\left( j/N\right) \right\} _{j=1}^{N}.
\label{EQ:ZZtilde}
\end{align}%
Let $\bar{\xi}_{\cdot k}=n^{-1}\sum_{i=1}^{n}\xi _{ik}$, then
\begin{align*}
\frac{1}{n}& \sum_{i=1}^{n}Z_{i}(x+h)\frac{1}{n}\sum\limits_{i^{\prime
}=1}^{n}\widetilde{Z}_{i^{\prime }}(x)=\mathbf{B}(x)^{\top }\mathbf{V}%
_{n,p}^{-1}\left\{ \sum_{k=1}^{\infty }\phi _{k}(x+h)\bar{\xi}_{\cdot
k}\sum_{k^{\prime }=1}^{\infty }\langle B_{\ell ,p},\phi _{k^{\prime
}}\rangle _{N}\bar{\xi}_{\cdot k^{\prime }}\right\} _{\ell =1}^{J_{s}+p} \\
& =\mathbf{B}(x)^{\top }\mathbf{V}_{n,p}^{-1}\Bigg\{\sum_{k=1}^{\infty }\phi
_{k}(x+h)\bar{\xi}_{\cdot k}\sum_{k^{\prime }=1}^{\infty }\langle B_{\ell
,p},\phi _{k^{\prime }}\rangle _{N}\bar{\xi}_{\cdot k^{\prime }}\Bigg\}%
_{\ell =1}^{J_{s}+p}.
\end{align*}%
Now uniformly for $1\leq \ell \leq J_{s}+p,h\in \left[ 0,h_{0}\right]$, one
has
\begin{equation*}
\left\vert \sum_{k=1}^{\infty }\phi _{k}(x+h)\bar{\xi}_{\cdot
k}\sum_{k^{\prime }=1}^{\infty }\langle B_{\ell ,p},\phi _{k^{\prime
}}\rangle _{N}\bar{\xi}_{\cdot k^{\prime }}\right\vert \leq CJ_{s}^{-1}%
\mathrm{S},
\end{equation*}%
where $\mathrm{S}=\sum_{k=1}^{\infty }\left\Vert \phi _{k}\right\Vert
_{\infty }\sum_{k^{\prime }=1}^{\infty }\left\Vert \phi _{k^{\prime
}}\right\Vert _{\infty }\left\vert \bar{\xi}_{\cdot k}\bar{\xi}_{\cdot
k^{\prime }}\right\vert $, and according to (A5) in \cite{CYT12},
\begin{equation*}
\mathrm{E}\mathrm{S}\leq \sum_{k=1}^{\infty }\left\Vert \phi _{k}\right\Vert
_{\infty }\sum_{k^{\prime }=1}^{\infty }\left\Vert \phi _{k^{\prime
}}\right\Vert _{\infty }\max_{1\leq k\leq \infty }\mathrm{E}\left\vert \bar{%
\xi}_{\cdot k}\right\vert ^{2}\leq cn^{-1}\sum_{k=1}^{\infty }\left\Vert
\phi _{k}\right\Vert _{\infty }\sum_{k^{\prime }=1}^{\infty }\left\Vert \phi
_{k^{\prime }}\right\Vert _{\infty }.
\end{equation*}%
As Assumption (A4) guarantees that $\sum_{k=1}^{\infty }\left\Vert \phi
_{k}\right\Vert _{\infty }<\infty $, while $\left\Vert \mathbf{V}%
_{n,p}^{-1}\right\Vert \leq CJ_{s}$ for large $N$, one obtains that
\begin{equation}
\sup_{h\in \left[ 0,h_{0}\right] }\sup_{x\in \left[ 0,1\right] }\left\vert
\mathbf{B}(x)^{\top }\mathbf{V}_{n,p}^{-1}\left\{ \sum_{k=1}^{\infty }\phi
_{k}(x+h)\bar{\xi}_{\cdot k}\sum_{k^{\prime }=1}^{\infty }\langle B_{\ell
,p},\phi _{k^{\prime }}\rangle _{N}\bar{\xi}_{\cdot k^{\prime }}\right\}
_{\ell =1}^{J_{s}+p}\right\vert ={\scriptstyle{\mathcal{O}}}_{p}(n^{-1/2}).
\label{EQ:k_inf kxi_kxi}
\end{equation}%
The lemma holds. $\blacksquare $ \medskip

\textsc{Proof of Lemma \ref{LEM:maxGaussian}. } Note that
\begin{align*}
\Pr \left( \max_{1\leq i\leq n}\left\vert \frac{W_{i}}{\sigma _{i}}%
\right\vert >a\sqrt{\log n}\right) &\leq \sum_{i=1}^{n}\Pr\left( \left\vert
\frac{W_{i}}{\sigma _{i}}\right\vert >a\sqrt{\log n}\right) \leq 2n\left\{
1-\Phi \left( a\sqrt{\log n}\right) \right\} \\
&<2n\frac{\phi \left( a\sqrt{\log n}\right) }{a\sqrt{\log n}}\leq 2n\phi
\left( a\sqrt{\log n}\right) =\sqrt{2/\pi}n^{1-a^{2}/2},
\end{align*}%
for $n\rightarrow \infty$, $a>2$, which proves (\ref{EQ:maxGaussiantailprob}). The lemma
follows by applying Borel-Cantelli Lemma with choice of $a>2$. $\blacksquare$

\setcounter{chapter}{8} \renewcommand{\theequation}{C.\arabic{equation}} %
\renewcommand{\thesection}{C.\arabic{section}} \renewcommand{%
\thesubsection}{C.\arabic{subsection}} \renewcommand{\thetheorem}{C.%
\arabic{theorem}} \renewcommand{\thelemma}{C.\arabic{lemma}} %
\renewcommand{\theproposition}{C.\arabic{proposition}} \renewcommand{%
\thecorollary}{C.\arabic{corollary}} \renewcommand{\thefigure}{C.%
\arabic{figure}} \renewcommand{\thetable}{C.\arabic{table}} %
\setcounter{table}{0} \setcounter{figure}{0} \setcounter{equation}{0} %
\setcounter{theorem}{0} \setcounter{lemma}{0} \setcounter{proposition}{0} %
\setcounter{section}{0} \setcounter{subsection}{0}

\section*{C. More Simulation Results and Findings}

\renewcommand{\thesubsection}{\thesection.\arabic{subsection}}

\section{A simulation study to evaluate the knots selection methods}

In this section, we conduct a simulation study to evaluate the performance of the knots selection methods proposed in Section \ref{subsec:knots select} in the main paper. The setting of the simulation is the same as in Section \ref{subsec:general-eg} in the main paper. For model fitting, the mean function is estimated by cubic splines, and the number of knots of the splines, $J_s$, is selected using either the formula-based method (Formula) and the GCV method (GCV) described in Section \ref{subsec:knots select}.  Each simulation is repeated $500$ times.

Figure \ref{FIG:hist} below shows the frequency bar plot of the GCV-selected $J_{s}$ over $500$ replications, where  the black triangles indicate the number of knots suggested using the formula given in Section \ref{subsec:knots select}. From Figure \ref{FIG:hist}, one sees that on average the GCV method tends to select a slightly larger number of knots than the formula method does, but both methods provide similar results as shown in Tables \ref{TAB:AMSE-SCB-general} and \ref{TAB:hetero-AMSE-SCB-genaral} in the main paper. The GCV method is indeed more time-consuming than the formula method. For example, in scenario $N=50$ and $\sigma_{\epsilon}=0.1$ of Table \ref{TAB:AMSE-SCB-general} , it takes 50 seconds for the formula and 9 minuses for GCV selected method, respectively.

\begin{figure}[tbp]
\centering
\subfigure[]{
\hspace*{0in} \includegraphics[trim={2cm 0.5cm 2cm 0cm}, width=3.5in, height=3.5in]{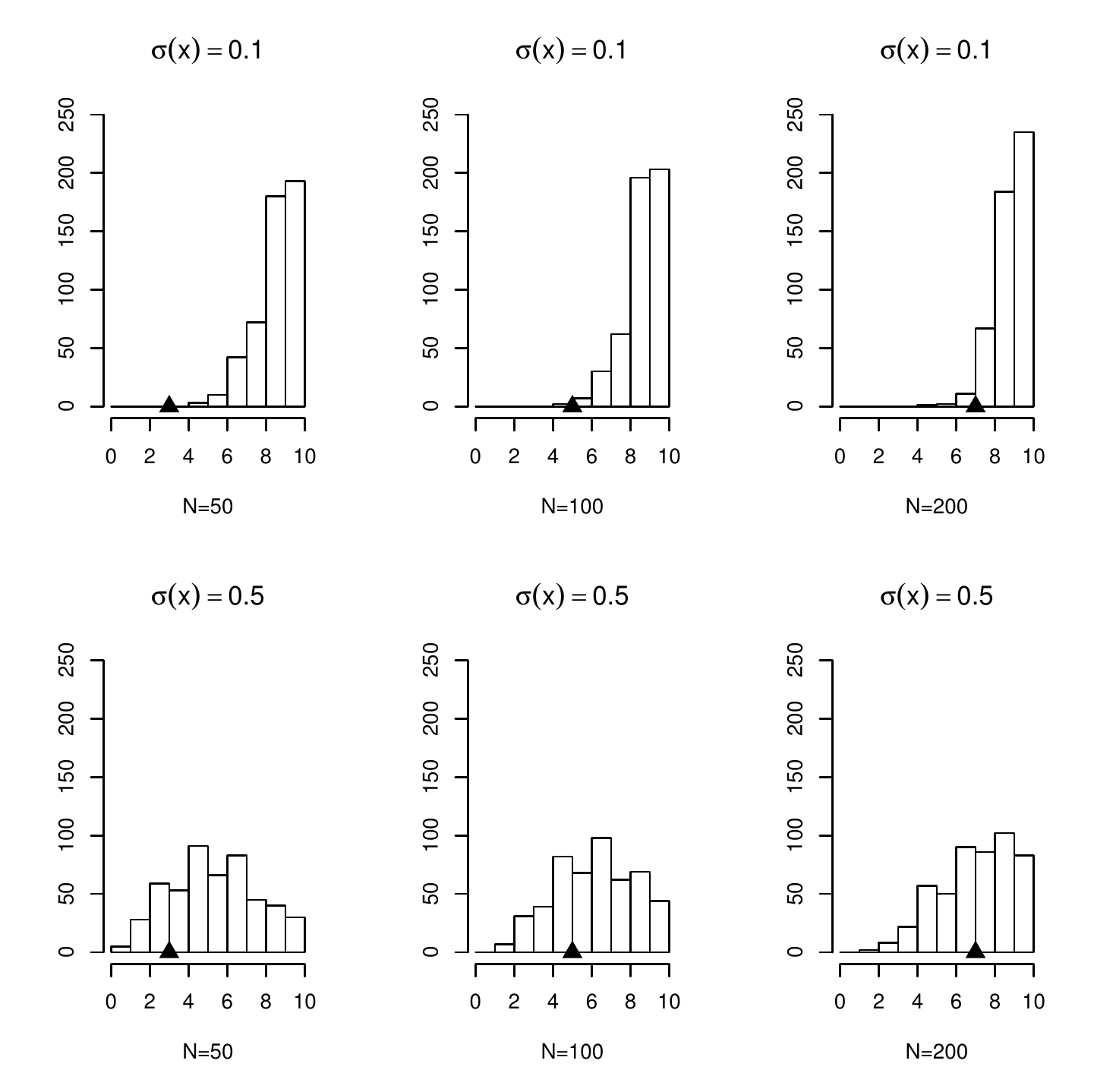}}
\par
\subfigure[]{
\hspace*{0in} \includegraphics[trim={2cm 0.5cm 2cm 0cm}, width=3.5in, height=3.5in]{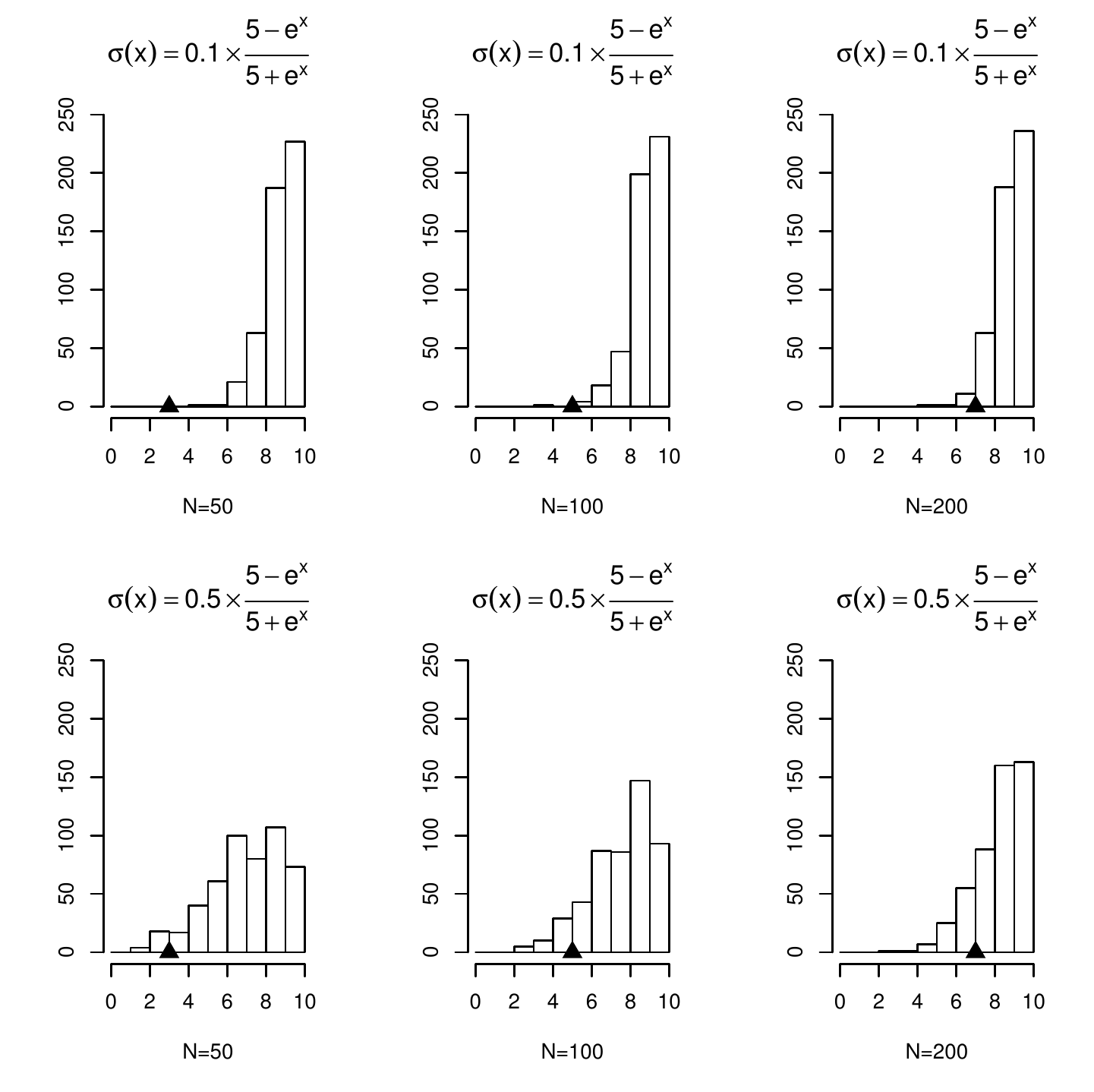}}
\caption{Bar graphs of the GCV selected number of knots  in $500$ replications with (a) homogeneous and (b) heteroscedastic errors. Black triangles indicate the number of knots suggested by the formula.}
\label{FIG:hist}
\end{figure}


\section{More results for spatial covariance models}

Tables \ref{TAB:SSAMSE-SCB-spatial-2}--\ref{TAB:SSAMSE-SCB-formula-2} report some simulation results based on the spatial covariance model presented in Section \ref{subsec:spatial} in the main paper. Specifically, we report the simulation results based on the data generated from the model with the heteroscedastic errors: $\sigma (x)=\sigma_{\epsilon }\left\{ 5+\exp \left( x\right) \right\} ^{-1}\left\{ 5-\exp\left( x\right) \right\} $ for M1, and $\sigma (x)=\sigma _{\epsilon}\left\{ 30+\exp \left( x/2\right) \right\} ^{-1}\left\{ 30-\exp \left(x/2\right) \right\} $ for M2 and M3. The number of curves $n=\lfloor0.8N\rfloor $ with $N=50$, $100$ and $200$, and the noise levels are $\sigma_{\epsilon}=0.1,~0.5$. The mean function is estimated by cubic splines, i.e., $p=4$, with the number of knots selected using the formula method.

The AMSE of the covariance estimators $\widehat{C}$ and $\widetilde{C}$ are reported in columns 3--4 of Table \ref{TAB:SSAMSE-SCB-spatial-2}. The performance of the two estimators is very similar. Columns 5 and 7 present the empirical coverage rate {CR}, i.e., the percentage of the true curve $C(\cdot)$ entirely covered by the SCB, based on $95\%$ and $99\%$ confidence levels, respectively. As the sample size increases, the coverage probability of the SCB becomes closer to the nominal level. Columns 3--4 of Table \ref{TAB:SSAMSE-SCB-formula-2} present the AMSEs of $\widehat{G}^{\mathrm{PROP}}(x,x^{\prime })$ and $\widehat{G}^{\mathrm{TPS}}(x,x^{\prime })$. The results of AMSEs indicate that $\widehat{G}^\mathrm{PROP}$ is more accurate than $\widehat{G}^{\mathrm{TPS}}$, while $\widehat{G}^{\mathrm{TPS}}$ usually gives larger AMSE. Columns 5--12 of Table \ref{TAB:AMSE-SCB-formula} report the CR and WD of SCE-I and SCE-II. One sees that the CRs of SCE-I are much closer to the nominal levels than those of SCE-II, and increasing the sample size helps to improve the CR of the SCEs to their nominal levels. One also observes the widths of the SCE-I are much narrower than those of the SCE-II.

\begin{table}[tbp]
\caption{Simulation results based on heteroscedastic errors with  $\sigma (x)=\sigma _{\epsilon}\left\{5+\exp \left( x\right) \right\} ^{-1}\left\{5-\exp \left(x\right) \right\} $ for M1, and $\sigma (x)=\sigma _{\epsilon}\left\{30+\exp \left( x/2\right) \right\} ^{-1}\left\{30-\exp \left(x/2\right) \right\} $ for M2 and M3: AMSE of the estimators $\protect\widehat{C}$, $\protect\widetilde{C}$; CR (outside/inside of the parentheses is based on $\protect\widehat{C}$, $\protect\widetilde{C}$), and the WD of the asymptotic SCBs based on $\protect\widehat{C}$.}
\label{TAB:SSAMSE-SCB-spatial-2}
\renewcommand*{\arraystretch}{0.5}
\par
\centering
\resizebox{150mm}{40mm}{
\begin{tabular}{ccccccccccc}
\toprule
\multirow{3}{*}{$\sigma _{\epsilon}$} & \multirow{3}{*}{Model} & \multirow{3}{*}{$N$} & \multicolumn{2}{c}{AMSE}
&  & \multicolumn{5}{c}{SCB} \\ \cline{4-5}\cline{7-11}
&  &  & \multirow{2}{*}{$\widehat{C}$} & \multirow{2}{*}{$\widetilde{C}$} &  &
\multicolumn{2}{c}{$95\%$} &  & \multicolumn{2}{c}{$99\%$} \\ \cline{7-8} \cline{10-11}
&  &  &  &  &  & CR & WD &  & CR & WD \\
\midrule

\multirow{9}{*}{$0.1$} & \multirow{3}{*}{M1} & $50 $ & $0.082$ & $0.081$ &  & $0.916(0.920)$ & $1.37$ &  & $0.958(0.966)$ & $1.68$ \\
& & $100$ & $0.040$ & $0.040$ &  & $0.922(0.926)$ & $0.99$ &  & $0.974(0.978) $ & $1.21$ \\
& & $200$ & $0.019$ & $0.018$ &  & $0.940(0.952)$ & $0.72$ &  & $0.980(0.986) $ & $0.87$ \\  \cline{2-11}
& \multirow{3}{*}{M2} & $50 $ & $0.095$ & $0.096$ &  & $0.906(0.904)$ & $1.44$ &  & $0.950(0.954)$ & $1.78$ \\
& & $100$ & $0.048$ & $0.049$ &  & $0.922(0.926)$ & $1.05$ &  & $0.980(0.976)$ & $1.30$ \\
& & $200$ & $0.022$ & $0.022$ &  & $0.958(0.958)$ & $0.76$ &  & $0.992(0.994)$ & $0.94$ \\  \cline{2-11}
& \multirow{3}{*}{M3} & $50$ & $0.109$ & $0.109$ &  & $0.904(0.908)$ & $1.50$ &  & $0.954(0.956)$ & $1.86$ \\
& & $100$ & $0.055$ & $0.055$ &  & $0.922(0.928)$ & $1.09$ &  & $0.976(0.978)$ & $1.35$ \\
& & $200$ & $0.025$ & $0.025$ &  & $0.960(0.958)$ & $0.79$ &  & $0.988(0.990)$ & $0.98$ \\
\midrule

\multirow{3}{*}{$0.5$} & \multirow{3}{*}{M1} & $50 $ & $0.082$ & $0.081$ &  & $0.898(0.918)$ & $1.38$ &  & $0.962(0.970)$ & $1.69$ \\
& & $100$ & $0.040$ & $0.040$ &  & $0.918(0.924)$ & $0.99$ &  & $0.976(0.980)$ & $1.21$ \\
& & $200$ & $0.019$ & $0.018$ &  & $0.948(0.952)$ & $0.72$ &  & $0.982(0.986)$ & $0.88$ \\  \cline{2-11}
& \multirow{3}{*}{M2} & $50$ & $0.096$ & $0.096$ &  & $0.904(0.908)$ & $1.45$ &  & $0.948(0.958)$ & $1.79$ \\
& & $100$ & $0.048$ & $0.049$ &  & $0.916(0.926)$ & $1.06$ &  & $0.974(0.980)$ & $1.30$ \\
& & $200$ & $0.022$ & $0.022$ &  & $0.960(0.956)$ & $0.77$ &  & $0.990(0.994)$ & $0.94$ \\   \cline{2-11}
& \multirow{3}{*}{M3} & $50$ & $0.110$ & $0.109$ &  & $0.904(0.908)$ & $1.51$ &  & $0.948(0.958)$ & $1.87$ \\
& & $100$ & $0.055$ & $0.055$ &  & $0.910(0.924)$ & $1.10$ &  & $0.976(0.976)$ & $1.36$ \\
& & $200$ & $0.025$ & $0.025$ &  & $0.960(0.958)$ & $0.79$ &  & $0.988(0.990)$ & $0.98$ \\
\bottomrule
\end{tabular}}
\end{table}

\begin{table}[tbp]
\renewcommand*{\arraystretch}{0.5}
\caption{Simulation results based on heteroscedastic errors with $\sigma (x)=\sigma _{\epsilon}\left\{5+\exp \left( x\right) \right\} ^{-1}\left\{5-\exp \left(x\right) \right\} $ for M1, and $\sigma (x)=\sigma _{\epsilon}\left\{30+\exp \left( x/2\right) \right\} ^{-1}\left\{30-\exp \left(x/2\right) \right\} $ for M2 and M3: AMSE of the estimators $\widehat{G}^{\mathrm{PROP}}(\cdot,\cdot)$, $\widehat{G}^{\mathrm{TPS}}(\cdot,\cdot)$; CR and WD of SCE-I and SCE-II.}
\label{TAB:SSAMSE-SCB-formula-2}
\centering
\vspace{0.1cm}
\resizebox{150mm}{40mm}{
\begin{tabular}{ccc ccc ccc ccc ccc}
\toprule
\multirow{3}{*}{$\sigma _{\epsilon}$} & \multirow{3}{*}{Model} & \multirow{3}{*}{$N$} & \multicolumn{2}{c}{AMSE} & & \multicolumn{4}{c}{SCE-I} & & \multicolumn{4}{c}{SCE-II} \\ \cline{4-5} \cline{7-10} \cline{12-15}
& & & \multirow{2}{*}{$\widehat{G}^{\mathrm{PROP}}$} & \multirow{2}{*}{$\widehat{G}^{\mathrm{TPS}}$} & & \multicolumn{2}{c}{$95\%$} & \multicolumn{2}{c}{$99\%$} & & \multicolumn{2}{c}{$95\%$} & \multicolumn{2}{c}{$99\%$} \\
\cline{7-10} \cline{12-15}
&  &  &  &  &  & CR & WD & CR & WD & & CR & WD & CR & WD\\

\midrule

\multirow{9}{*}{$0.1$} &  \multirow{3}{*}{M1} & $50$ & $0.079$ & $0.121$ & & $0.916$ & $1.40$ & $0.958$ & $1.71$ & & $0.722$ & $2.07$ & $0.838$ & $2.55$\\
&  & $100$ & $0.039$ & $0.061$ & & $0.922$ & $1.01$ & $0.974$ & $1.24$ & & $0.850$ & $1.64$ & $0.952$ & $2.03$\\
&  & $200$ & $0.018$ & $0.032$ & & $0.940$ & $0.73$ & $0.980$ & $0.90$ & & $0.880$ & $1.19$ & $0.964$ & $1.46$\\
\cline{2-15}

& \multirow{3}{*}{M2} & $50$ & $0.097$ & $0.150$ & & $0.906$ & $1.50$ & $0.950$ & $1.86$ & & $0.710$ & $2.08$ & $0.830$ & $2.56$\\
&  & $100$ & $0.048$ & $0.071$ & & $0.922$ & $1.10$ & $0.980$ & $1.35$ & & $0.786$ & $1.63$ & $0.900$ & $2.00$\\
&  & $200$ & $0.022$ & $0.036$ & & $0.958$ & $0.79$ & $0.992$ & $0.98$ & & $0.930$ & $1.16$ & $0.976$ & $1.43$\\
\cline{2-15}

& \multirow{3}{*}{M3} & $50$ & $0.114$ & $0.153$ & & $0.904$ & $1.57$ & $0.954$ & $1.95$ & & $0.720$ & $2.15$ & $0.816$ & $2.64$\\
&  & $100$ & $0.057$ & $0.074$ & & $0.922$ & $1.15$ & $0.976$ & $1.42$ & & $0.856$ & $1.52$ & $0.944$ & $1.86$\\
&  & $200$ & $0.026$ & $0.039$ & & $0.960$ & $0.83$ & $0.988$ & $1.03$ & & $0.878$ & $1.09$ & $0.956$ & $1.33$\\

\midrule
\multirow{9}{*}{$0.5$} & \multirow{3}{*}{M1} & $50$ & $0.079$ & $0.129$ & & $0.902$ & $1.41$ & $0.964$ & $1.73$ & & $0.730$ & $2.10$ & $0.834$ & $2.57$\\
&  & $100$ & $0.039$ & $0.064$ & & $0.920$ & $1.02$ & $0.980$ & $1.25$ & & $0.810$ & $1.65$ & $0.918$ & $2.03$\\
&  & $200$ & $0.018$ & $0.035$ & & $0.942$ & $0.74$ & $0.984$ & $0.90$ & & $0.914$ & $1.21$ & $0.980$ & $1.49$\\
\cline{2-15}

& \multirow{3}{*}{M2} & $50$ & $0.097$ & $0.143$ & & $0.904$ & $1.51$ & $0.948$ & $1.86$ & & $0.674$ & $2.10$ & $0.802$ & $2.57$\\
&  & $100$ & $0.048$ & $0.071$ & & $0.916$ & $1.10$ & $0.972$ & $1.35$ & & $0.774$ & $1.63$ & $0.894$ & $2.01$\\
&  & $200$ & $0.022$ & $0.036$ & & $0.960$ & $0.79$ & $0.990$ & $0.98$ & & $0.924$ & $1.16$ & $0.970$ & $1.43$\\
\cline{2-15}

& \multirow{3}{*}{M3} & $50$ & $0.114$ & $0.149$ & & $0.900$ & $1.58$ & $0.948$ & $1.96$ & & $0.696$ & $2.12$ & $0.820$ & $2.60$\\
&  & $100$ & $0.057$ & $0.075$ & & $0.910$ & $1.15$ & $0.974$ & $1.42$ & & $0.830$ & $1.52$ & $0.930$ & $1.87$\\
&  & $200$ & $0.026$ & $0.038$ & & $0.960$ & $0.83$ & $0.988$ & $1.03$ & & $0.876$ & $1.09$ & $0.944$ & $1.33$\\
\bottomrule
\end{tabular}}
\end{table}
\setcounter{chapter}{9} \renewcommand{\theequation}{D.\arabic{equation}} %
\renewcommand{\thesection}{D.\arabic{section}} \renewcommand{%
\thesubsection}{D.\arabic{subsection}} \renewcommand{\thetheorem}{D.%
\arabic{theorem}} \renewcommand{\thelemma}{D.\arabic{lemma}} %
\renewcommand{\theproposition}{D.\arabic{proposition}} \renewcommand{%
\thecorollary}{D.\arabic{corollary}} \renewcommand{\thefigure}{D.%
\arabic{figure}} \renewcommand{\thetable}{D.\arabic{table}} %
\setcounter{table}{0} \setcounter{figure}{0} \setcounter{equation}{0} %
\setcounter{theorem}{0} \setcounter{lemma}{0} \setcounter{proposition}{0} %
\setcounter{section}{0} \setcounter{subsection}{0}

\section*{D. Additional Real Data Analysis}
\label{APP:realdata}
\section{Biscuit Dough Piece Data}

In the following, we apply the methodology to a ``biscuit dough piece data",
which is an experiment involved varying the composition of biscuit dough
pieces by measuring the quantitative near infrared reflectance (NIR)
spectroscopy. Quantitative NIR spectroscopy plays an important role in
analyzing such diverse materials as food and drink, pharmaceutical products,
and petrochemicals. The NIR spectrum of a sample of, say, wheat flour is a
continuous curve measured by modern scanning instruments at hundreds of
equally spaced wavelengths. Then the information contained in the curve can
be used to predict the chemical composition of the sample. For a full
description of the experiment, see \cite{OFMD84}. This dataset is available
in the \textsf{R} package ``fds"({\texttt{https://cran.r-project.org/web/
packages/fds/fds.pdf}}), which contains several subsets such as ``nirc"
(calibration) and ``nirp" (prediction).

We focus on the calibration set ``nirc", which contains $40$ doughs and $700$
point NIR spectra for each dough. According to \cite{BFV01}, the observation
number 23 in the calibration set appears as an outlier. Thus we remove it
and take the other samples (doughs). Hence, there are measurements on $n=39$
samples, where for each sample $N=700$ spectral was recorded every $2$
nanometre (nm) with wavelength being measured on $[1100,2498]$. Denote by $%
Y_{ij}$ the spectral of the $i$th sample at the wavelength $x_{j}$, $%
j=1,\ldots ,N$ and $i=1,\ldots ,n$.

\begin{figure}[tbp]
\begin{center}
\includegraphics[width=3.5in,height=2.2in]{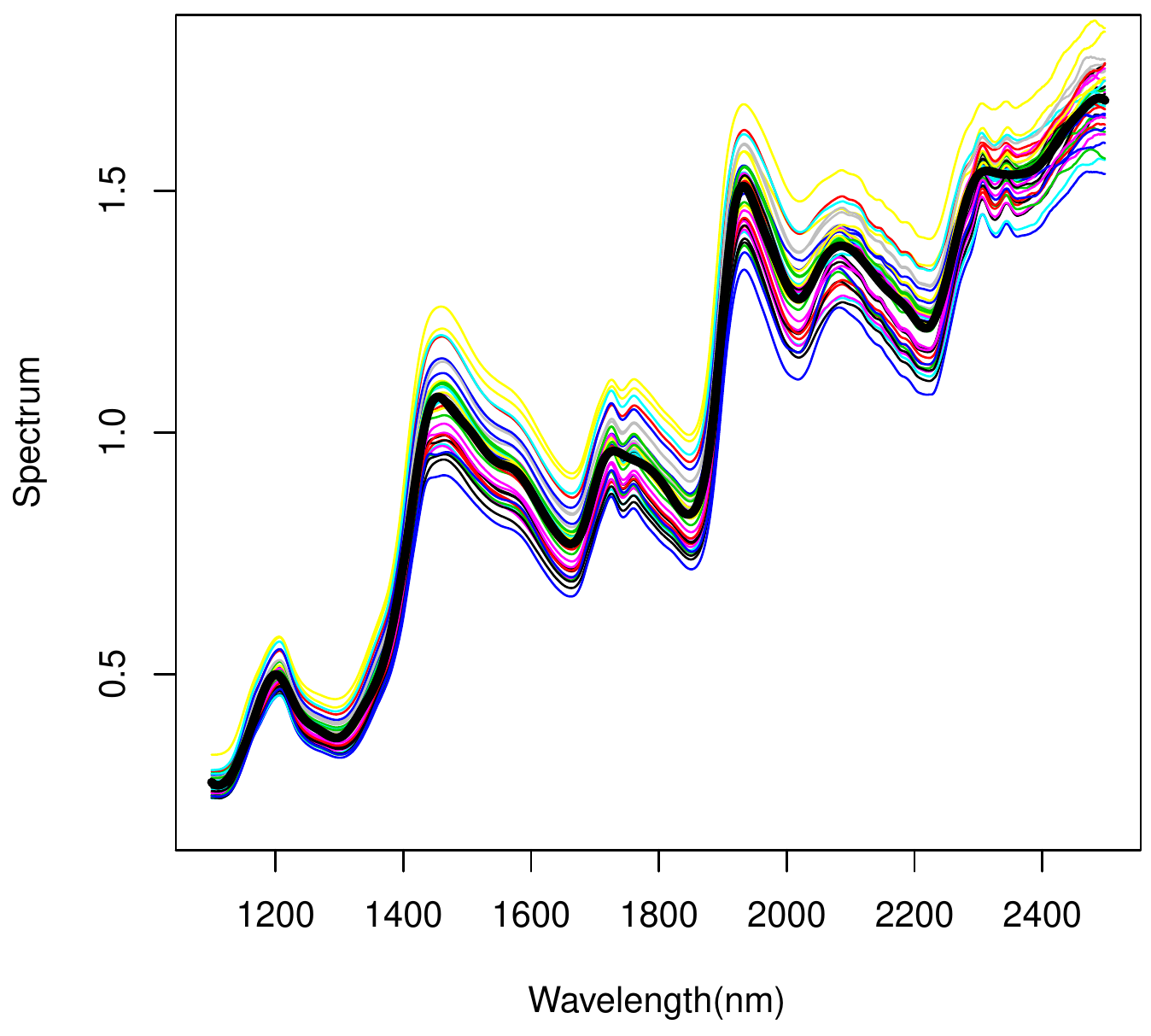} \vskip -.2in
\caption{Biscuit dough pieces data curves (thin solid lines) with their mean
function estimator (thick solid line).}
\label{Fig: Bis_mean and level}
\end{center}
\end{figure}

\begin{figure}[tbp]
\begin{center}
\subfigure[]{\label{fig:subfig:a}
\includegraphics[width=2.75in, height=2in]{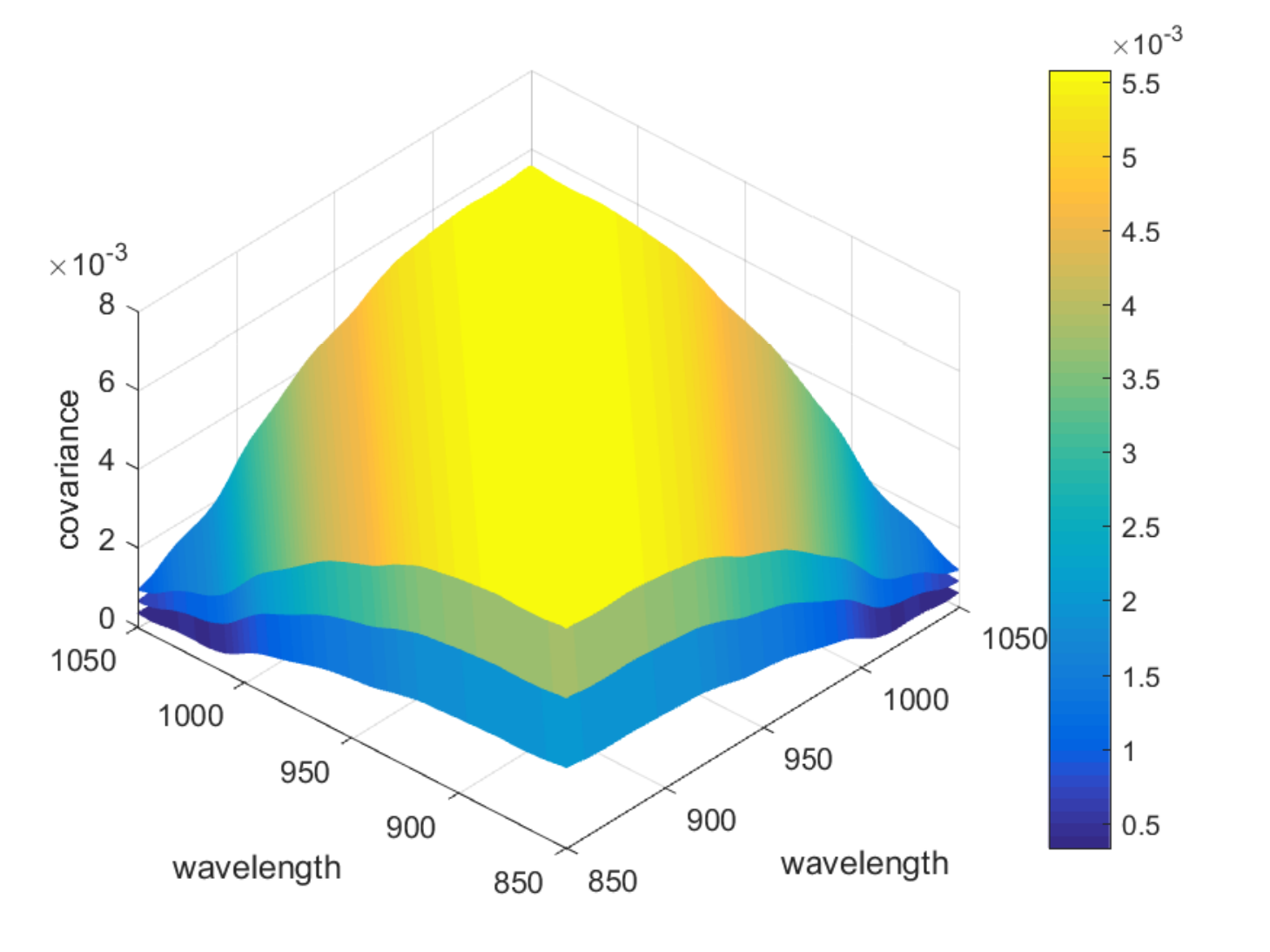}}
\hspace{0.2in}
\subfigure[]{\label{fig:subfig:b}
\includegraphics[width=2.75in, height=2in]{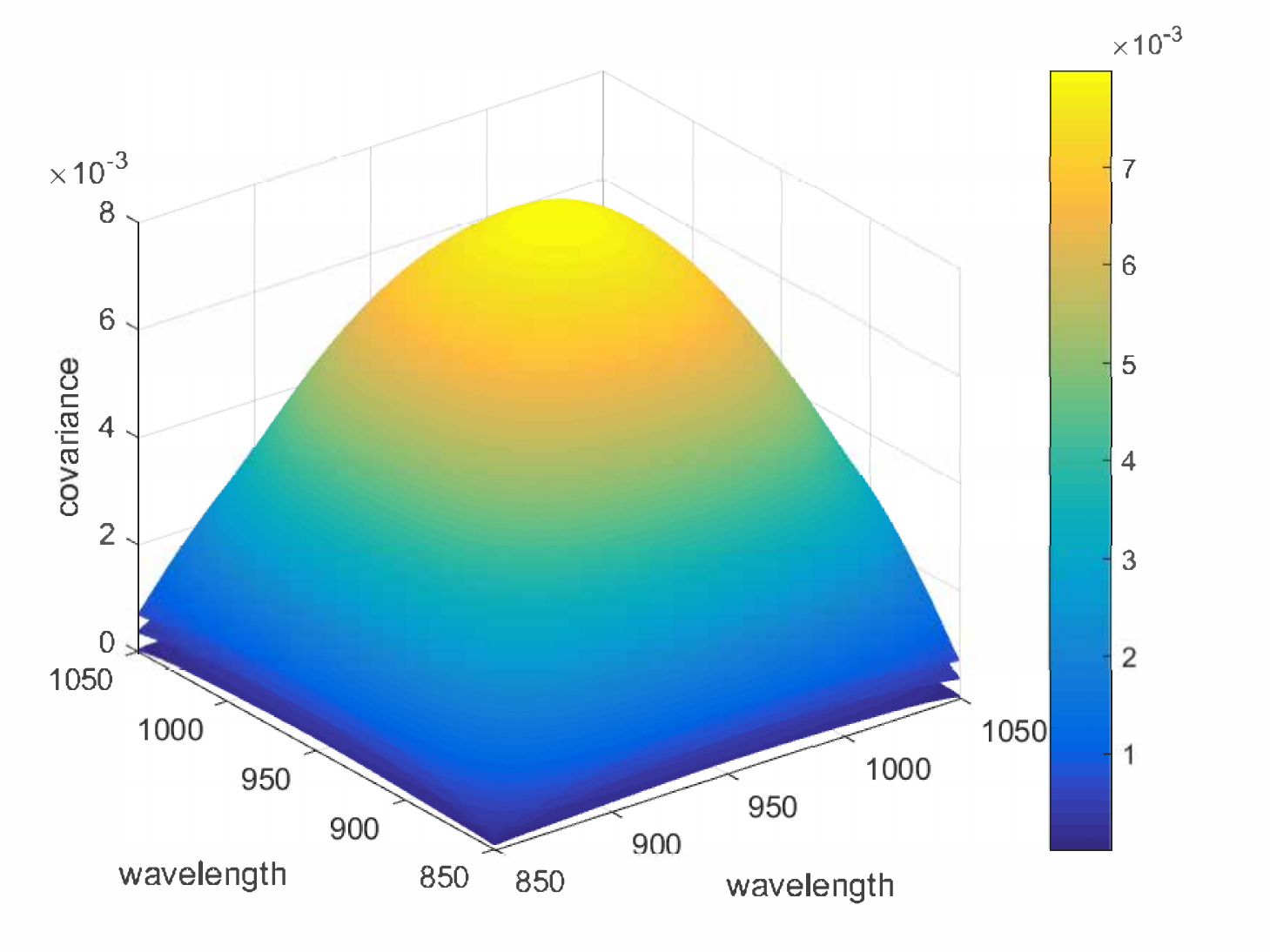}}
\end{center} \vskip -.25in
\caption{Biscuit dough piece data plots: (a) covariance estimator $\protect%
\widehat{G}^\mathrm{PROP}$ and its 95\% SCE;
(b) covariance estimator $\protect\widehat{G}^{TPS}$ and its 95\% SCE.}
\label{Fig: Bis_SCE}
\end{figure}

Figure \ref{Fig: Bis_mean and level} displays NIR spectra curves together
with their estimated mean curve, and ``Wavelength (nm)" is plotted on $x$%
-axis and ``Spectrum" on the $y$-axis. An illustrative 3D plot of SCE of the
Biscuit dough piece data is depicted is in Figure \ref{Fig: Bis_SCE}: (a)
based on $\widehat{G}^\mathrm{PROP}$ (middle) and its 95\% SCE (upper and
lower); (b) based on $\widehat{G}^\mathrm{TPS}$ (middle) with its $95\%$
asymptotic confidence envelop (up and below).

\bibliographystyle{asa}
\bibliography{references}

\end{document}